\documentclass[%
 aip,
 amsmath,amssymb,
preprint,%
author-numerical,%
]{revtex4-1}

\usepackage{float}
\floatstyle{boxed} 

\usepackage{graphicx}
\usepackage{epstopdf, epsfig}
\usepackage[dvipsnames]{xcolor}

\usepackage{tikz,siunitx} % for legends and etc
\usepackage{pgfplots}

\usepackage{longtable}
\usepackage{mwe} % for the image to be in the same page
\setcitestyle{authoryear,square,semicolon,sort}
\usepackage{amsmath, amssymb}
\usepackage{mathrsfs}
\usepackage{placeins}
\usepackage{upgreek}
\usepackage{dirtytalk}

\usepackage{xkeyval,xcolor}% http://ctan.org/pkg/{xkeyval,xcolor}
\usepackage{natbib}
\newcommand\cites[1]{\citeauthor{#1}'s\ (\citeyear{#1})}
\usepackage[utf8]{inputenc} % To be able to write ~ in the first place

\DeclareUnicodeCharacter{00A0}{~}

\def\mathclap#1{\text{\hbox to 0pt{\hss$\mathsurround=0pt#1$\hss}}}
\makeatletter
\newlength{\sfp@hseplen}\newlength{\sfp@vseplen}
\define@cmdkey{subfigpos}[sfp@]{pos}[ul]{}% \sfp@pos
\define@cmdkey{subfigpos}[sfp@]{font}[\small]{}% \sfp@font
\define@cmdkey{subfigpos}[sfp@]{vsep}[1\baselineskip]{\setlength{\sfp@vseplen}{\sfp@vsep}}% \sfp@vsep
\define@cmdkey{subfigpos}[sfp@]{hsep}[35pt]{\setlength{\sfp@hseplen}{\sfp@hsep}}% \sfp@hsep
\newcommand{\subfigimg}[3][,]{%
  \setkeys{Gin,subfigpos}{pos,font,vsep,hsep,#1}% Set (default) keys
  \setbox1=\hbox{\includegraphics{#3}}% Store image in box
  \ifnum\pdfstrcmp{\sfp@pos}{ul}=0% UPPER LEFT placement of subfig label
    \leavevmode\rlap{\usebox1}% Print image
    \rlap{\hspace*{\sfp@hsep}\raisebox{\dimexpr\ht1-\sfp@vsep}{\sfp@font{#2}}}% Print label
    \phantom{\usebox1}% Insert appropriate spacing
  \else\ifnum\pdfstrcmp{\sfp@pos}{ur}=0% UPPER RIGHT placement of subfig label
    \leavevmode\usebox1% Print image
    \llap{\raisebox{\dimexpr\ht1-\sfp@vsep}{\sfp@font{#2}}\hspace*{\sfp@hsep}}% Print label
  \else\ifnum\pdfstrcmp{\sfp@pos}{lr}=0% LOWER RIGHT placement of subfig label
    \leavevmode\usebox1% Print image
    \llap{\raisebox{\sfp@vsep}{\sfp@font{#2}}\hspace*{\sfp@hsep}}% Print label
  \else% Assume LOWER LEFT placement of subfig label
    \leavevmode\rlap{\usebox1}% Print image
    \rlap{\hspace*{\sfp@hseplen}\raisebox{\sfp@vsep}{\sfp@font{#2}}}% Print label
    \phantom{\usebox1}% Insert appropriate spacing
  \fi\fi\fi
}
\makeatother

\begin{document}

\preprint{AIP/123-QED}

\title[]{Aeroacoustic investigation of airfoil at near stall conditions}
% Force line breaks with \\

\author{Prateek Jaiswal}
\email{jaiswalprateek@protonmail.com}
 \altaffiliation[]{Department of Mechanical Engineering, University of Sherbrooke, Sherbrooke, QC, CA.}%Lines break automatically or can be forced with \\
 \author{Jose Rendón}
 \altaffiliation[]{Department of Mechanical Engineering, University of Sherbrooke, Sherbrooke, QC, CA.}

 %\author{Sidharth Krishnan Kalyani}
 %\altaffiliation[]{Department of Mechanical Engineering, University of Sherbrooke, Sherbrooke, QC, CA.} 
 
\author{St{\'e}phane Moreau}%
\altaffiliation[]{Department of Mechanical Engineering, University of Sherbrooke, Sherbrooke, QC, CA.}%\\This line break forced with \textbackslash\textbackslash
%

%\author{Daniele Ragni}
% \homepage{https://www.tudelft.nl/en/ae/}
%\affiliation{Faculty of Aerospace Engineering, Delft University of Technology, Delft 2629HS, The Netherlands}%

%\author{Stefan Pr\"{o}bsting}
% \homepage{http://en.sjtu.edu.cn/academics/schools/the-school-of-naval-architecture-ocean-and-civil-engineering/}
 %\affiliation{Faculty of Aerospace Engineering, Delft University of Technology, Delft 2629HS, The Netherlands}%
%\affiliation{Department of Naval Architecture and Ocean Engineering, Shanghai Jiao Tong University, Shanghai 200240, China}%

\date{\today}% It is always \today, today,

\begin{abstract}
This paper presents a detailed aeroacoustic investigation of a Controlled-Diffusion airfoil at near stall condition. The study aims at answering two research questions: identify the flow mechanism responsible for separation noise for an airfoil near stall conditions and whether the noise is generated by a dipole for airfoil close to stall and can be quantified by Amiet's diffraction theory. The study uses synchronized PIV, RMP and far-field microphone measurements to perform experiments at two chord based Reynolds numbers of %140,000 and 245,000
about 150,000 and 250,000. The results show that when the airfoil is placed at a higher angle of attack, such as $15^{\circ}$, strong amplification of flow disturbance is seen, resulting in the rolling up of the shear layer in the aft-region of the airfoil, forming %the rollers.
large coherent structures. While these rollers play a central role in the increase in noise due to flow separation, the flapping of shear layer does not contribute to the separation noise. The present study conclusively shows that separation noise is dipolar in nature, and that the quadrupolar contribution for low-speed airfoils at near-stall conditions can be neglected. However, the increase in flow disturbances measured close to the trailing-edge of the airfoil implies that the assumption of small amplitude disturbance is no longer valid, which is the central premise of the thin linearized airfoil theory. Outside the frequency range at which flow separation operates, Amiet's theory %should be 
is able to predict the far-field noise even at high angles of attack.
\end{abstract}

\maketitle

%\begin{quotation}
%The ``lead paragraph'' is encapsulated with the \LaTeX\ 
%\verb+quotation+ environment and is formatted as a single paragraph before the first section heading. 
%(The \verb+quotation+ environment reverts to its usual meaning after the first sectioning command.) 
%Note that numbered references are allowed in the lead paragraph.
%
%The lead paragraph will only be found in an article being prepared for the journal \textit{Chaos}.
%\end{quotation}

\section*{Nomenclature}

{\renewcommand\arraystretch{1.0}
\noindent\begin{longtable*}{@{}l  l@{}}
$C$ & {Airfoil} chord\\
$c_0$ & Speed of sound\\
$C_p$ & Mean pressure {coefficient} \\
$C_{prms}$ & Root-mean-square of the wall-pressure {coefficient} \\
$\overline{E_{11}}$  &  Pre-multiplied turbulent energy spectra\\
$ H $    & Boundary layer shape factor\\
$M_{\infty}$ & Inlet Mach number\\
$p_\infty$ & Inlet static pressure\\
{$p_{rms}$} & root-mean-square of the wall pressure \\
$p^{\prime}_a$ & Far-field acoustic pressure\\
$p^{\prime}_w$ & Fluctuating wall-pressure\\
$R_{ij}$ & Second order two-point zero time delay correlation \\
$Re_{c}$ & Reynolds number based on the chord\\
$S_{pp}$ & Far-field acoustic power spectral density\\
$u_{i}$ & Fluctuating velocity component\\
$U_{c}$ & Convective speed of wall-pressure fluctuations\\
$U_{\infty}$ & Inlet velocity\\
$U_e$ & Boundary layer edge velocity\\
$U_1,U_2,U_3$ & Mean velocity in trailing edge reference frame\\
$\overline{u_1u_1},\overline{u_2u_2},\overline{-u_1 u_2}$ & Root-mean-square of velocity fluctuations in trailing edge reference frame\\
{${-\rho~\overline{u_1 u_2}_{max}}$} & {maximum Reynolds shear stress} \\
$V_x,V_y$ & Mean velocity in wind tunnel reference frame\\
$x,y,z$ & Wind tunnel coordinate system\\
$x_{1},x_{2},x_{3}$ & Coordinate system aligned with the airfoil trailing edge\\ 
$x'_{1},x'_{2},x'_{3}$ & Coordinate system aligned with the airfoil leading edge\\ 
%$\alpha_g$ & Geometric angle of attack\\
%$S_{0}$ & $\sqrt{\beta_{2}^{2}x_{1}'^{2} + \beta_{1}^{2}x_{2}'^{2} + \beta_{0}^{2}x_{3}'^{2}}$, position of the observer corrected for compressibility effects\\
%$X_{1}',X_{2}',X_{3}'$ & coordinates normalised by the half-chord\\ 
%$U_{1}$ & projection of mean flow speed on $x_{1}'$ axis\\
%$U_{2}$ & projection of mean flow speed on $x_{2}'$ axis\\
$\delta_{95}$ & Boundary layer thickness based on $95\%$ of $U_e$\\
$\delta^{*}$ & Boundary layer displacement thickness\\
$\Lambda$ & Dimensionless radiation ratio \\
%$\delta()$ & Dirac function\\
$\theta$ & Boundary layer momentum thickness\\
%$\mu$ & dynamic viscosity\\
%$\nu$ & kinematic viscosity\\
%$\upi_{0}$ & wall-parallel-integrated wavenumber spectral density of wall-pressure fluctuations\\
%$\upi$ & wavenumber cross-spectral density of wall-pressure fluctuations\\
$\rho$ & Constant air density\\
%$\sigma^{2}$ & variance of a homogeneous turbulence field\\
%$\Pi_c$ & Cole's wake strength parameter\\
%$\beta_c$ & Clauser's parameter  \\
%$\phi$ & wall-pressure auto-spectrum (Goody)\\
\end{longtable*}

% short intro to why its important
%\end{widetext}

\section{Introduction}

Airfoil trailing-edge noise is dominant in a host of engineering applications. Several of the distinct mechanisms, 
which are referred as airfoil self-noise, are related to scattering of pressure gust past the airfoil trailing edge. Among them, noise due to flow separation is found to be dominant at high angles of attack, where large scale flow separation may occur. This is particularly the case for some wind turbine architectures, such as the H-Darrieus type wind turbine \citep{venkatraman2023numerical}. Therefore, accurate models are needed during the pre-design phase to estimate acoustic noise by such machines. To achieve this a better understanding of the noise generation mechanism is needed. Nevertheless, only few comprehensive aeroacoustic studies have been performed, for airfoils placed at high angles of attack~\citep{moreau2009back,lacagnina2019mechanisms,zang2021experimental,kalyani2022flow,Raus2022}. As such, the overall objectives of the present manuscript are to identify the dominant flow mechanism(s) responsible for separation noise, and test the applicability of diffraction theory \citep{amiet1976noise} to predict noise at high angles of attack.  

Numerically, Moreau and co-workers had performed several high-fidelity incompressible simulations for airfoil at high incidence almost a decade ago \citep{christophe2008,moreau2008trailing,stall}. In these simulations, an isolated airfoil installed in an open-jet anechoic wind tunnel (the test configuration) was simulated as opposed to the full scale wind turbine \citep{venkatraman2023numerical}. As such only the noise due to the boundary-layer and its separation were studied. The far-field noise was quantified using both acoustic analogies \citep{curle1955influence,williams1970aerodynamic} and \cites{amiet1976noise} diffraction theory. \cite{christophe2008} reported over-prediction of the wall-pressure by the LES, and the far-field acoustic spectra estimated by \cites{amiet1976noise} model to be 10 dB higher than the measurements. In particular, this disagreement was present only at the low frequency, where the noise due to separation is expected to be the dominant mechanism. Similarly, the semi-empirical models for far-field noise based on \cites{amiet1976noise} theory, referred to as MODA \citep{bertagnolio2017semi}, have been shown to yield poor results. However, the reason for this disagreement when predicting separation noise with  \cites{amiet1976noise} model is unknown and requires further investigation.

More recently, compressible simulations have been performed by \cite{turner2022quadrupole} to quantify the individual contributions of equivalent source type (dipole and quadrupole) from low-speed airfoils in near stall conditions. They achieve this by subtracting noise estimated by the solid formulation of \cites{williams1969sound} acoustic analogy from the noise estimated by the permeable formulation. \cite{turner2022quadrupole} show that the noise contribution by quadrupole sources is %non-negligible
significant, when an airfoil is placed at high incidence. However, while the porous formulation is complete, the solid formulation ignores the correlation between the dipole and quadrupole noise sources. This can lead to spurious directivity patterns as already demonstrated by \cite{spalart2019differences}. Nevertheless, it is important to quantify the individual contributions of various equivalent source types that may contribute to far-field noise.   

While equivalent noise sources are an important metric in aeroacoustics research, they are by no means unique. This is because the multipole expansion \citep{goldstein1976aeroacoustics} dictates that one equivalent image source can be replaced by another. For instance, a quadrupole can be expressed as two dipoles that are of equal strength but in phase opposition. As such correct identification of equivalent noise source cannot by itself describe or confirm the precise flow mechanism behind separation noise. As such it is imperative to perform a detailed flow quantification and analysis to understand the noise mechanism. Previously, \cite{brooks1989airfoil} hypothesized that airfoil separation noise results from the interaction between turbulent structures in the shear layer and the airfoil trailing-edge, as separated {structures} are convected past the airfoil, resulting in significant pressure fluctuations. However, previous experiments were unable to accurately identify the noise mechanism, as flow-field measurements were unavailable. 

More recently, using PIV and synchronized wall-pressure and hot-wire measurements, \cite{lacagnina2019mechanisms} identified three possible distinct noise generation mechanisms to explain noise generation by an airfoil close to stall. Importantly, all of these mechanisms were linked to instabilities in the shear layer and were localized in a region within the separated shear layer away from the wall. The separated shear layer may not only result in a substantial increase in the contribution of quadrupole noise \citep{turner2022quadrupole}, but may also invalidate the unsteady Kutta condition. This is because the latter relies on the flow leaving the airfoil trailing edge smoothly. Furthermore, separation noise is dominant for airfoil placed at high angles of attack. Therefore, the central premise of the thin-airfoil linearized theory may not hold for such cases because the amplitude of the induced disturbance by the flow separation may not be small. Evidently, changes may occur in resulting radiation ratio, and thus \cites{amiet1976noise} radiation factor may not be able to correctly quantify the hydrodynamic-to-acoustic conversion \citep[see figures 3 and 12 of][for instance]{roger2004broadband}. Therefore, in the present manuscript, we ask the question: can the separation noise be fully quantified using a dipole source, such as those outlined in Amiet's diffraction theory? If so, are there other possible mechanisms of noise generation that may explain noise generation due to an airfoil close to stall? Furthermore, is the mechanism behind the separation noise universal ?

To this end, aeroacoustic measurements have been performed in %University of Sherbrooke's 
the anechoic flow facility at Universit{\'e} de Sherbrooke. In particular, planar PIV measurement, wall-pressure and far-field acoustic measurements have been achieved. For the present study, a Controlled-Diffusion (CD) airfoil is used. These measurements have been performed at a fixed geometric angle of attack of $15^{\circ}$. For the CD airfoil at this angle of attack, flow separation near the leading-edge region was reported by \cite{christophe2008}. As such, the present aeroacoustic investigation is performed to understand noise due to flow separation, for an airfoil that is close to stall conditions \citep{kalyani2022flow}. Comparing the flow and pressure characteristics between the present case and that reported earlier, where the boundary-layer is fully attached near the trailing-edge of the airfoil \citep[see][for instance]{Wu2019,jaiswal2020use,wu2020noise}, is expected to elucidate the true contribution of separation noise. % To this end, section \ref{} delineates the experimental methods and instrumentation, section \ref{} gives the aero-acoustics results obtained, finally discussion and conclusions are made in section 
  
%Open jet facilities are well suited for airfoil self-noise studies compared with any other flow facilities as concluded by \cite{stephane-2003}.
%The UDES facility were the far field acoustic, surface pressure fluctuations and $C_p$ were measured is a closed loop anechoic open jet facility. The size of the anechoic room is $7 \times 5.5 \times 4 m^3$. The wind tunnel is driven by two centrifugal pump. The opening of the section has a 

\section{Experimental Set-up and Instrumentation:} \label{sec:1}

% write about the anechoic room
The aero--acoustic measurements were performed in the anechoic wind tunnel at Université de Sherbrooke (UdeS). The anechoic room, is about  $7 \times 5.5 \times 4$ m$^3$ in dimension. The open jet has a dimension of $50 \times 30$ cm$^2$, and can achieve a maximum velocity of $40$ m/s. As the temperature of the open jet can be controlled, all the measurements are performed at a constant free-stream density $\rho$.

% Write about the flow

% write about the mockup 
%71.1 × 35.5 cm2
% write about the airfoil

% airfoil geometry 

% instrumentation on the airfoil.

The CD airfoil is placed at a $15^{\circ}$ geometric angle of attack with the help of plexiglass plates of thickness 4.25~mm laser cut to reduce uncertainty in angle of attack while placing the airfoil and at the same time giving good optical access. All the measurements are performed at a free-stream velocity $U_\infty$ of 16~m/s and 28~m/s, which respectively corresponds to Mach numbers $M_\infty\equiv U_\infty/c_0 \simeq 0.05$ and $M_\infty \simeq 0.08$ ($c_0$ speed of sound) and Reynolds numbers based on the airfoil chord length $C$ and the free-stream velocity of $Re_c \simeq 1.5\times 10^5$ and $Re_c \simeq 2.5\times 10^5$. 

\subsection{Planar PIV measurements setup}

\begin{figure}[ht!]
\centering
  \includegraphics[width=0.8\textwidth]{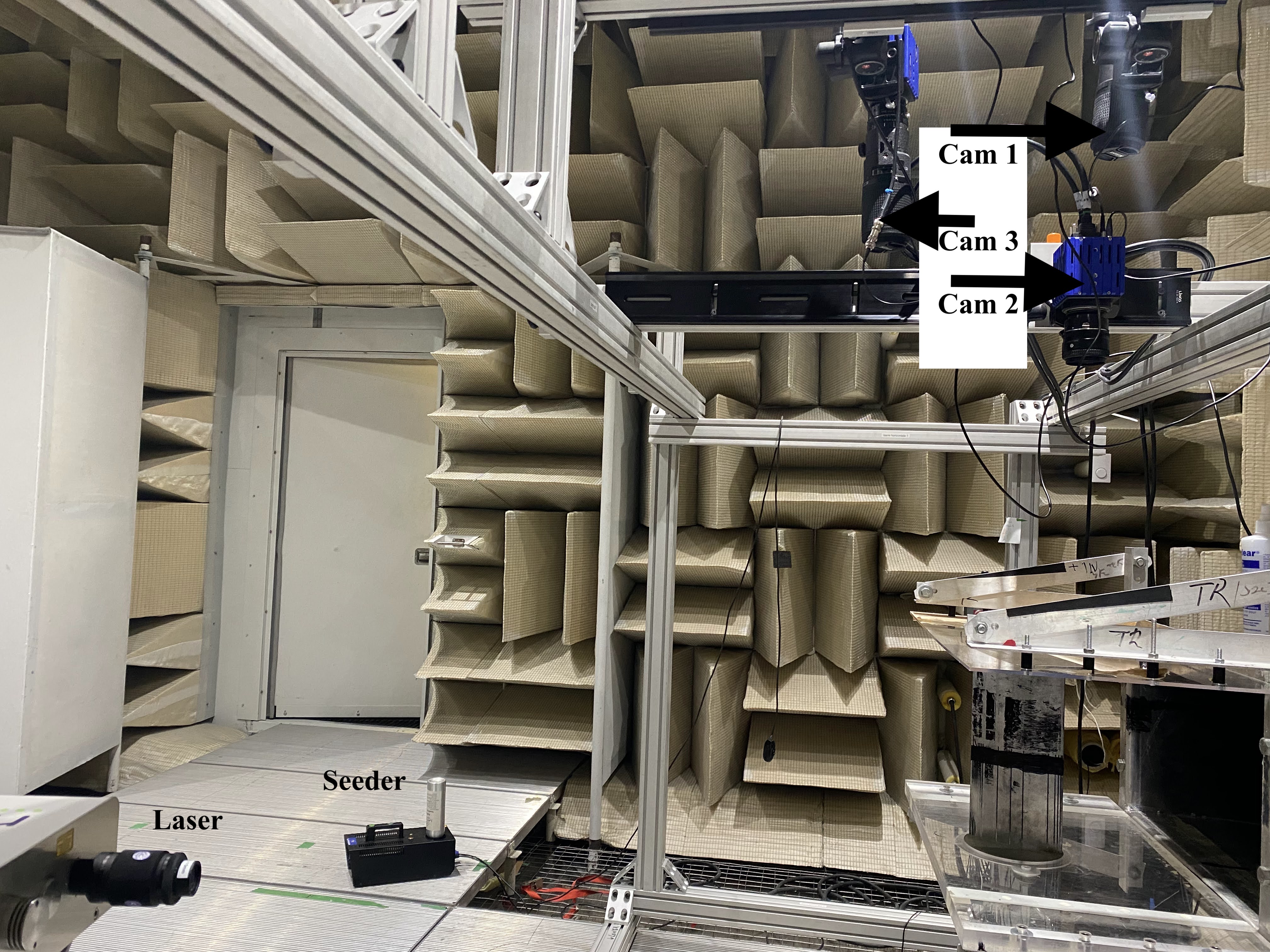}
\caption{Planar-PIV setup}
\label{set}       % Give a unique label
\end{figure}

Two-dimensional PIV measurements were performed on the suction side of the airfoil, as shown in figure~\ref{set}. Three sCMOS cameras, with a $5.5$ megapixel sensors each, were used to acquire images in a dual frame mode. A ND:YAG dual pulsed laser from Lavision was used for illumination. The light sheet for Planar PIV was generated with a set of spherical lenses and a diverging cylindrical lens with a focal length of -20~mm. Tracer particles of about 1~$\mu$m %\um$ 
were generated to seed the flow. The images were recorded for each case at an acquisition frequency of 2~Hz. Inter-frame time was increased until the cross-correlation coefficient remains between $0.6-0.9$. The resulting inter-frame time meant particle image displacement of more than 20 pixels was achieved in the free stream. This ensures low relative error ($\sim 0.5 \% $) in the estimation of particle image displacement. The data collected at a free-stream velocity of $U_{\infty}=16$~m/s were processed using Lavision's Davis 8 software while for the $U_{\infty}=28$~m/s case they were processed with the newer Davis 10 software. The final vector calculations were performed on %Compute Canada's cluster
the computer clusters from Digital Research Alliance of Canada. For %the
$U_{\infty}=16$~m/s case, a total of 11 passes were used for the multi-grid scheme, starting with an initial window of $128 \times 128$ pixels to the final window size of $4\times 4$ pixels. In the iterative multi-grid scheme, an overlap of 75~\% and an elliptical weighting (elongated in the mean-flow direction) is used. The final window size was about $0.0923 \times 0.0923$ mm$^2$. In contrast, for the $U_{\infty}=28$~m/s case, a reduced window overlap of 50~\% was used while keeping all the other parameters the same as for the $U_{\infty}=16$~m/s case. This was done to accelerate the vector calculations and to reduce the size of the final vector field. 

\begin{table}
\begin{center}
\caption{Parameters used for the Planar PIV boundary-layer measurements.}
\label{tab_piv_tonal}       % Give a unique label
\begin{tabular}{p{6.5cm}p{3.0cm}p{3.0cm}p{3.0cm}}
\hline\noalign{\smallskip}
Parameters & Leading-edge %\newline {boundary-layer \newline measurements} 
(M1) & Mid-chord %\newline boundary-layer \newline measurements 
(M2) & Trailing-edge %\newline {boundary-layer \newline measurements} 
(M3)  \\
\noalign{\smallskip}\hline\noalign{\smallskip}
Number of Images & $1800$ & $1400$ & $1800$ \\ 
%Window of \newline Interrogation (pixel$^2$) & $24\times24$ & $16\times16$ \\
Interrogation window [pixel$^2$] & $4\times4$ & $4\times4$ & $4\times4$ \\
Lenses focal Length [mm] & $200$ & $50$ & $200$  \\ 
%FOV [mm] &  $22.7$  & $ 19.3$ & $ 19.3$  \\ 
Final window size [mm$^2$] & $0.11\times 0.11$ & $0.32\times0.32$ & $0.09\times0.09$\\ 
%Digital magnification [mm/pixel] & $36$ & $132$ & $44$ \\
%Maximum particle image \newline displacement (pixel) & $14$ & $13$   \\
Maximum particle image displacement [pixel] & $20$ & $24$ & $20$  \\
\hline 
\end{tabular}
\end{center}
\end{table}

\subsection{Steady wall-pressure measurements}

\begin{figure*}[ht!]
\centering
  \includegraphics[width=150mm]{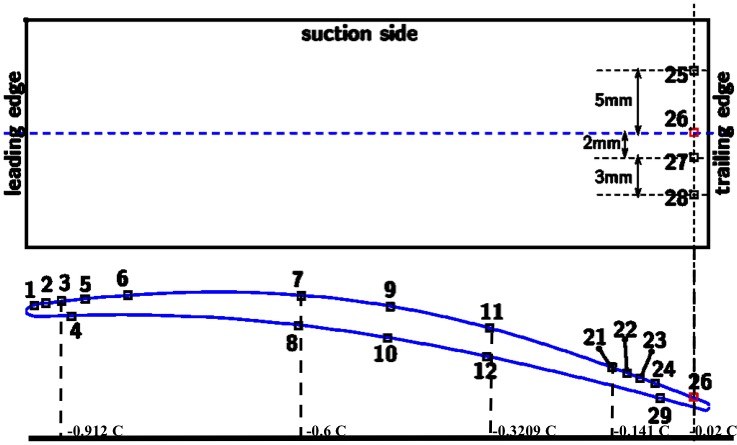}
\caption{Location of pinholes on the CD airfoil. Some of the pinhole locations on the suction side of the airfoil have been indicated.}
\label{rmp_locations}       % Give a unique label
\end{figure*}

Figure~\ref{rmp_locations} shows the pinholes located along the chord of the airfoil, which are used to measure the mean wall-pressure coefficient. %of wall pressure. 
There are in total 21 probes on both suction and pressure sides of the airfoil, 18 of them are placed in the streamwise direction and the last 3 in the spanwise direction.
Pinholes on the pressure side of the airfoil are labelled as $4$, $8$, $10$, $12$ and $29$, whereas those on the suction side are 1 to 6 at the leading edge, 7, 9 and 11 at mid-chord and 21 to 28 at the trailing edge (see figure \ref{rmp_locations}). The differential pressure is measured using an array of miniature amplified low pressure sensors in order to get a full reading along the airfoil chord~\citep{Neal:diss}. These miniature amplified low pressure sensors have an accuracy of $0.25 \%$ Full Scale (FSS), which is between $1244.2-248.84$ Pa. %pascals. 
Details on the setup and acquisition can be found in \cite{Prateek:diss}. In the current paper, the wall differential pressure are normalized by inlet free stream dynamic pressure, which yields the mean pressure coefficient %of pressure 
$C_p\equiv (p-p_\infty)/(0.5\,\rho\,U_\infty^2)$ with $p_\infty$ the inlet pressure.

\subsection{Unsteady wall-pressure measurements}

The pinholes on %the suction-side 
both sides of the airfoil are %equipped with 
are also connected to Remote Microphone Probes (RMP) to record unsteady static wall-pressure measurements~\citep{perennes_rmp,moreau2005effect}. For the present set of experiments, Knowles FG $23329$-P$07$ miniature microphones were used. %There are in total 21 probes on the suction side of the airfoil, 18 of them are placed in the streamwise direction and the last 3 in the spanwise direction. 
These microphones have a flat response over a large range of frequency ($0.1-10$~kHz), and have a nominal sensitivity of $22.4$~mV/Pa. The pinhole diameter of $0.5$~mm ensures spectral averaging is avoided well beyond $10$~kHz \citep{grasso2019analytical}. As these microphones are connected remotely to the pinhole, a correction in phase and magnitude is needed. This is achieved following the methodology outlined by \cite{jaiswal2020use}. 

\subsection{Hot wire measurements}

Hot wire anemometry (HWA) %was 
is used to investigate the spectral content of the velocity disturbance over the airfoil. The HWA %were performed by placing the 
probe is placed directly above %the 
RMP 26 ($x/C=0.98$), %over 
on the suction side of airfoil. The hot wire measurements were performed using a TSI $1210-\rm{T}~1.5$ single wire probe. The probe consists of a platinum quoted tungsten wire with a $0.0038$ mm diameter and a $1.27$ mm length, which satisfies the recommended wire length-to-diameter ratio of 200 \citep{ligrani1987}. The hot wire probe was connected to a TSI IFA $300$ anemometer operating in Constant Temperature Anemometry (CTA) mode. The output signals of this anemometer were recorded with $25600$ kHz acquisition frequency using a NI 9234 24 bit module. In order to attenuate any unwanted parasitic noise, a low pass filter of $1000$~Hz was applied. Based on previous wall-pressure measurements \citep{moreau2005effect} %we do not expect to see any 
no substantial contributions of velocity disturbances beyond this frequency is expected for the $15^{\circ}$ angle-of-attack case. Furthermore, the HWA were performed only at $U_\infty=16$~m/s. %case. 
The total recording time for each of these point-wise measurements was about 60~seconds. For more details on the setup the reader is referred to \cite{Prateek:diss}.

\subsection{Acoustic measurements}

Far-field acoustic pressure was measured using Integrated Circuit Piezoelectric (ICP) microphones with a $1/2$ inch diaphragm. The microphones are placed in the airfoil mid-chord plane. In total 8 microphones were placed on an circular arc around the airfoil at a distance of 1.21~m (or about 10 times the chord length) to ensure they are in an acoustic far-field location. The microphones were calibrated using a B\&K piston-phone, which ensures the calibration uncertainty is within 0.2~dB.   

\subsection{Synchronized %Planar-PIV 
measurements}

In order to relate the near-field velocity disturbance field to the resultant far-field acoustic noise, synchronized velocity-pressure measurements have been performed as previously done at a lower 5$^\circ$ angle-of-attack~\citep{jaiswal2022experimental}. Furthermore, the wall-pressure measurements were also synchronized to study the footprint of velocity disturbances on the wall. To obtain acoustic directivity pattern caused by diffraction of unsteady gust, the far-field microphones were synchronized with the RMPs. The near-field and far-field pressure measurements are time resolved compared to PIV measurements, which has a limited time resolution. As such, the acquisition frequency for all the measurements performed are set to powers of two. In particular, the PIV measurements were performed at 2~Hz while unsteady near and far-field pressure were recorded at an acquisition frequency of 65536~Hz (or $2^{16}$~Hz). The synchronization between PIV and pressure measurements %was 
is achieved using the procedure outlined by \cite{henning2008investigation}, where further details on the implementation can be found. 

\section{Results}

To ensure that the flow facility and installation do not dictate overall flow dynamics \citep[see][for instance]{moreau2005effect,wu2016numerical}, the mean wall-pressure coefficient {has first} been compared. The results in %four 
two different facilities, in which the CD airfoil has been tested in within a 50~cm wide jet, show an overall good agreement over most of the airfoil chord, $C$, as shown in figure~\ref{Cp}. $(x,y)$ represents the fixed laboratory reference frame at the airfoil midspan, $x$ being parallel to the jet axis and oriented with the flow. The origin of the reference frame is taken at the airfoil trailing edge. Repeatability tests at UdeS have also been achieved~\citep{kalyani2022flow}.

\begin{figure*}[h!]
  \centering
  \begin{tabular}{@{}p{0.45\linewidth}@{\quad}p{0.45\linewidth}@{}}
    \subfigimg[width=75 mm,pos=ul,vsep=21pt,hsep=42pt]{(a)}{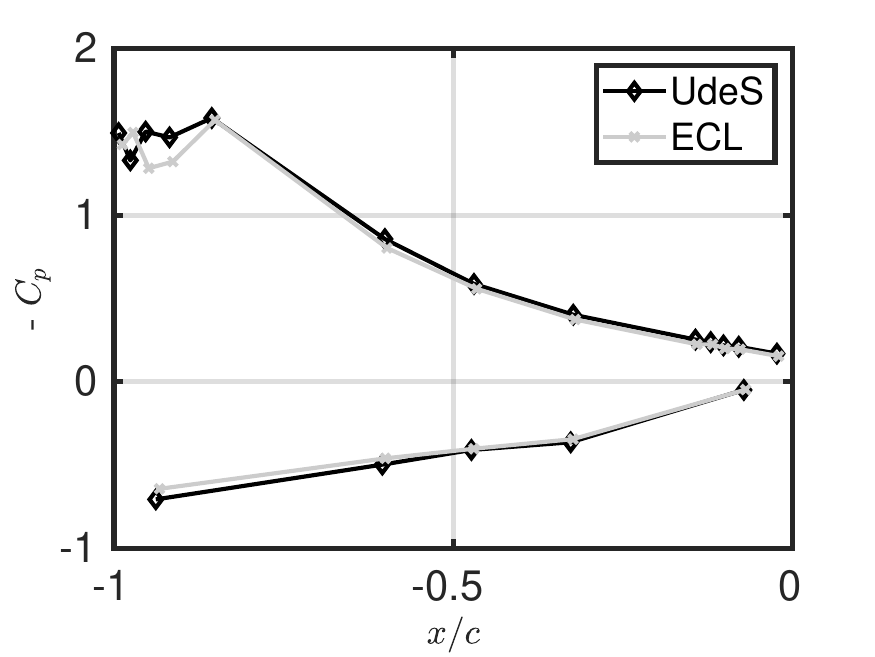} &
    \subfigimg[width=70 mm,pos=ul,vsep=21pt,hsep=42pt]{(b)}{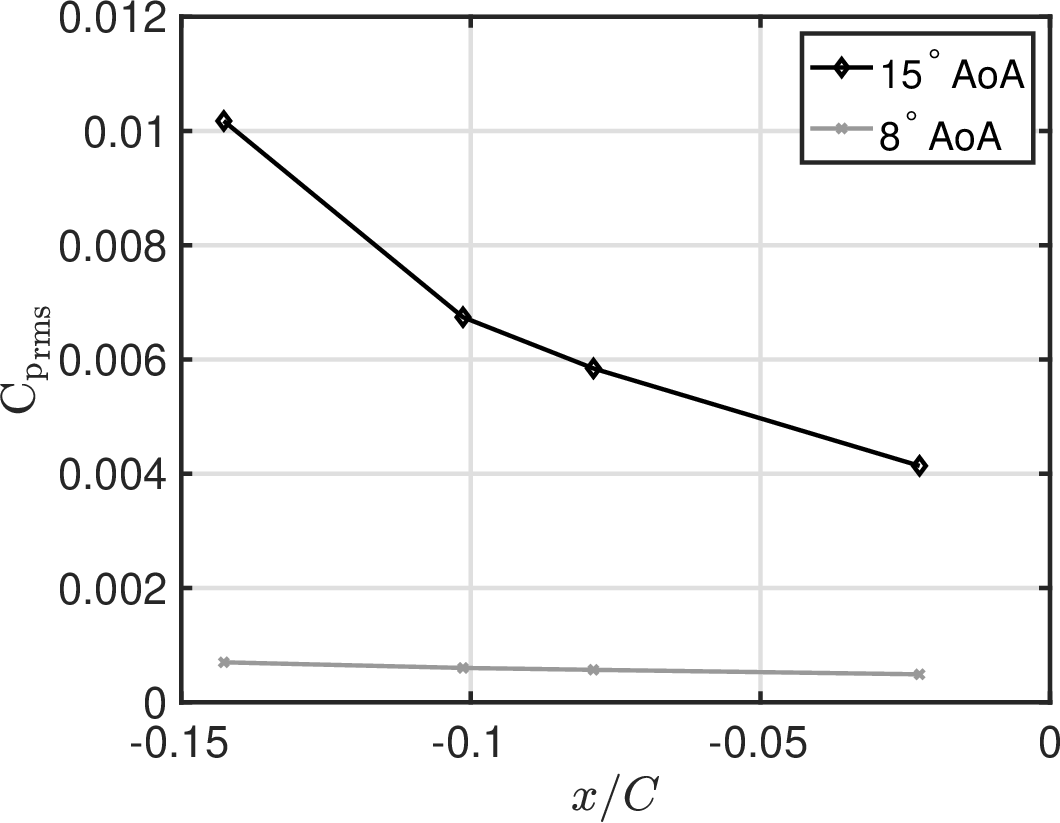} \\
  \end{tabular}
  \caption{%Negative m
 (a) Mean wall-pressure coefficient ({$-C_p$}) %for the CD airfoil 
 in two different anechoic wind tunnels; %facilities; : solid grey line with cross symbols, %Anechoic wind tunnel facility at 
 %Ecole Centrale de Lyon. %; solid black line with open diamond symbol, %Anechoic wind tunnel facility at 
 %UdeS. 
 (b) r.m.s of wall-pressure coefficient ({$C_{prms}$}) for %the trailing-edge sensors 
 RMPs 21, 23, 24 and 26 at $U_{\infty}=16$~m/s. %The case for $8^\circ$ degree incidence 
% shown as grey solid grey line with cross symbols while solid black line with open diamond symbol for $15$ degree incidence.
%solid grey line with cross symbols, $8^\circ$; solid black line with open diamond symbol, $15^\circ$.  
}
\label{Cp}       % Give a unique label
\end{figure*}

Previous experimental and numerical studies on this airfoil~\citep{moreau2005effect,christophe2008trailing,kalyani2022flow} have reported an increase in low-frequency noise, when it is placed at a high angle of attack. This observation is confirmed by the far-field microphone measurements shown in figure~\ref{ff_spectra}, which shows an overall increase in low frequency sound pressure levels when comparing the $8^\circ$ and $15^\circ$ cases for the two Reynolds numbers. As shown in figure~\ref{ff_spectra}, this is also consistent with the previous measurements by \cite{moreau2005effect} (open symbols).
This noise increase is most likely %possibly 
linked to an overall increase in r.m.s levels of wall pressure close to the trailing-edge region, as shown in figure~\ref{Cp}~(b). As the overall goal of the present manuscript is to identify flow mechanisms responsible for separation noise and to test the applicability of diffraction theory \citep{amiet1976noise}, the cause (flow disturbances) to the effect (far-field noise) will be established with the help of the latter. 

\begin{figure*}[h!]
\centering
  \includegraphics[width=75mm]{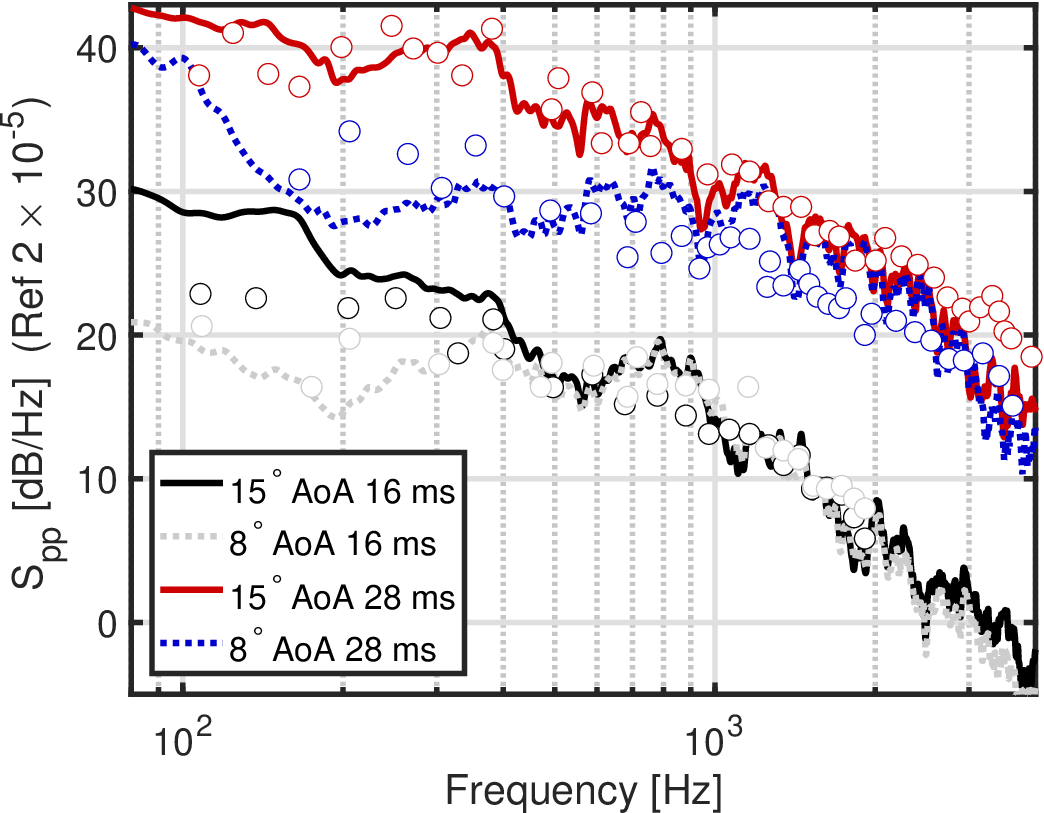}
\caption{Sound pressure level %measured 
at %${\sim 1.21}$
1.2~m %by a microphone perpendicular to 
from the airfoil trailing edge %, and located 
on the suction side. Open circles for ECL measurements \citep{moreau2005effect} %of the airfoil. Legends: Dotted blue and grey lines correspond to CD airfoil placed at 8$^{\circ}$ angle-of-attack case for inlet velocities of 28 and 16 [m/s] respectively. Solid red and black lines correspond to CD airfoil placed at 15$^{\circ}$ angle-of-attack case for inlet velocities of 28 and 16 [m/s] respectively.
}
\label{ff_spectra}       % Give a unique label
\end{figure*}

%According to \cite{amiet1976noise}, the induced pressure, which cancels the unsteady pressure jump at the trailing-edge, propagates to far-field as noise. The unsteady pressure jump results from the incident pressure field, and are themselves induced by the turbulent boundary layer \citep{grasso2019analytical,jaiswal2020use}. 
Amiet's model and its extension~\citep{amiet1976noise, roger_moreau,moreau2009back,Roger2012} relies on Curle's analogy combined with a compressible linearized Euler model for the wall-pressure fluctuations on an infinitely thin flat plate seen as equivalent dipoles. The PSD of the far-field acoustic pressure at any observer located at $\mathbf{X}=(X_1,X_2,X_3)$, for any angular frequency $\omega$, generated by a flat plate of chord length $C$ and span $L$ then reads:
\begin{equation}
  S_{pp}(\mathbf{X},\omega)\,\approx\,\left(\frac{k\,C\,X_2}{4\pi S_0^2}\right)^2\frac{L}{2}\left|{\cal I}\left(\frac{\omega}{U_c},k\frac{X_3}{S_0}\right)\right|\Phi_{pp}(\omega)\,l_z\left(\omega,k\frac{X_3}{S_0}\right),
\label{amiet_final}
\end{equation}
where $k$ is the acoustic wave number, $S_0$ the corrected distance to the observer, ${\cal I}$ the analytical radiation integral (or acoustic transfer function) given in \cite{roger_moreau}, $U_c$ the streamwise convection velocity, $\Phi_{pp}$ the wall-pressure spectrum and $l_z$ the spanwise coherence length.

In summary, the wall-pressure field can be characterized by the PSD of wall-pressure fluctuations, the convection velocity and the spanwise correlation length. In order to explain the increase in the low frequency far-field acoustic spectra, the statistical description of the incident wall-pressure field will be explored in the next section. %that follows. 

\subsection{{Unsteady wall-pressure field}}

Figure \ref{Wall_pressure_spectra} shows %Power spectral density
PSD measurements using RMPs on the suction side of the airfoil along its chord. The first two probes located at the leading edge, show rapid decay in spectral energy %perhaps due to 
most likely because of the laminar nature of the boundary layer. The humps and peaks observed in RMP 3 probe ($x/C \simeq 0.09 $) can be linked to boundary-layer instabilities \citep{jaiswal2020use}, which are present due to the existence of a Laminar Separation Bubble (LSB). RMP~5 ($x/C \simeq 0.15 $) onward the wall-pressure spectra decay is much slower than it was for the first three probes, suggesting a possible turbulent re-attachment. Near the mid-chord region (RMP~9), the wall-pressure statistics  almost attains a $-5$ slope at high-frequencies, suggesting a mean attached turbulent boundary layer.  

\begin{figure*}[ht!]
  \centering
  \begin{tabular}{@{}p{0.45\linewidth}@{\quad}p{0.45\linewidth}@{}}
    \subfigimg[width=75 mm,pos=ur,vsep=21pt,hsep=42pt]{(a)}{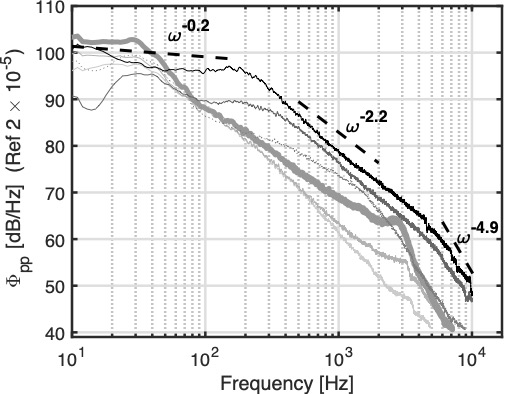} &
    \subfigimg[width=75 mm,pos=ur,vsep=21pt,hsep=42pt]{(b)}{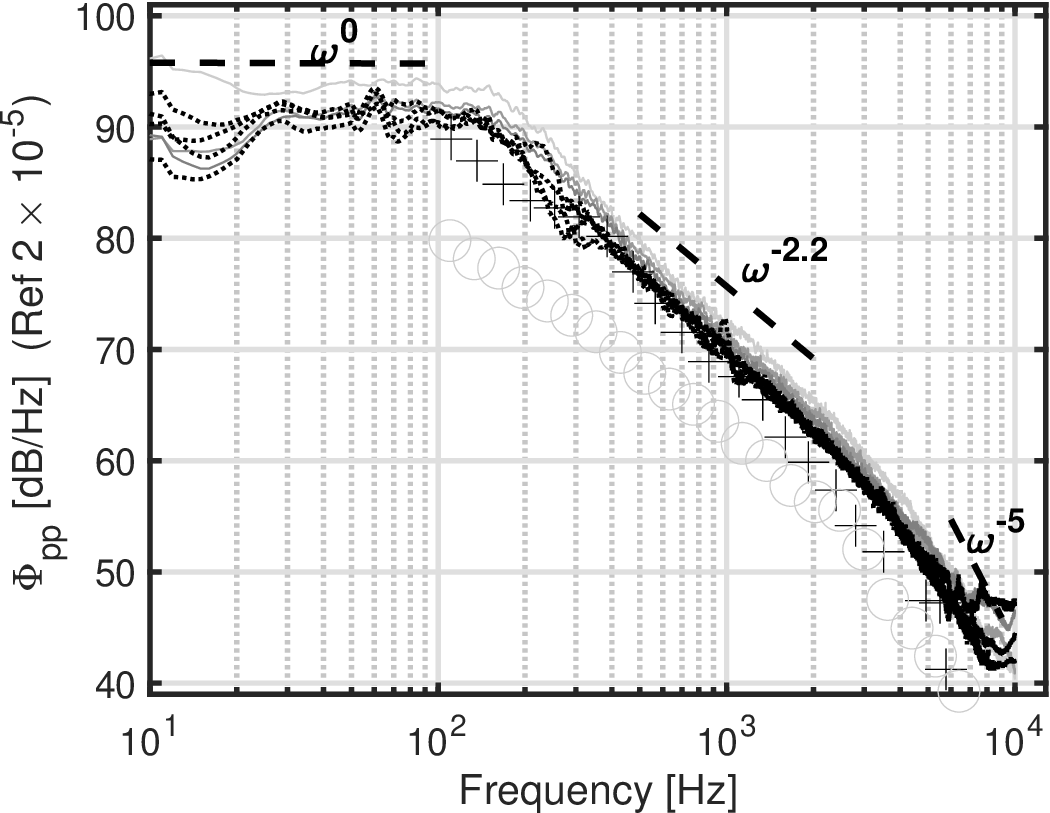} \\
  \end{tabular}

\caption{PSD of wall-pressure fluctuations at $Re_c=150000$ on the airfoil suction side (color transition from gray to black with increase in $x/C$): %side of the airfoil. 
(a) RMP 1 to 9 (thick solid line to highlight RMP 3); (b) RMP 21-28 (black dotted lines for spanwise sensors. Black plus for LES \citep{christophe2008}, and grey circles for ECL measurements \citep{moreau2005effect}). %Legend: Color transition from gray to black signals increase in $x/C$ in each sub-figure. In sub-figure (a) thick solid line is used to highlight RMP 3 sensor. (b) Spanwise sensors are colored with black dotted lines.
}
%\end{center}
\label{Wall_pressure_spectra} 
\end{figure*}

%At 
On the aft part of the airfoil, %chord, 
an almost constant gradient in the wall-pressure spectra, of $f^{-2.2}$, emerges in the mid-frequency range. Similar observations were made by \cite{zang2021experimental} who used the NACA-65-410 airfoil for their study at high angles of attack, and by~\cite{Raus2022} on the oscillatory NACA-0012 airfoil in the similar light-stall flow regime. In contrast, at low frequencies, the wall-pressure spectrum becomes flat to an extent that its slope is near zero for probes beyond RMP 9. As such, the classical $f^{2}$ \citep{goody2004empirical} scaling at low frequency is not observed. It is hypothesized that this change in slope is more linked to the presence of %an open-jet
the jet, which predominantly contributes to the low frequency and interacts more with the airfoil at high angle of attacks. At higher frequencies, a constant spectral slope of $f^{-5}$ emerges, which is consitent with previous studies made on the CD airfoil \citep{Prateek:diss}. Overall, the spectra become statistically similar beyond %$\sim 0.85C$ 
0.85~$c$. To quantify the effects of mean pressure gradient on wall-pressure fluctuations, differences in %spectral levels are calculated by subtracting the Power Spectral Density 
PSD between the airfoil at $8^{\circ}$ and $15^{\circ}$ angles of attack at $Re_{c}\simeq 150000$ %. The differences, expressed in [dB/Hz], 
are plotted in figure \ref{diff_pressure} for the trailing-edge sensors $21-25$ only between %$\sim 10-1000$ 
10 and 1000~Hz. %The figure clearly shows an 
An increase in spectral content is clearly shown for the %15-degree angle of attack 
$15^{\circ}$  case compared to the 8$^{\circ}$ %angle of attack 
case \citep{Prateek:diss}. %\cite{lacagnina2019mechanisms} associated this increase in wall-pressure spectral content with flow separation.

\begin{figure*}[ht!]
\centering
  \includegraphics[width=85mm]{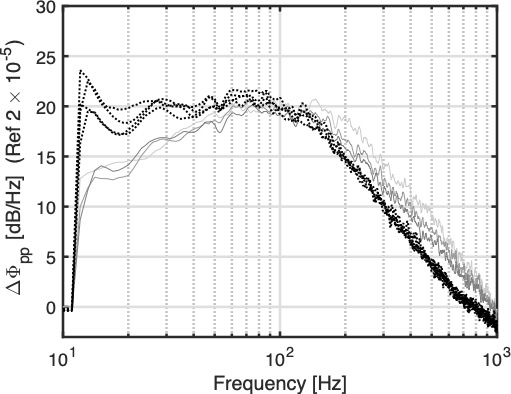}
\caption{%Differences in %Power spectral density 
%power spectra density 
PSD differences between the $15^{\circ}$ and $8^{\circ}$ %angle of attack 
cases %. The difference is expressed in [dB/Hz] for all the trailing-edge sensors from $\sim x/C=0.85$ onward, between 
for RMPs 21 and 25. Color transition from gray to black signals increase in $x/C$.}
\label{diff_pressure}       % Give a unique label
\end{figure*}

\begin{figure*}[ht!]
\centering
  \includegraphics[width=85mm]{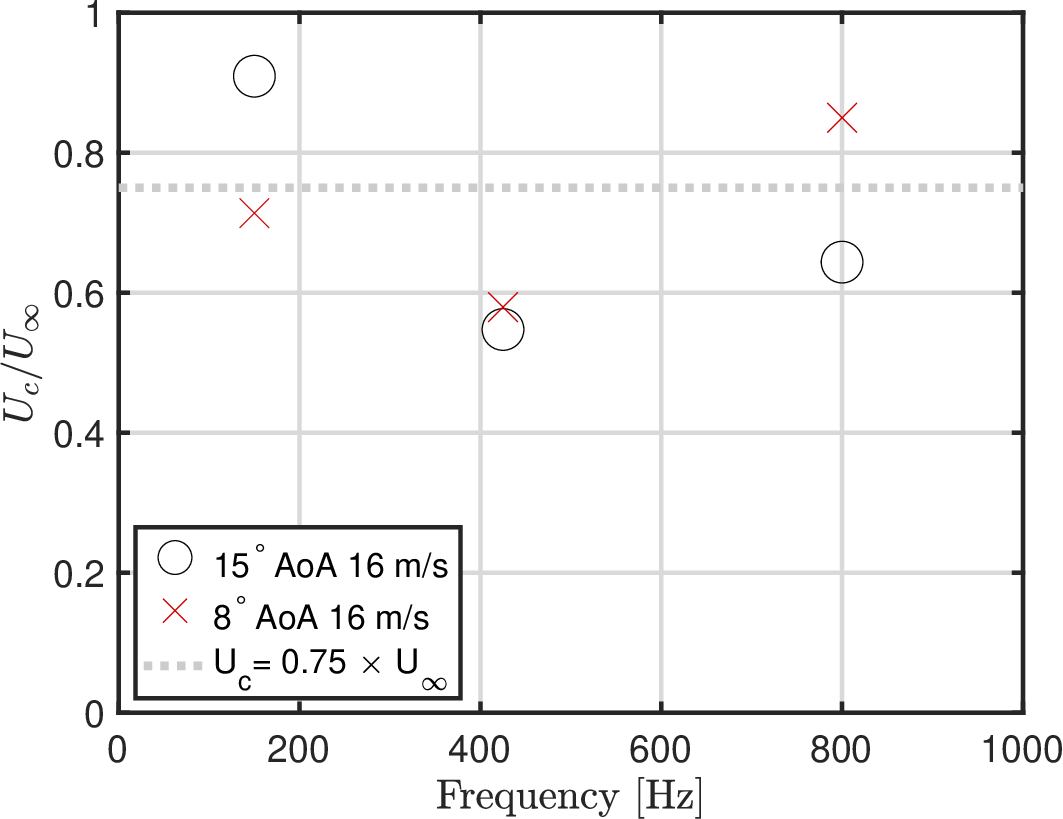}
\caption{Convection velocity ($U_c$) %of pressure gust 
measured between RMPs 23 and 24. Dotted grey line is \cites{moreau2005effect} mean estimate of $U_c$ for $15^{\circ}$ angle of attack and 16~m/s case. %close to the trailing-edge of the CD airfoil. Legend: 
%Black circles correspond to the CD airfoil placed at $15^{\circ}$ degree angle of attack and 16~ms, Red cross are for $8^{\circ}$ degree angle of attack and 16~ms. Dotted grey line correspond to $U_c=0.72 \times U_{\infty}$. 
}
\label{convection_velocity}       % Give a unique label
\end{figure*}

The second quantity of interest, the convection velocity $U_c$, was estimated by correlation analysis between two RMPs, 23 and 24, which are separated by a finite streamwise distance (%$\sim 0.02 \times $C
about 0.02~$C$). This was performed at several band-passed frequencies to obtain the convection velocity at frequencies where the separation noise dominates.~The results obtained are shown in figure \ref{convection_velocity}. Open circles represent the $15^{\circ}$ angle-of-attack case, while cross symbols stand for the $8^{\circ}$ angle-of-attack case. The dashed line refers to the mean convection velocity (0.75~$U_\infty$) estimated by~\cite{moreau2005effect} from the phase slope of two sensors at the trailing edge at $15^{\circ}$ and 16 m/s. The present results are therefore consistent with the previous ECL measurements and also with the estimate provided by~\cite{grasso2019analytical} based on the direct numerical simulation~\citep{wu2020noise} for the $8^{\circ}$ case (0.72~$U_\infty$). 
The lower value at 400~Hz is also consistent with that reported by~\cite{kalyani2022flow} for this frequency range (0.52~$U_\infty$).
%\textbf{~Although it is consistent with the previous measurements by \cite{moreau2005effect}, the final value of convection velocity is slightly lower in the present case. This could be due to an
Note that the observed variations can be caused by the uncertainty in the estimation of $U_c$, which should be a function of total recording time of the signal as shown in the appendix. 
%As such an uncertainty based on recording time is estimated in \textbf{Appedix}}. 
In the low frequency ranges %$\le 600$
below 400~Hz, an increase in convection velocity is observed for the $15^{\circ}$ angle-of-attack case compared to the $8^{\circ}$ angle-of-attack case. This result is quite surprising, as an increase in adverse pressure gradient leads to a decrease in convection velocity \citep[see, for instance,][]{schloemer1967effects}. Therefore, this observation will be addressed in the subsequent sections. Nevertheless, at higher frequencies %$\ge 700$ [Hz]
beyond 400~Hz, the convection velocity for the $15^{\circ}$ angle-of-attack case does become lower than that for the $8^{\circ}$ angle-of-attack case. Finally, the convection velocity decreases with an increase in frequencies because at low frequency, only the large eddies contribute to the pressure gusts \citep[see, for instance,][]{schloemer1967effects}. At higher frequencies, the contribution from smaller eddies, which are close to the wall, becomes significant, resulting in lower convection velocities. Therefore, \cite{schloemer1967effects}'s observation regarding the frequency dependence of convection velocity is valid for both angles of attack. %compared.

Lastly, quantifying the spanwise correlation length is anything but straightforward. \cite{corcos_JFM1964}, under the assumption that the normalized cross-power spectral density can be represented by two separate dimensionless variables $\omega \Delta{x_1} / U_c$ and $\omega \Delta{x_3} / U_c$, showed that the two-dimensional coherence function can be written as follows:
 \begin{equation} \label{seperation}
  \gamma(\Delta{x_1},\Delta{x_3},\omega) = \frac{ \Phi_{pp}(\omega, \Delta{x_1}, \Delta{x_3})}{\Phi_{pp}(\omega, 0, 0)}  = A(\omega \Delta{x_1} / U_c) \,\, B(\omega \Delta{x_3} / U_c) \,\,  {\rm e}^{-i \omega \Delta{x_1} / U_c}
\end{equation}
The magnitude-squared coherence in the spanwise direction, is obtained by multiplying $\gamma(0,\Delta{x_3},\omega)$ by its complex conjugate, and is plotted in figure \ref{Wall_pressure_correl}~(a). As can be seen in the latter, the coherence goes to zero beyond 1000~Hz, which makes the estimation of the  associated length scales impossible. %that in the current set of experiments estimation of the spanwise coherence and its associated length scales is not possible beyond 1000~Hz. 
More importantly, a hump centred around $\sim 100$~Hz is observed in the values of $\gamma^2$. These results are consistent with the previous measurements by \cite{moreau2005effect} shown as symbols in figure \ref{Wall_pressure_correl}~(a). Even though similar results were also reported by~\cite{kalyani2022flow} beyond 100~Hz, the oscillatory behavior observed in their figure~4 below 100~Hz is caused by a too short signal length as shown in the appendix. 

\begin{figure*}[ht!]
  \centering
  \begin{tabular}{@{}p{0.45\linewidth}@{\quad}p{0.45\linewidth}@{}}
    \subfigimg[width=75 mm,pos=ul,vsep=21pt,hsep=42pt]{(a)}{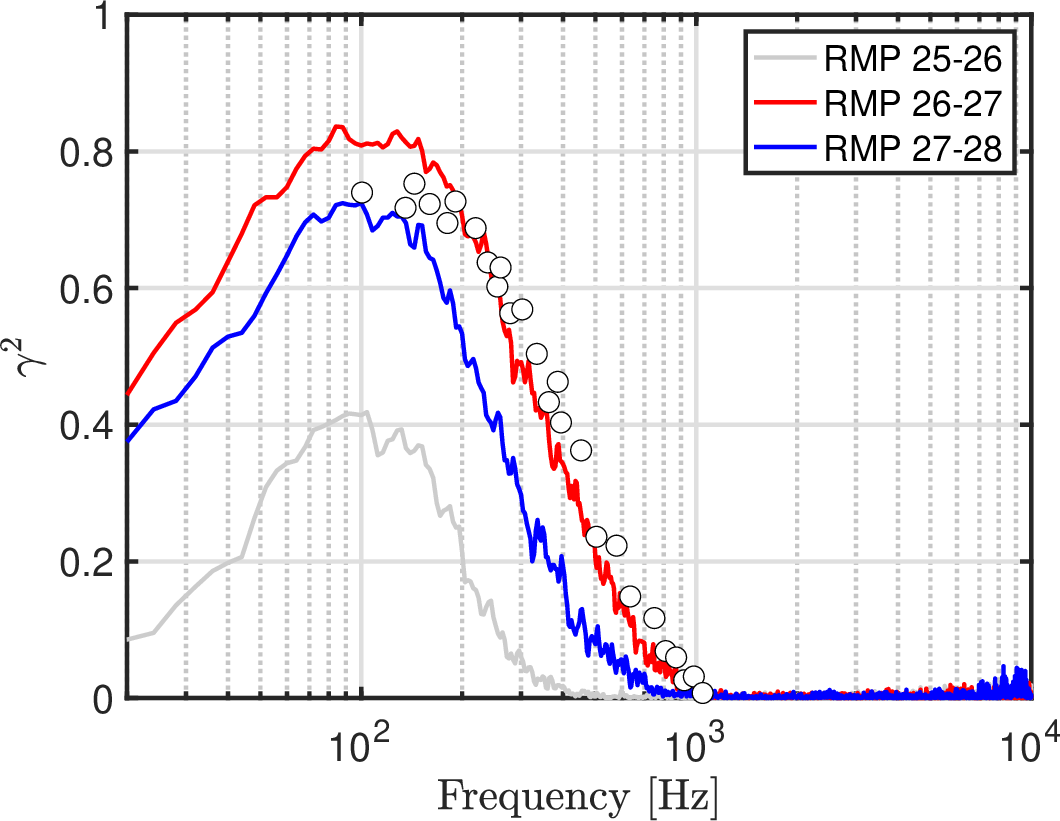} &
    \subfigimg[width=75 mm,pos=ul,vsep=21pt,hsep=42pt]{(b)}{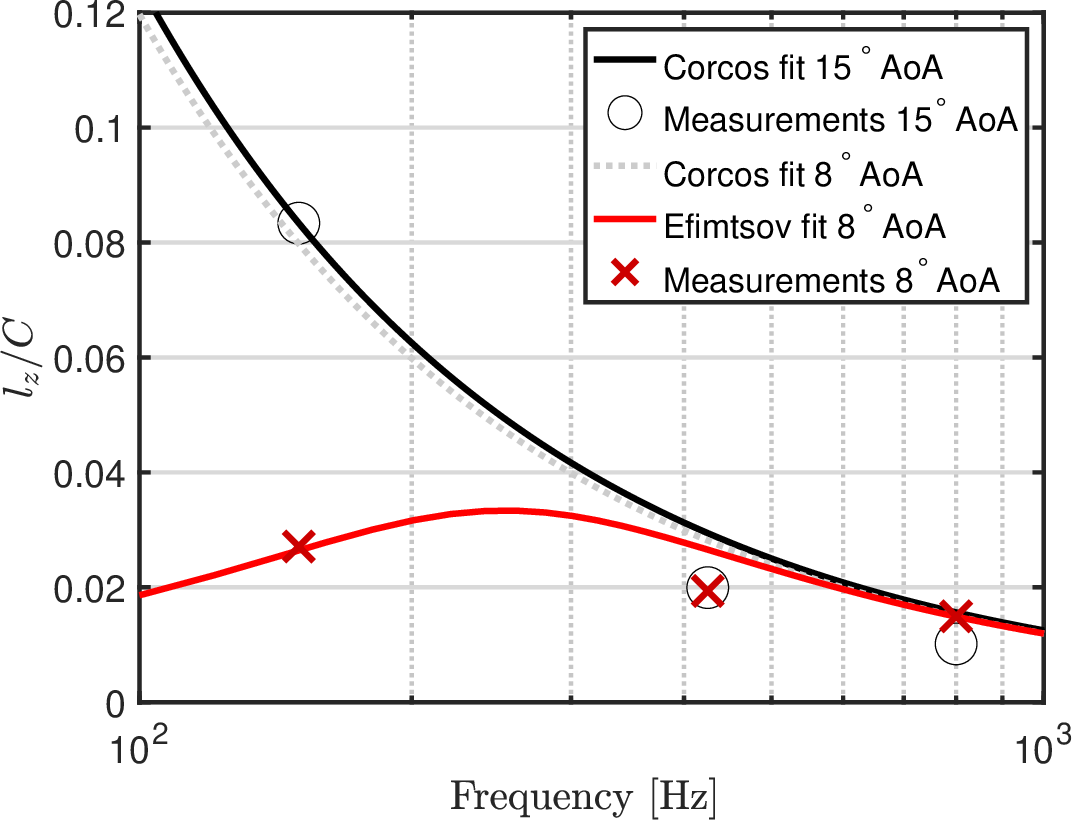} \\
  \end{tabular}

\caption{Analysis of the spanwise wall-pressure at $U_{\infty}=16$~m/s (a) Magnitude squared coherence ($\gamma^2$). Black circles for ECL measurements \citep{moreau2005effect} for $\gamma^2$ between RMP 26-27. %Solid red line for $\gamma^2$ between RMP 26-27, Solid blue line for $\gamma^2$ between RMP 26-25, and solid grey line for $\gamma^2$ between RMP 25-27. 
(b) Estimation of spanwise correlation lengths.}
%for the wall-pressure fluctuations at . Black circles correspond to measured values $l_z$ for the CD airfoil placed at $15^{\circ}$ degree angle of attack and 16~ms. Solid black line is the Corcos model fit for the same case. Red cross correspond to measured values $l_z$ for the CD airfoil placed at $8^{\circ}$ degree angle of attack and 16~ms, while solid red line represent modified Efimtsov's model, and dotted gray the Corcos model.}
%
%\end{center}
\label{Wall_pressure_correl} 
\end{figure*}

%
%Here $\Phi_{pp}$, is the single sided wall-pressure frequency cross spectral density. 
\cite{corcos_JFM1964} was also first to recognize the exponential decay nature of the wall-pressure correlation separated by a finite distance, as also evidenced in figure \ref{Wall_pressure_correl}~(a). Invoking observation of exponential decay of correlation made by \cite{corcos_JFM1964}, the function $A$ and $B$ can be written as:

\begin{equation} \label{functions}
A(\omega \Delta{x_1} / U_c) = {\rm e}^{- \omega\, b_2 \Delta{x_1} / U_c} \quad {\rm and} \quad B(\omega \Delta{x_3} / U_c) = {\rm e}^{- \omega\, b_1 \Delta{x_3} / U_c}
\end{equation}
where, $b_1$ and $b_2$ are %the 
fitting parameters. 
Under the assumption of zero streamwise separation, the normalized cross-power spectral density can then be written as follows:

\begin{equation} \label{spanwise}
\gamma(0,\Delta{x_3},\omega) = {\rm e}^{- \omega\,b_1 \Delta{x_3} / U_c}
\end{equation}

The spanwise correlation length $l_{z}(f)$ can be estimated by:

 \begin{equation} \label{spanwiselenght}
l_{z}(\omega) =  \int_{0}^{\infty}  \gamma(0,\Delta{x_3},\omega) \Delta{x_3}
\end{equation}

Plugging equation \eqref{spanwise} in equation \eqref{spanwiselenght} yields:

 \begin{equation} \label{span_length}
l_{z}(\omega) = b_1\,U_c/ \omega  
\end{equation}
%
%where $a_1$ = $1/b_1$.
%

The reader should be cautioned that Corcos's model (equation \eqref{span_length}) can lead to nonphysical values of spanwise correlation length, as it relies on the assumption that the convection velocity is independent of frequency. Nevertheless, it provides a reasonable estimation of the correlation length and has been used in the past by several authors \citep{roger2004broadband}. Therefore, Corcos's model (equation \eqref{span_length}) was used to estimate the spanwise correlation length. However, as the frequency at which the model should be used is unclear, the frequency was arbitrarily chosen.  The resulting lengths are shown as the solid black and broken grey lines for the $15^{\circ}$ angle-of-attack and the $8^{\circ}$ angle-of-attack cases, respectively. The resulting values for the constant $b_1$ are $1.37$ and $1.34$ for the $15^{\circ}$ and the $8^{\circ}$ angle-of-attack cases, respectively. Although the estimate of Corcos's model predicts high-frequency attenuation in spanwise correlation length, it over-predicts it at low frequency for the $8^{\circ}$ angle-of-attack case. This can be corrected by using \cite{efimtsov1982characteristics} model (solid red line), which takes the boundary-layer thickness ($\delta$) and friction velocity ($u_{\tau}$) into account to re-scale correlation length in the low-frequency range. The three empirical constants were set to $1.34$, $19.5$, and $13.5$ in order to estimate the spanwise correlation length with \cite{efimtsov1982characteristics} model.

 In order to experimentally estimate the spanwise correlation length, the spanwise coherence between several spanwise sensors (RMPs $25-28$) near the trailing edge ($x/C=0.98$) was calculated. The estimated values of the real part of the coherence were fitted with an exponential decay function for a given frequency.  %$f_0$. 
 The exponential decay function was chosen based on observations by \cite{corcos_JFM1964}. Finally, the correlation length $l_{z}$ %($l_{z}(f_0)$) 
 was obtained by combining equations \eqref{spanwise} and \eqref{spanwiselenght}. The resulting values of the correlation lengths ($l_{z}(f)$) are represented by symbols (crosses and circles) in figure \ref{Wall_pressure_correl}~(b). As in the case of convection velocity, the spanwise correlation length also increases for the $15^{\circ}$ angle of attack compared to the $8^{\circ}$ angle of attack case. As wall pressure is an imprint of turbulent flows convecting over the surface, what flow structures can explain such an increase in convection velocity? To answer this question, velocity field measurements were carried out using PIV and will be discussed in the following section.
  
%Corcos's model (equation \eqref{span_length}) suggests unreasonable high values of $\Lambda_{p}(f)$ as frequency goes to $0$. To correct this anomaly, \cite{efimtsov1982characteristics} proposed a new model. Furthermore, \cites{efimtsov1982characteristics} model also takes the boundary-layer thickness ($\delta$) and friction velocity ($u_{\tau}$) into account. The \cites{efimtsov1982characteristics} model for wall-pressure correlation length is given by the following equation:

 %\begin{equation} \label{efimtsov}
%\Lambda_{p}(f) = \delta {\bigg[   \frac{\big(  a_1 2 \pi f \delta  \big)}{U_c}  +  \frac{ a_2 }{ \big(   ( 2 \pi f \delta / u_{\tau} )^2 +   (  a_2/a_3 )^2  \big)}   \bigg]}^{-1/2}
%\end{equation}

%here, $a_2$ and $a_3$ are additional semi-empirical constants. The values of these constants at low Mach number ($M< 0.1$) are $548$ for $a_2$ and $13.5$ for $a_3$ \citep[see][for instance]{graham1997comparison}.

\subsection{{Velocity measurements}}
\begin{figure*}[ht!]
  \centering
  \begin{tabular}{@{}p{0.45\linewidth}@{\quad}p{0.45\linewidth}@{}}
    \subfigimg[width=80 mm,pos=ul,vsep=21pt,hsep=42pt]{(a)}{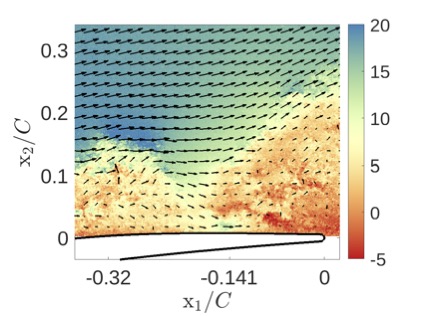} &
    \subfigimg[width=80 mm,pos=ul,vsep=21pt,hsep=42pt]{(b)}{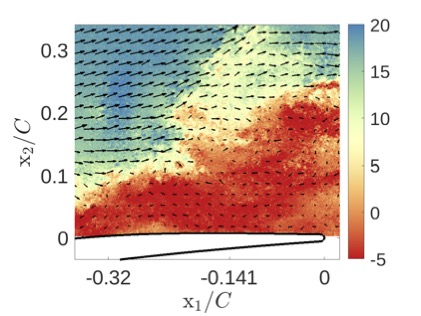} \\
  \end{tabular}

\caption{Snapshots of instantaneous %of 
wall-parallel velocity at two time instants.}
%\end{center}
\label{Inst_velo} 
\end{figure*}

Two snapshots of wall-parallel velocity are plotted in figure \ref{Inst_velo}. As can be seen, large rollers, similar to that reported by \cite{jaiswal2022experimental} at 5$^\circ$ incidence, are observed, which evidences the presence of large coherent structures. This is also consistent with the flow topology seen by~\cite{christophe2008} in their LES at 15$^\circ$. These structures are typically induced by instability within the separated shear layer. At some instant, even a fully separated boundary-layer is observed as shown in figure \ref{Inst_velo}~(b). These instantaneous flow fields confirm large scale separation and passage of coherent rollers at the trailing edge, which are reminiscent of Kelvin-Helmholtz flow type.

\begin{figure*}[ht!]
  \centering
  \begin{tabular}{@{}p{0.5\linewidth}@{\quad}p{0.5\linewidth}@{}}
    \subfigimg[width=80 mm,pos=ul,vsep=21pt,hsep=38pt]{(a)}{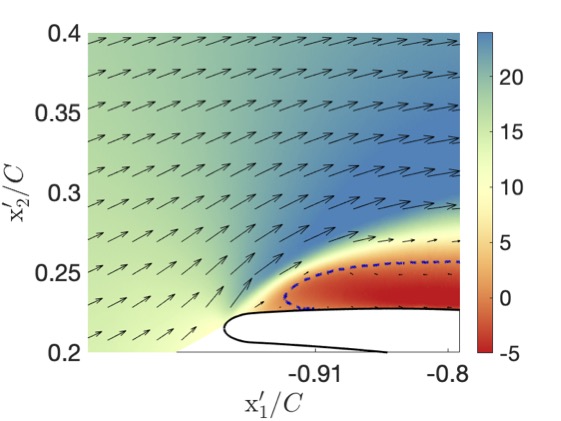} &
    \subfigimg[width=80 mm,pos=ul,vsep=21pt,hsep=38pt]{(b)}{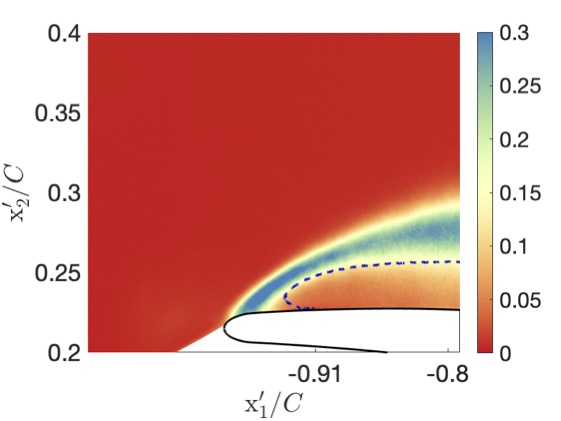} \\
     \subfigimg[width=80 mm,pos=ul,vsep=21pt,hsep=38pt]{(c)}{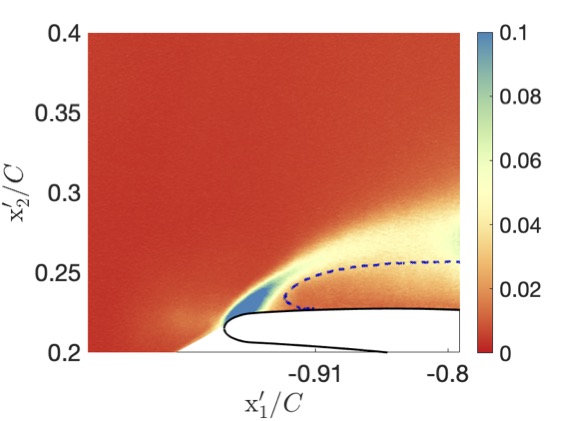} &
    \subfigimg[width=80 mm,pos=ul,vsep=21pt,hsep=38pt]{(d)}{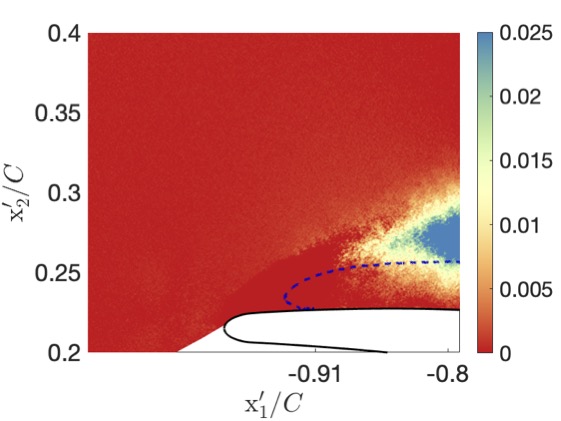} \\
  \end{tabular}
\caption{Contours of velocity statistics over the airfoil leading edge (blue dashed lines are the iso-contours of zero wall-parallel velocity $U_{\infty}$): (a) mean wall-parallel velocity $U_1$; (b) $\frac{\overline{u_1 u_1}}{U^2_{\infty}}$; (c) $\frac{\overline{u_2 u_2}}{U^2_{\infty}}$; (d) $-\frac{\overline{u_1 u_2}}{U^2_{\infty}}$. Coordinate system is aligned with the leading edge. %In the plots, the dashed blue lines are the iso-contours of zero wall-parallel velocity. ${\rm{U}_{\infty}}$
%Legends: (a) Mean wall-parallel velocity U$_1$ contours. (b) Contours of $\frac{\overline{u_1 u_1}}{U^2_{\infty}}$. (c) $\frac{\overline{u_2 u_2}}{U^2_{\infty}}$ (d) $\frac{\overline{-u_1 u_2}}{U^2_{\infty}}$.  
}
\label{fig:velo_stats_LE}
\end{figure*}

Figures \ref{fig:velo_stats_LE} and \ref{fig:velo_stats_MD} show the mean boundary-layer statistics recorded by the first and the second camera, respectively. Figure \ref{fig:velo_stats_LE} is plotted with respect to an observer sitting on the leading-edge of the airfoil while in figure \ref{fig:velo_stats_MD}, the coordinate system of the velocity field is aligned with the wind-tunnel axis. As evidenced in table \ref{tab_piv_tonal}, the spatial resolution achieved by the first camera is about three times higher than that of the second camera. Thus, further spatial filtering is expected in the results shown in figure \ref{fig:velo_stats_MD}. Figure~\ref{fig:velo_stats_LE} shows that, in a time-averaged sense, the mean flow becomes separated from RMP~3 (%X/C=0.09
$x/C=0.09$) onward. This is consistent with figure \ref{Cp}, which shows a plateau in $C_p$ between RMP~3 %(X/C=0.09) 
and RMP 6. More importantly, the separated region shown by %dashed blue 
the black dashed lines in figures~\ref{fig:velo_stats_LE} and \ref{fig:velo_stats_MD} has a negligible r.m.s value of velocity disturbances (all Reynolds stresses close to zero). This confirms the laminar nature of the time-averaged separated flow region, and in the literature, it is commonly referred to as the %Laminar Separation Bubble (LSB). 
LSB. The presence of such an LSB is characteristic of the flow past the CD airfoil at $Re_c=150000$ and is consistent with the finding of \cite{christophe2008}, who also reported the presence of a LSB when the CD airfoil is placed at a 15$^\circ$ incidence. The LSB near the leading-edge of the airfoil seems to deflect the mean flow away from the airfoil, which provides a possible explanation for a drop in mean-loading reported in figure \ref{Cp}. The deflected mean flow and resultant flow acceleration near the leading-edge, at the point of inception of the LSB, can be evidenced from an increase in the length of arrows in figure \ref{fig:velo_stats_LE}. In a time-averaged sense, the LSB seems to cover at least 30\% of the airfoil chord. However, due to the limited field-of-view, the exact extent of the LSB could not be quantified.
\begin{figure*}[ht!]
  \centering
  \begin{tabular}{@{}p{0.5\linewidth}@{\quad}p{0.5\linewidth}@{}}
    \subfigimg[width=80 mm,pos=ul,vsep=21pt,hsep=38pt]{(a)}{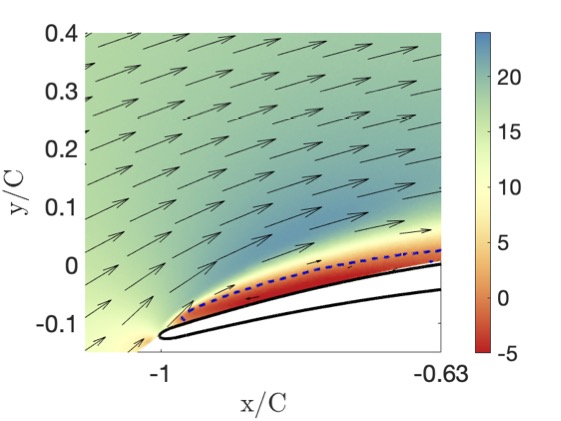} &
    \subfigimg[width=80 mm,pos=ul,vsep=21pt,hsep=38pt]{(b)}{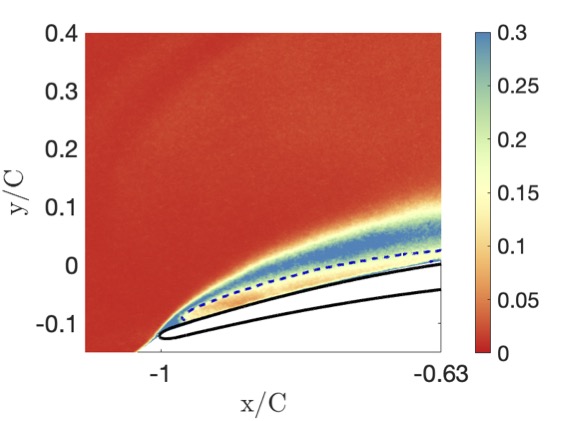} \\
     \subfigimg[width=80 mm,pos=ul,vsep=21pt,hsep=38pt]{(c)}{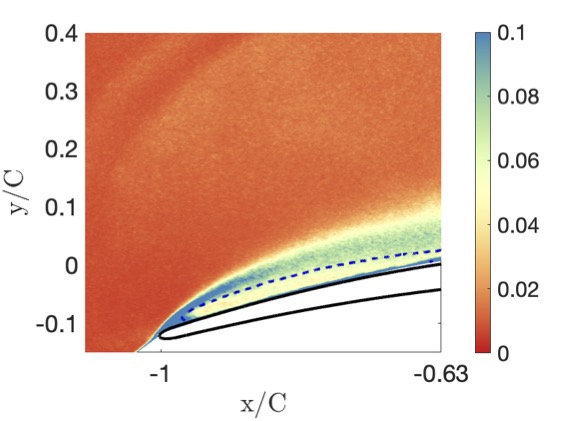} &
    \subfigimg[width=80 mm,pos=ul,vsep=21pt,hsep=38pt]{(d)}{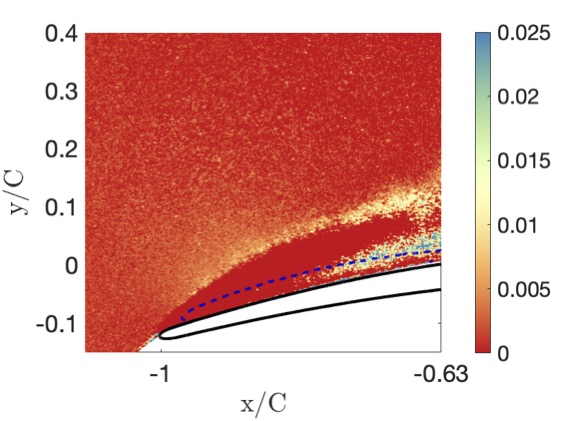} \\
  \end{tabular}
\caption{Contours of velocity statistics over the airfoil leading edge (blue dashed lines are the iso-contours of zero wall-parallel velocity $U_{\infty}$): (a) mean wall-parallel velocity $U_1$; (b) $\frac{\overline{u_1 u_1}}{U^2_{\infty}}$; (c) $\frac{\overline{u_2 u_2}}{U^2_{\infty}}$; (d) $-\frac{\overline{u_1 u_2}}{U^2_{\infty}}$. Coordinate system is aligned with the leading edge. %In the plots, the dashed blue lines are the iso-contours of zero wall-parallel velocity. ${U_{\infty}}$
%Legends: (a) Mean wall-parallel velocity U$_1$ contours. (b) Contours of $\frac{\overline{u_1 u_1}}{{\rm{U^2}_{\infty}}}$. (c) $\frac{\overline{u_2 u_2}}{{\rm{U^2}_{\infty}}}$ (d) $\frac{\overline{-u_1 u_2}}{{\rm{U^2}_{\infty}}}$.  
}
\label{fig:velo_stats_MD}
\end{figure*}
Overall, the mean flow topology presented in figures \ref{fig:velo_stats_LE} and \ref{fig:velo_stats_MD} show that the flow at the leading-edge region of the CD airfoil at 15$^\circ$ angle of attack and $Re_c\simeq 150000$ is laminar in nature. The LSB ensures the flow transition that occurs only after $x/C > 0.4$ as found in the previous LES~\citep{christophe2008}.   

\begin{figure*}[ht!]
  \centering
  \begin{tabular}{@{}p{0.5\linewidth}@{\quad}p{0.5\linewidth}@{}}
    \subfigimg[width=80 mm,pos=ul,vsep=21pt,hsep=38pt]{(a)}{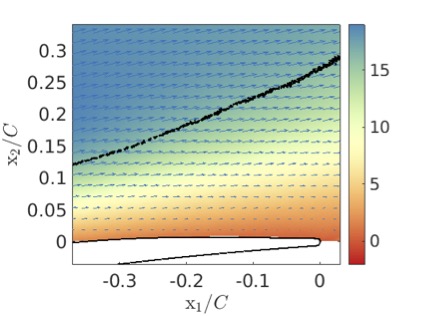} &
    \subfigimg[width=80 mm,pos=ul,vsep=21pt,hsep=38pt]{(b)}{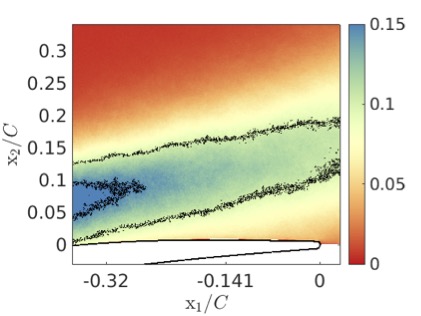} \\
     \subfigimg[width=80 mm,pos=ul,vsep=21pt,hsep=38pt]{(c)}{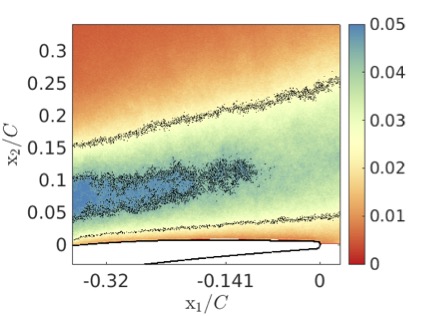} &
    \subfigimg[width=80 mm,pos=ul,vsep=21pt,hsep=38pt]{(d)}{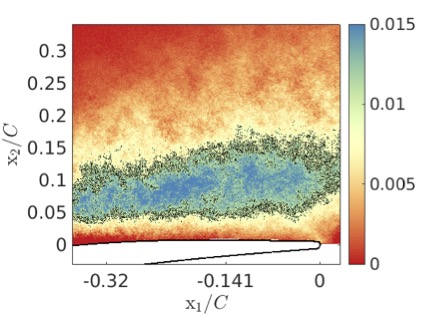} \\
  \end{tabular}
\caption{Contours of velocity statistics over the airfoil trailing edge for $U_{\infty}=16$~m/s: (a) mean wall-parallel velocity $U_1$ (black line corresponds to free-stream inlet velocity $U_{\infty}$), (b) $\frac{\overline{u_1 u_1}}{U^2_{\infty}}$ (black dashed lines indicate iso-values of $0.15$ and $0.1$); (c) $\frac{\overline{u_2 u_2}}{U^2_{\infty}}$, black dashed lines indicate iso-values of $0.045$ and $0.025$; (d) $-\frac{\overline{u_1 u_2}}{U^2_{\infty}}$, black dashed line indicates iso-values of $0.01$. Coordinate system is aligned with the trailing edge. %Legends: (a) Mean wall-parallel velocity U$_1$ contours, black contour line corresponds to free-stream inlet velocity ${\rm{U}_{\infty}}$. (b) Contours of $\frac{\overline{u_1 u_1}}{{\rm{U^2}_{\infty}}}$, dashed black lines indicate iso-contours values of $0.15$ and $0.1$. (c) $\frac{\overline{u_2 u_2}}{{\rm{U^2}_{\infty}}}$, dashed black lines indicate iso-contours values of $0.045$ and $0.025$. (d) $\frac{\overline{-u_1 u_2}}{{\rm{U^2}_{\infty}}}$, dashed black line indicate iso-contour values of $0.01$.   
}
\label{fig:velo_stats}
\end{figure*}

The mean boundary-layer statistics recorded by the third camera are shown in figure \ref{fig:velo_stats} for the case when the airfoil is placed at a $15^{\circ}$ angle of attack with an inlet velocity of 16~m/s. Figure \ref{fig:velo_stats}~(a) shows the mean wall-parallel velocity. Despite the large-scale flow separations observed in figure \ref{Inst_velo}, the boundary-layer near the trailing edge is fully attached in a time-averaged sense. As such, in the present pre-stall noise study, the time-averaged flow near the trailing-edge of the CD airfoil is different from the one reported by \cite{lacagnina2019mechanisms}, who reported a separated time-averaged flow near the trailing edge. The black dotted line is the iso-contour of the inlet velocity free-stream velocity, which roughly corresponds to the overall extent of the boundary layer. The Reynolds stress tensor terms, $\overline{u_1 u_1}/U^2_{\infty}$, $\overline{u_2 u_2}/U^2_{\infty}$, and $-\overline{u_1 u_2}/U^2_{\infty}$, are shown in figures \ref{fig:velo_stats} (b), (c), and (d), respectively. Compared to the leading-edge region, the disturbances (quantified by r.m.s of velocities) close to the trailing-edge are substantially higher, which implies that the flow transitions to a fully turbulent boundary-layer somewhere between 40 and 65\% of the chord. Higher levels of r.m.s velocities are the sources of far-field noise \citep{williams1970aerodynamic}. In particular, elevated regions of r.m.s velocity do not have a clear peak but a broad region of elevated intensity. This is typical of flows that experience the presence of shear-layer instabilities \citep{jaiswal2022experimental}.    

\begin{figure*}[ht!]
  \centering
  \begin{tabular}{@{}p{0.5\linewidth}@{\quad}p{0.5\linewidth}@{}}
    \subfigimg[width=80 mm,pos=ul,vsep=21pt,hsep=38pt]{(a)}{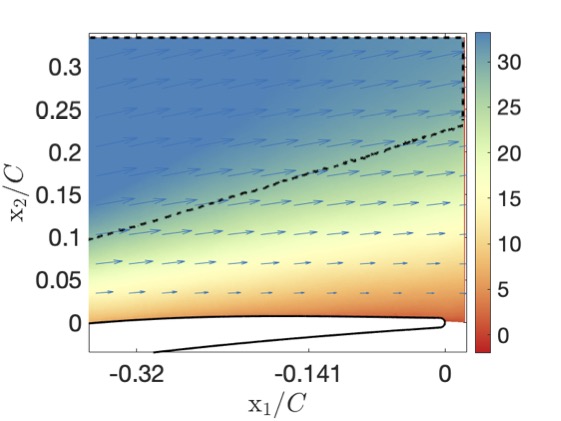} &
    \subfigimg[width=80 mm,pos=ul,vsep=21pt,hsep=38pt]{(b)}{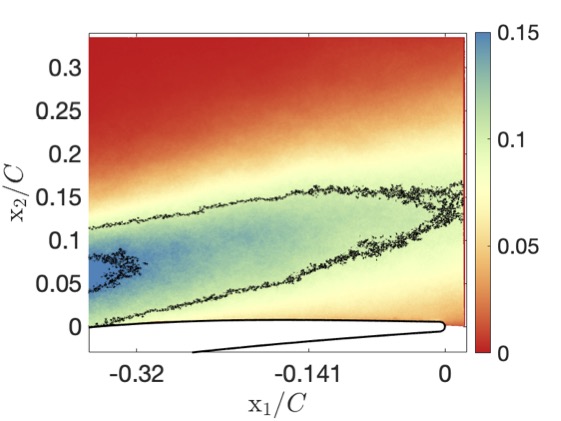} \\
     \subfigimg[width=80 mm,pos=ul,vsep=21pt,hsep=38pt]{(c)}{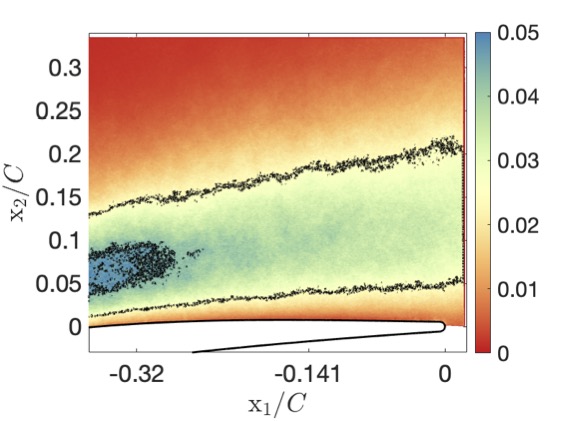} &
    \subfigimg[width=80 mm,pos=ul,vsep=21pt,hsep=38pt]{(d)}{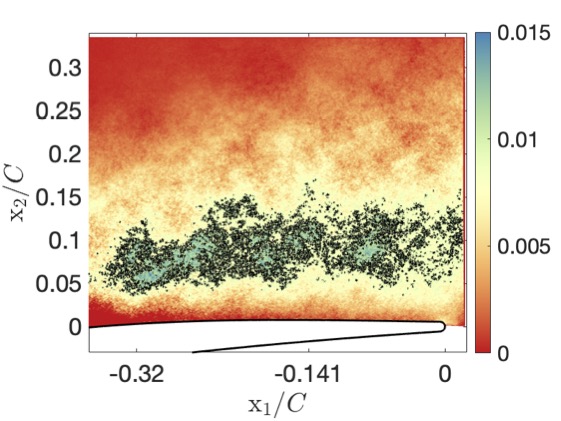} \\
  \end{tabular}
\caption{Contours of velocity statistics over the airfoil trailing edge for $U_{\infty}=28$~m/s: (a) mean wall-parallel velocity $U_1$ (black line corresponds to free-stream inlet velocity $U_{\infty}$), (b) $\frac{\overline{u_1 u_1}}{U^2_{\infty}}$ (black dashed lines indicate iso-values of $0.15$ and $0.1$); (c) $\frac{\overline{u_2 u_2}}{U^2_{\infty}}$, black dashed lines indicate iso-values of $0.045$ and $0.025$; (d) $-\frac{\overline{u_1 u_2}}{U^2_{\infty}}$, black dashed line indicates iso-values of $0.01$. Coordinate system is aligned with the trailing edge. %Legends: (a) Mean wall-parallel velocity U$_1$ contours, black contour line corresponds to free-stream inlet velocity ${\rm{U}_{\infty}}$. (b) Contours of $\frac{\overline{u_1 u_1}}{{\rm{U^2}_{\infty}}}$, dashed black lines indicate iso-contours values of $0.15$ and $0.1$. (c) $\frac{\overline{u_2 u_2}}{{\rm{U^2}_{\infty}}}$, dashed black lines indicate iso-contours values of $0.045$ and $0.025$. (d) $\frac{\overline{-u_1 u_2}}{{\rm{U^2}_{\infty}}}$, dashed black line indicate iso-contour values of $0.01$.   
}
\label{fig:velo_stats_28ms}
\end{figure*}

%\begin{figure*}[h!]
%  \centering
%  \begin{tabular}{@{}p{0.5\linewidth}@{\quad}p{0.5\linewidth}@{}}
 %   \subfigimg[width=75 mm,pos=ul,vsep=21pt,hsep=38pt]{(a)}{RMP21_meanu.eps} &
 %   \subfigimg[width=75 mm,pos=ul,vsep=21pt,hsep=38pt]{(b)}{RMP21_urms.eps} \\
 %    \subfigimg[width=75 mm,pos=ul,vsep=21pt,hsep=38pt]{(c)}{RMP21_vrms.eps} &
 %   \subfigimg[width=75 mm,pos=ul,vsep=21pt,hsep=38pt]{(d)}{RMP21_uvrms.eps} \\
 % \end{tabular}
%\caption{Comparison of velocity profiles at RMP 21 ($x/C=0.8582$). Legend: Solid black and red lines correspond to airfoil placed at $15^{\circ}$ angle-of-attack and at an inlet velocity of 16 and 28 [m/s] respectively. Dotted grey line corresponds to airfoil placed at $15^{\circ}$ angle-of-attack and at an inlet velocity of 16 [m/s].  }
%\label{fig:mean_velocity_profiles}
%\end{figure*}

\begin{figure*}[ht!]
  \centering
  \begin{tabular}{@{}p{0.5\linewidth}@{\quad}p{0.5\linewidth}@{}}
    \subfigimg[width=75 mm,pos=ul,vsep=21pt,hsep=38pt]{(a)}{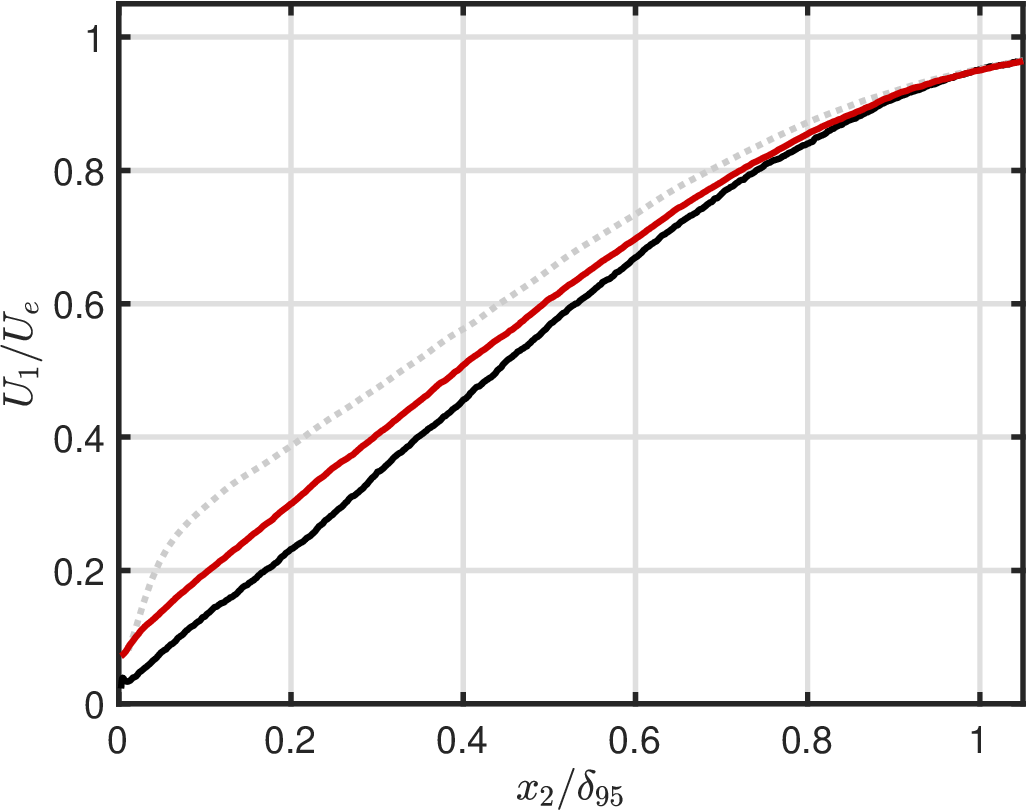} &
    \subfigimg[width=75 mm,pos=ul,vsep=21pt,hsep=38pt]{(b)}{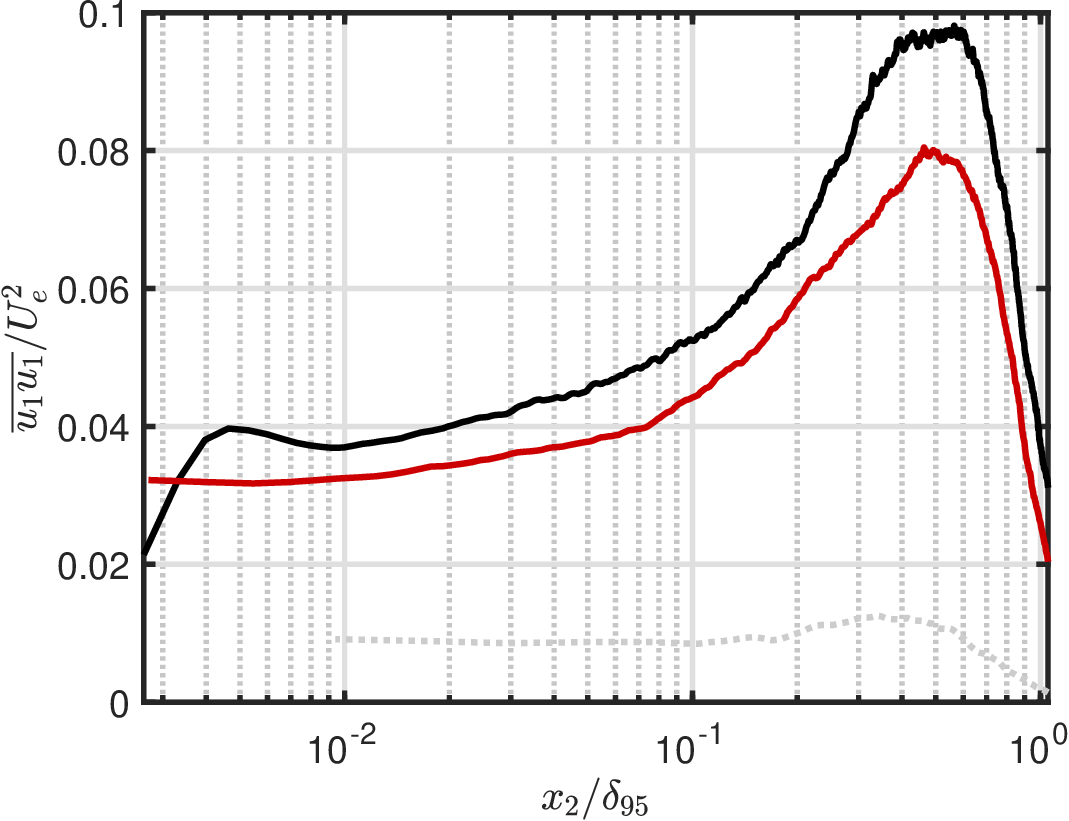} \\
     \subfigimg[width=75 mm,pos=ul,vsep=21pt,hsep=38pt]{(c)}{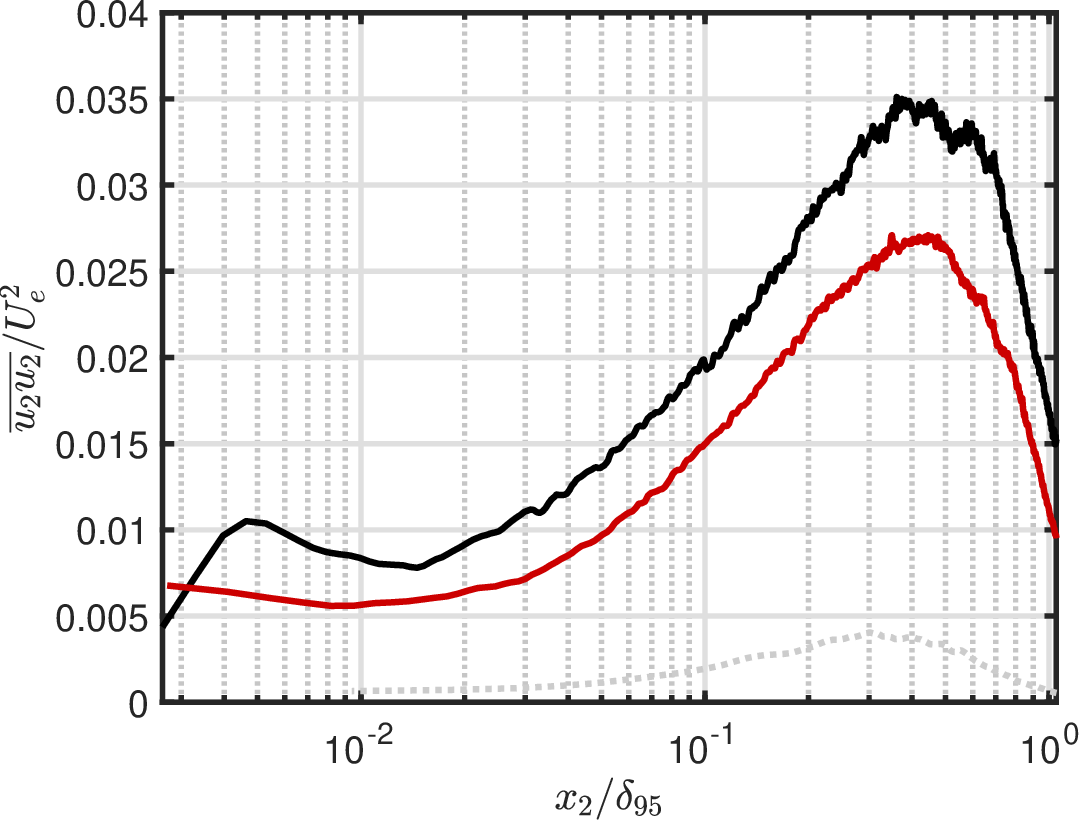} &
    \subfigimg[width=75 mm,pos=ul,vsep=21pt,hsep=38pt]{(d)}{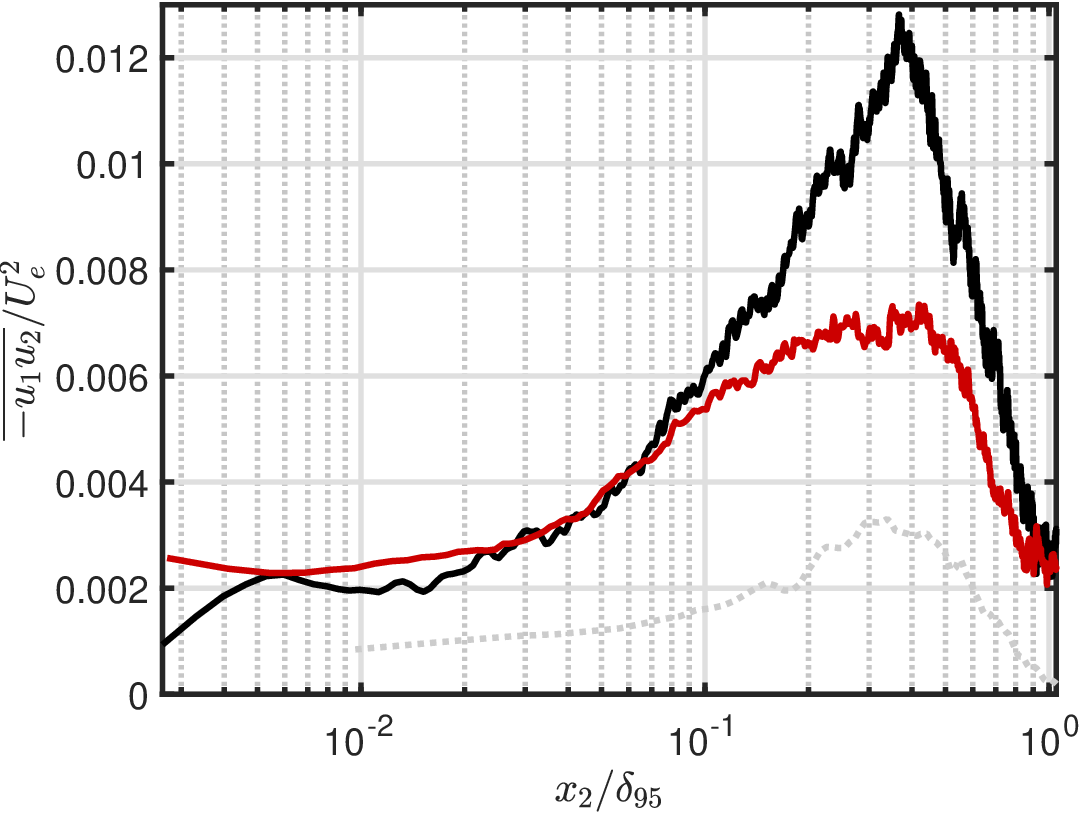} \\
  \end{tabular}
\caption{Comparison of velocity profiles at RMP 26 ($x/C=0.98$). Legend: Solid black and red lines correspond to airfoil placed at $15^{\circ}$ angle-of-attack and at an inlet velocity of 16 and 28~m/s respectively. Dotted grey line corresponds to airfoil placed at $8^{\circ}$ angle-of-attack and at an inlet velocity of 16~m/s.  }
\label{fig:mean_velocity_profiles}
\end{figure*}

\begin{table*}
% table caption is above
\caption{Boundary layer %integral 
%% THEY ARE NOT ALL INTEGRAL
parameters at RMP 21 ($x/C=0.8582$)}
\label{Integral_values_21}       % Give a unique label
\begin{tabular}{p{1.5cm}p{1cm}p{1cm}p{1cm}p{1cm}p{1cm}p{1cm}p{2cm}p{2cm}p{2cm}}
\hline\hline\noalign{\smallskip}
AOA & % $\frac{X}{C}$ 
{$U_{\infty}$} \newline [m/s] & $U_e$ [m/s] & $\delta_{95}$ \newline [mm] & $\delta^*$ \newline [mm] & $\theta$ \newline [mm] & $H$ & {Re$_{\theta}$} & $-\overline{u_1 u_2}_{max}$ \newline [m$^2$/s$^2$] & $\frac{p_{rms}}{-\rho~\overline{u_1 u_2}_{max}}$ %& \revPOF{ $\frac{p_{rms}}{-\rho  \overline{u_1 u_2}_{max}}$} 
\\
\noalign{\smallskip}\hline\noalign{\smallskip}
\textcolor{black}{$8^{\circ}$} & $16$ & $18.7$  & $4.41$ & $1.36$ & $0.82$ &$ 1.65$ & {997} & $0.77$ & $ 2.6$
\\
\noalign{\smallskip}
\textcolor{black}{$15^{\circ}$} & $16$ & $17.81$  & $28.97$ & $12.73$ & $5.06$ &$ 2.51$ & {5847} & $4.01$ & $3.08$   %& \revPOF{$-$} 
\\
\noalign{\smallskip}
\textcolor{black}{$15^{\circ}$} & $28$ & $32.8$  & $28.61$ & $11.53$ & $5.25$ &$ 2.19$ & {11188} & $8.68$ & $-$  
\\
\noalign{\smallskip}\hline\hline\noalign{\smallskip}
\end{tabular}
\end{table*}

\begin{table*}
% table caption is above
\caption{Boundary layer %integral 
%% THEY ARE NOT ALL INTEGRAL
parameters at RMP 26 ($x/C=0.98$)}
\label{Integral_values}       % Give a unique label
\begin{tabular}{p{1.5cm}p{1cm}p{1cm}p{1cm}p{1cm}p{1cm}p{1cm}p{2cm}p{2cm}p{2cm}}
\hline\hline\noalign{\smallskip}
AOA & % $\frac{X}{C}$ 
{$U_{\infty}$} \newline [m/s] & $U_e$ [m/s] & $\delta_{95}$ \newline [mm] & $\delta^*$ \newline [mm] & $\theta$ \newline [mm] & $H$ & {Re$_{\theta}$} & $-\overline{u_1 u_2}_{max}$ \newline [m$^2$/s$^2$] & $\frac{p_{rms}}{-\rho~\overline{u_1 u_2}_{max}}$ %& \revPOF{ $\frac{p_{rms}}{-\rho  \overline{u_1 u_2}_{max}}$} 
\\
\noalign{\smallskip}\hline\noalign{\smallskip}
\textcolor{black}{$8^{\circ}$} & $16$ & $17.44$  & $6.34$ & $2.42$ & $1.19$ &$ 2.03$ & {1350} & $1$ & $ 2.77$
\\
\noalign{\smallskip}
\textcolor{black}{$15^{\circ}$} & $16$ & $17.15$  & $34.88$ & $16.16$ & $6.03$ &$ 2.67$ & {6712} & $3.76$ & $2.22$   %& \revPOF{$-$} 
\\
\noalign{\smallskip}
\textcolor{black}{$15^{\circ}$} & $28$ & $31.88$  & $33.78$ & $14.35$ & $6.088$ &$ 2.35$ & {12577} & $7.46$ & $-$  
\\
\noalign{\smallskip}\hline\hline\noalign{\smallskip}
\end{tabular}
\end{table*}

In order to understand the impact of the Reynolds number, the results of the measurements performed at 28~m/s are plotted in figure \ref{fig:velo_stats_28ms}. Upon comparison with figure \ref{fig:velo_stats}, it shows similar overall behavior in the measured velocity field in the trailing-edge region. The overall length of the boundary layer is similar to the 16~m/s case at $x_2 \simeq 0.32$~$C$ and close to the trailing-edge region ($x_2 \simeq 0.02$~$C$), as shown in tables \ref{Integral_values_21} and \ref{Integral_values}. Furthermore, for the 28~m/s case, the turbulence intensity appears to be much lower than in the 16~m/s case, resulting in more localized levels of iso-contours in figure \ref{fig:velo_stats_28ms} compared to those in figure \ref{fig:velo_stats}. This is especially true for the cross-term $-\overline{u_1 u_2}/U^2_{\infty}$. 

A more quantitative comparison can be obtained by looking at the velocity profiles near the airfoil trailing edge, as shown in Figure \ref{fig:mean_velocity_profiles}. {The velocity profile at RMP 26 ($x/C=0.98$) shows when the CD airfoil placed at $15^{\circ}$ and 16~m/s flow incidence and inlet velocity respectively, the near wall mean velocity is reduced compared to the inlet velocity. Similar observations have been made by \cite{Caiazzo} (see figure 4), who reported a decrease in the near wall mean velocity as the mean pressure gradient increases.} As such, we expect the boundary layer to grow faster in the streamwise direction near the trailing-edge region for the 28~m/s case compared with the 16~m/s case. This faster growth of the boundary layer for the 28~m/s case is captured in the shape factor, which remains smaller at both RMP 21 and RMP 26 locations compared to 16~m/s %. Compare 
(see the values in Tables \ref{Integral_values_21} and \ref{Integral_values}, for instance). Yet, %they 
both velocity cases have higher values of the shape factor compared to the case when the airfoil is fixed at $8^{\circ}$ angle of attack and %the 
16~m/s. %case. 
Notably, a higher value of the shape factor indicates flow close to separation \citep[see Figure 10 of][]{sanjose2019modal}. Therefore, as the flow speed increases, the probability of flow separation decreases. Nevertheless, the overall boundary layer extent is similar for the 28~m/s and 16~m/s cases, as evidenced %by 
in Tables \ref{Integral_values_21} and \ref{Integral_values}. As such, the Reynolds number based on the momentum thickness (Re$_{\theta}$) for the airfoil placed at $15^{\circ}$ angle-of-attack is substantially higher than that of the $8^{\circ}$ case near the trailing-edge region.
The profiles of velocity statistics, namely $\overline{u_1 u_1}/U^2_\infty$, $\overline{u_2 u_2}/U^2_\infty$, and $-\overline{u_1 u_2}/U^2_\infty$, for the two velocity cases at $15^{\circ}$ angle of incidence and the case when the airfoil is placed at $8^{\circ}$ angle of attack and $U_\infty=16$~m/s are compared in figures \ref{fig:mean_velocity_profiles} (b-d). Generally, the velocity statistics are normalized with the friction velocity to remove any Reynolds number ($Re_{\tau}$) based effects. However, the overall goal of plot \ref{fig:mean_velocity_profiles} (b-d) is to demonstrate the levels of velocity disturbances with respect to the inlet velocity $U_{\infty}$. Such a scaling inherently shows the applicability of thin-airfoil linearized theory, which assumes that the velocity disturbances are small compared to the inlet velocity $U_{\infty}$. While the peak levels of velocity statistics scale with the boundary-layer thickness $\delta_{95}$, in absolute units (for instance in meters) they are much further away from the wall compared to the 8$^{\circ}$ angle-of-attack and 16~m/s case. More importantly, the profiles confirm that the r.m.s levels of velocity disturbances are elevated for the CD airfoil placed at 15$^{\circ}$ angle-of-attack at 16~m/s case compared to the rest. With the exception of the wall-parallel disturbances ($\overline{u_1 u_1}/U^2_\infty$) for the 15$^{\circ}$ angle-of-attack at 16~m/s case, the disturbances are at least an order of magnitude smaller in the rest of the cases tested. {Previous studies \citep[see figure 9 of][for instance]{Caiazzo} have reported an increase in r.m.s levels of velocity disturbances for wall bounded flows subjected to mean adverse pressure gradients. }

%If one assumes the validity of \cites{townsend1980structure} outer-layer hypothesis then the friction velocity should be considerably higher for the $15^{\circ}$ angle of attack cases. For the $15^{\circ}$ angle-of-attack and $16$ [m/s] case a dual hump in the velocity disturbance profiles can also be seen.  %Although the precise reason for this is unknown, it may be  characteristic of Kelvin-Helmholtz flow instability \citep{jaiswal2022experimental}, which can induce roller structures shown in figure \ref{Inst_velo}.

\begin{figure*}[ht!]
\centering
  \begin{tabular}{@{}p{0.45\linewidth}@{\quad}p{0.45\linewidth}@{}}
    \subfigimg[width=90 mm,pos=ul,vsep=21pt,hsep=42pt]{}{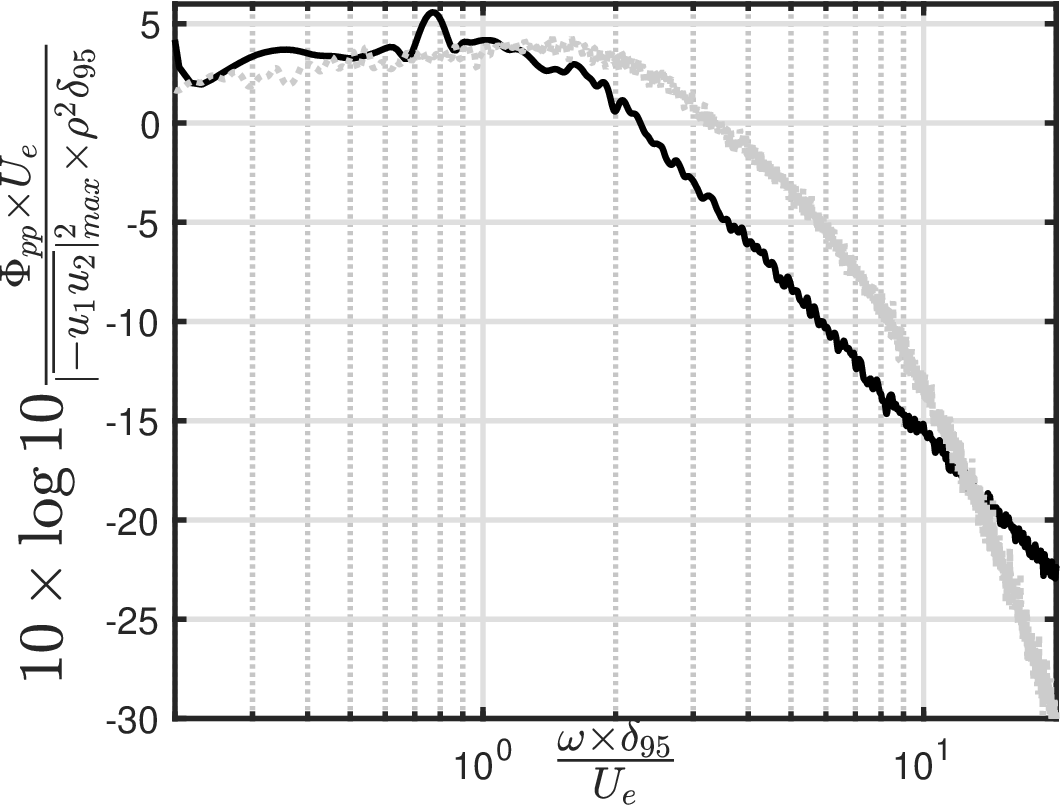} \\
  \end{tabular}
\caption{Power spectral density of the wall-pressure fluctuations at RMP $26$ at fixed inlet velocity $U_{\infty}=16$~m/s. The PSD has been normalized by the maximum value of the square of~$\overline{-u_1 u_2}$, the square of the free-stream density ${\rho}$, the local edge velocity $U_e$ and the boundary-layer thickness $\delta_{95}$. Legends: Dotted grey line for CD airfoil at an incidence of $8^{\circ}$ while solid black line represents CD airfoil at an incidence of $15^{\circ}$.}
\label{PSD_norm}       % Give a unique label
\end{figure*}

Recently,~\cite{pargal2023non} showed that normalizing the wall-pressure spectra by the square of the maximum value of the Reynolds stress, denoted by ${|\overline{-u_1 u_2}}|^2_{max}$, leads to a collapse in low-frequency spectra over a broad range of cases for boundary-layer flows subjected to arbitrary mean pressure gradient. This normalization holds true because, as first shown by~\cite{na1998structure}, %demonstrated that 
the term $p_{rms}/(-\rho~\overline{u_1 u_2}_{max})$ falls between 2 and 3 for boundary-layer flows. %Following, which several researchers 
This was later confirmed by~\citep{abe2017reynolds,le2020measurements} %have shown this to be the case 
for canonical boundary-layer flows, and more recently by \cite{Caiazzo} for flows past an airfoil. These observations are confirmed %by 
in~tables \ref{Integral_values_21} and \ref{Integral_values} for the present case. Small deviations from the aforementioned values can be ascribed to measurement uncertainty, and the presence of open jet, which predominantly contributes to low frequency wall-pressure spectra and which is absent in the aforementioned data \citep{pargal2023non,abe2017reynolds,le2020measurements,Caiazzo}. More importantly, when the scaling proposed by \cite{pargal2023non} is used to scale the wall-pressure spectra in Figure \ref{PSD_norm}, a collapse in the low-frequency range is achieved. This collapse is remarkable because the wall-pressure spectra exhibit a difference of $20$~dB, as shown in Figure \ref{diff_pressure}, corresponding to an order of magnitude difference in wall-pressure fluctuations.   

\begin{figure*}[ht!]
\centering
  \begin{tabular}{@{}p{0.45\linewidth}@{\quad}p{0.45\linewidth}@{}}
    \subfigimg[width=90 mm,pos=ul,vsep=21pt,hsep=42pt]{}{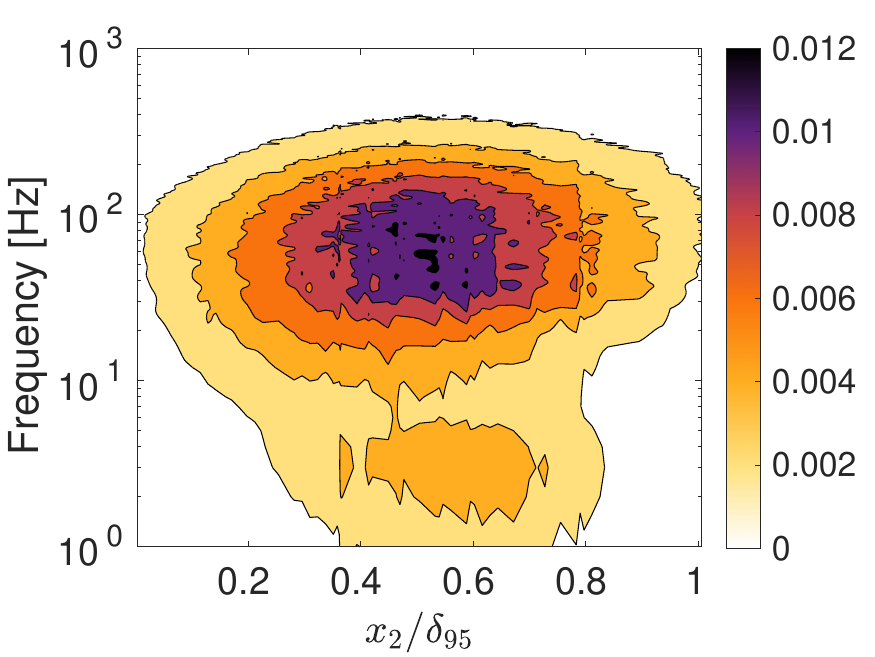} \\
  \end{tabular}
\caption{Premultiplied 1-D velocity energy spectra $\overline{E_{11}}$  $\left(f \times E_{11}/U_{e}^2\right)$ as a function of frequency, over the airfoil at RMP 26.}
\label{HWA_spectra}       % Give a unique label
\end{figure*}

As the PIV velocity measurements were not time-resolved, additional single wire measurements were performed. As mentioned, single HWA measurements were performed close to the trailing-edge (RMP 26) of the airfoil. These measurements were done for airfoil placed at $15^{\circ}$ angle of attack and fixed inlet velocity of $U_{\infty}=16$~m/s. Figure \ref{HWA_spectra} shows the pre-multiplied spectrogram $f \times E_{11}/U_{e}^2$. The plot shows that the pre-multiplied energy spectrum peaks at about 100~Hz, away from the wall ($ 0.4-0.6 \times \delta_{95}$), approximately the location where the peak in r.m.s. of velocity fluctuations was reported in figure \ref{fig:mean_velocity_profiles}. 

In summary, figures \ref{fig:mean_velocity_profiles} to \ref{HWA_spectra} show that large scale flow disturbances may be present and confirm the instantaneous snapshots in figure \ref{Inst_velo}. These large structures are in turn responsible for elevated %of 
levels of r.m.s. %of 
velocity fluctuations and the peak in the pre-multiplied spectrogram. As such, modal decomposition could be useful to understand the hierarchy and the organization of velocity disturbances close to the trailing edge.

\section{Modal analysis}
The Proper Orthogonal Decomposition (POD) \citep{holmes2012turbulence} was employed to uncover the modes present in the velocity disturbance field. One benefit of using POD is that, unlike linear stability analysis, it does not require velocity disturbances to be small. In the present paper, POD was carried out using the snapshot approach of the algorithm developed by \cite{sirovich1987turbulence}. For more information, please refer to the monograph by \cite{holmes2012turbulence}. The modal energy distribution of the measured velocity field is shown in figure \ref{Energy_modes}. The spatial POD modes are used to identify the spatial organization of the velocity disturbance field and their associated energy levels ($\rm{E}_r$), and are plotted in figure \ref{fig:spatial_modes}. In the present manuscript, only the spatial modes associated with the vertical velocity disturbances (${E_{22}}$) are used because they are the principal drivers of wall-pressure fluctuations and far-field acoustics \citep{jaiswal2020use}. Figure \ref{Energy_modes}~(b) clearly shows that the first 12 modes contribute to approximately 40\% of the total energy, although the cumulative energy for the 16~m/s case appears to be slightly lower compared to the 28~m/s case. The relative contributions for the 16~m/s and 28~m/s are shown for the first 12 modes. As can be seen, the relative energies of modal pairs 3-4, 5-6, and 11-12 appear to be similar and may form a modal pair. However, upon inspection, it was found not to be the case %. %For example, 
(see figure \ref{fig:spatial_modes} for example). 
Yet, the spatial organization appears to be similar between the 16~m/s and 28~m/s cases. Moreover, in these figures, the dashed black lines that represent the time-averaged location where the wall-parallel velocity is equal to the free-stream velocity $U_{\infty}$ show that the spatial modes are distributed across the boundary layer. In contrast, modal decomposition performed by \cite{lacagnina2019mechanisms} had shown that the spatial modes are uniquely present outside the time-averaged extent of the shear layer. In fact, the spatial distribution of the velocity disturbance field looks similar to the instantaneous %image 
field in figure~\ref{Inst_velo}~(a), and it could be due to the passage of coherent structures, and it may correspond to the disturbance at the frequency range of $80-300$~Hz. More importantly, the spatial extent or wavelength of this modal pattern (mode 3) closely corresponds to the peak in the pre-multiplied spectrogram (figure \ref{HWA_spectra}). As such, it may be responsible for the hump in the low-frequency wall-pressure (figure \ref{diff_pressure}) and far-field acoustic spectra (figure \ref{ff_spectra}). To verify this, the correlation between mode 3 and band-passed pressure will be performed next. 

%A second low frequency peak can also be seen in the figure \ref{HWA_spectra}. Although its peak frequency range corresponds to the region where VLSM  

\begin{figure*}[ht!]
  \centering
  \begin{tabular}{@{}p{0.45\linewidth}@{\quad}p{0.45\linewidth}@{}}
    \subfigimg[width=75 mm,pos=ur,vsep=21pt,hsep=42pt]{(a)}{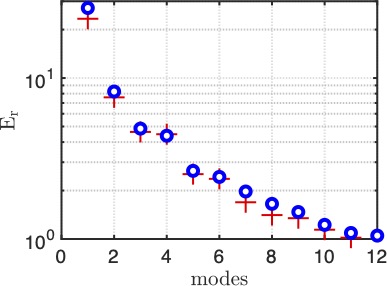} &
    \subfigimg[width=75 mm,pos=ur,vsep=21pt,hsep=42pt]{(b)}{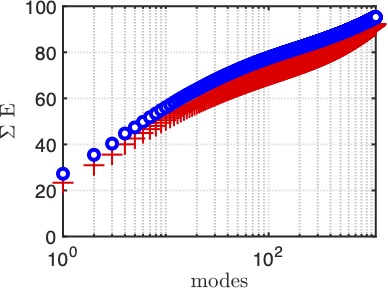} \\
  \end{tabular}

\caption{POD based modal decomposition of the measured velocity field in the trailing-edge region (camera 3) for an airfoil placed at $15^\circ$ %degrees 
angle of attack. (a) Relative energy of individual POD modes; (b) Cumulative sum of POD modes. Legends: Red cross for inlet velocity of $U_{\infty}=16$~m/s while the blue circles represent inlet velocity of $U_{\infty}=28$~m/s.} 
%\end{center}
\label{Energy_modes} 
\end{figure*}

\begin{figure*}[ht!]
  \centering
  \begin{tabular}{@{}p{0.5\linewidth}@{\quad}p{0.5\linewidth}@{}}
    \subfigimg[width=80 mm,pos=ul,vsep=21pt,hsep=38pt]{(a)}{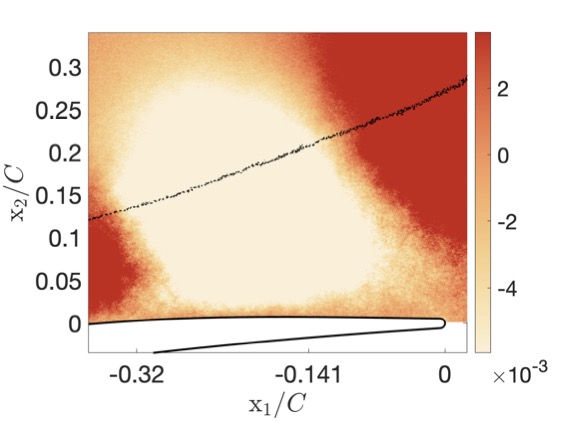} &
    \subfigimg[width=80 mm,pos=ul,vsep=21pt,hsep=38pt]{(b)}{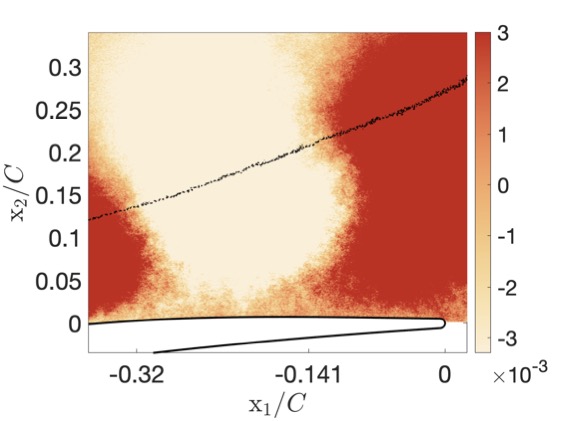} \\
     \subfigimg[width=80 mm,pos=ul,vsep=21pt,hsep=38pt]{(c)}{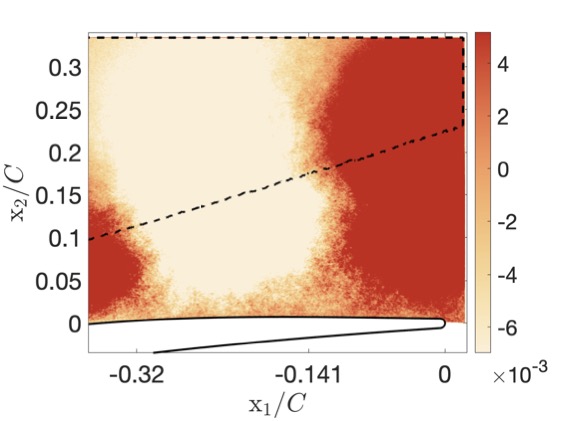} &
    \subfigimg[width=80 mm,pos=ul,vsep=21pt,hsep=38pt]{(d)}{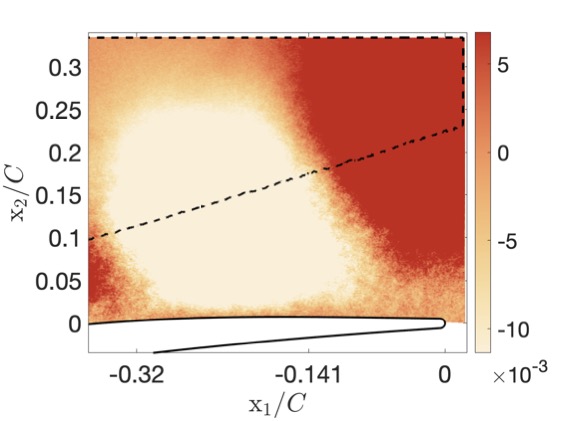} \\
  \end{tabular}
\caption{Modes of the vertical velocity disturbances, $E_{22}$, measured at the trailing-edge of the airfoil. Coordinate system is aligned with the trailing edge. Left plots (a) and (c) correspond to mode 3 while %figures on the 
right ones are for mode 4. Top figures %on the top 
(a) and (b) correspond to %the case with inlet velocity equal to 
$U_{\infty}=16$ m/s while bottom figures %on the bottom 
correspond to $U_{\infty}=28$ m/s. Black contour lines show the corresponding mean free-stream inlet velocity $U_{\infty}$.}
\label{fig:spatial_modes}
\end{figure*}

\begin{figure*}[ht!]
  \centering
  \begin{tabular}{@{}p{0.45\linewidth}@{\quad}p{0.45\linewidth}@{}}
   \subfigimg[width=75 mm,pos=ur,vsep=21pt,hsep=42pt]{(a)}{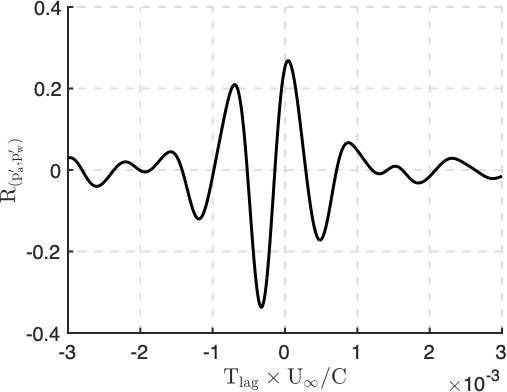} &
    \subfigimg[width=75 mm,pos=ur,vsep=21pt,hsep=42pt]{(b)}{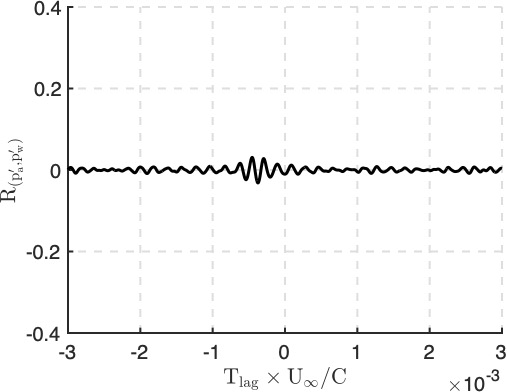} \\
  \end{tabular}

\caption{Cross-correlation between filtered wall-pressure and far-field pressure for $15^{\circ}$ angle-of-attack and $U_{\infty}$=16 m/s. (a) Cross-correlation for pressure signals filtered between 100-300 [Hz]; (b) Cross-correlation for pressure signals filtered between 600-2100 [Hz].}
%\end{center}
\label{Correlation} 
\end{figure*}

\section{Correlation analysis}

Having characterized the velocity and pressure field, the manuscript will now attempt to delineate the correlation between these two quantities of interest. Correlation between flow quantities are quantified using Pearson's correlation coefficient. Pearson correlation coefficient at two different locations $(x_1,x_2,x_3)$ and $({x_1}^{\prime},{x_2}^{\prime},{x_3}^{\prime})$ is denoted by{:}% equation (\ref{correlation_twopoint}).
{\begin{equation} 
\textcolor{black}{R_{\zeta\chi}({x_1},{x_1}^{\prime},x_2,{x_2}^{\prime},x_3,{x_3}^{\prime}) =   \frac{\overline{{\zeta_{}}(x_1,x_2,x_3) {\chi_{}}({x_1}^{\prime},{x_2}^{\prime},{x_3}^{\prime}) }}{\sqrt{{\overline{\zeta(x_1,x_2,x_3})^2}} \times \sqrt{\overline{{\chi}({x_1}^{\prime},{x_2}^{\prime},{x_3}^{\prime})^2}}}}
  \label{two_point}         
\end{equation}}
where $ {\zeta}(x_1,x_2,x_3) $ and ${\chi}({x_1}^{\prime},{x_2}^{\prime},{x_3}^{\prime})$ are the fluctuating components of variables of interest.

The Pearson correlation method is used in pattern recognition %, and 
to quantify the similarity between patterns or features in data. For example, in time series data, %where 
the correlation between the values of two time series at different time points can be used to quantify the similarity between the patterns of the time series. However, correlation alone does not establish causation, as correlation cannot yield causal asymmetry and hence cannot separate the cause from the effect \citep{bossomaierintroduction}. As such, in the present section the overall goal is recognize pattern in velocity disturbance field that are {\it similar} to ones measured in time series of pressure signals recorded at the wall or at far-field locations. This can aid %one 
to identify velocity disturbance pattern associated with separation noise. The causality is inferred through \cites{amiet1976noise} equation \eqref{amiet_final} and Poisson's equation \citep[see][for instance]{grasso2019analytical}, which relates velocity disturbance to wall-pressure fluctuations.

\subsection{{Wall and far-field pressure correlation analysis}}

To identify patterns in measured time series of wall-pressure and far-field acoustic pressure the correlation, $R_{p'_w,p'_a}$, %between them was 
has been calculated. To segregate the separation noise, %the
both signals %were 
have been band-passed filtered between $80-300$~Hz and $600-2100$~Hz, where contributions from separation noise can be ignored (see figure \ref{HWA_spectra}). The results are shown in figures \ref{Correlation}~(a-b). A negative correlation between the wall-pressure fluctuations $p'_w$ and the far-field acoustic ones $p'_a$ is measured when these signals %were 
have been band-passed filtered between $80-300$. This can be caused by the passage of eddies at these frequencies near the trailing-edge and their diffraction in the form of acoustic pressure at a far-field location (see also figure \ref{HWA_spectra}). The phase opposition between near-field and far-field is due to the dipole nature of the source term. In contrast, for the band passed frequencies between $600-2100$~Hz, no meaningful correlation is obtained. This already suggests a significant low-frequency contribution of the surface noise sources caused by the largest turbulent coherent structures.    
%\subsection{{Velocity and wall-pressure correlation}}

%\subsection{{Velocity and far-field correlation analysis}}

\subsection{{Correlation between POD modes and  pressure}}

The temporal signals associated with mode 3, %was 
has been correlated with the band-passed filtered wall-pressure and far-field pressure signals. The frequency band for the separation noise %was 
has been chosen to be between $80-300$~Hz and $600-2100$~Hz, as in figure \ref{Correlation}. Once again the band passed filtering %was 
has been achieved using a zero-phase digital filtering, which conserves the phase. Figures \ref{fig:correlation_modes}~(a) and (b) show the correlation between the third mode ($R_{E_{22},p'_w}$) and the wall-pressure fluctuations measured by  RMP $26$ ($x/C=0.98$). In these plots, $T_f$ corresponds to the time of flight of an acoustic signal emitted at the trailing-edge of the airfoil to reach the far-field location where the noise is measured. The third mode of wall-normal velocity fluctuations ($E_{22}$) and the recorded wall-pressure signals $p'_w$ show a meaningful correlation only at the band-pass frequency range of $80-300$~Hz (figure \ref{fig:correlation_modes}~(a)), while the correlation drops to background noise levels for the higher frequency band $600-2100$~Hz (figure \ref{fig:correlation_modes}~(b)). Similarly for the correlation between mode 3 and the far-field acoustic pressure $p'_a$ meaningful results are only obtained for the lower frequency band $80-300$~Hz. The only difference is that it takes the near field hydrodynamic event a finite time to reach the far-field location, where the acoustic measurements are achieved. As such, before the cross-correlation was performed, the time series of the acoustic signal was shifted by the time of flight $T_f$. More importantly a phase opposition is seen in figure \ref{fig:correlation_modes}~(c)) between mode 3 and the far-field acoustic pressure. This is not surprising as the third mode of the wall-normal velocity disturbances ($E_{22}$) and the recorded wall-pressure signals are in phase (see figure \ref{fig:correlation_modes}~(c)) while the acoustic pressure and wall-pressure field are in phase opposition (figure \ref{Correlation}).

\begin{figure*}[ht!]
  \centering
  \begin{tabular}{@{}p{0.5\linewidth}@{\quad}p{0.5\linewidth}@{}}
    \subfigimg[width=75 mm,pos=ur,vsep=21pt,hsep=42pt]{(a)}{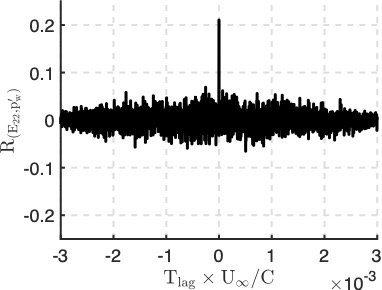} &
    \subfigimg[width=75 mm,pos=ur,vsep=21pt,hsep=42pt]{(b)}{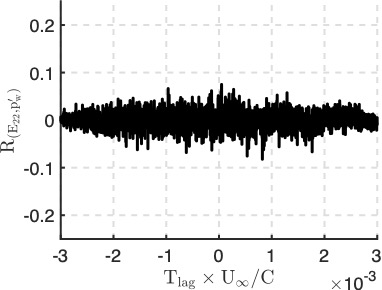} \\
     \subfigimg[width=75 mm,pos=ur,vsep=21pt,hsep=42pt]{(c)}{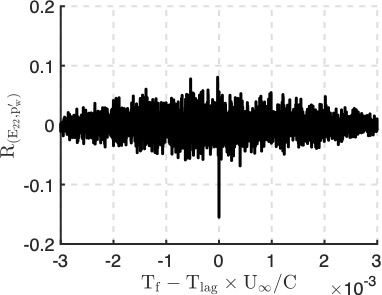} &
    \subfigimg[width=75 mm,pos=ur,vsep=21pt,hsep=42pt]{(d)}{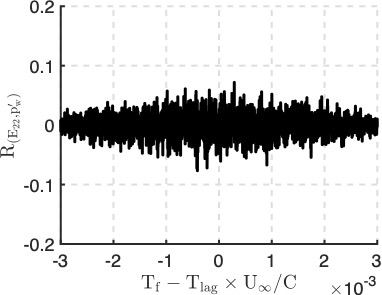} \\
  \end{tabular}
\caption{Cross-correlation between the third mode of wall-normal velocity fluctuations (E$_{22}$), filtered wall-pressure fluctuations $p'_w$ and far-field acoustic pressure $p'_a$. Legends: (\textit{a}) and (\textit{b}) $R_{E_{22},p'_w}$; (\textit{c}) and (\textit{d}) $R_{E_{22},p'_a}$. Left figures %on the left 
(\textit{a}) and (\textit{c}) correspond to pressure fluctuations band-passed filtered between $80-300$ [Hz] while right figures %on right 
correspond to pressure fluctuations band-passed filtered between $600-2100$ [Hz].} %$T_f$ corresponds to the time of flight of an acoustic signal emitted at the trailing-edge of the airfoil to reach at the far-field location where the noise is measured.}
\label{fig:correlation_modes}
\end{figure*}

To conclude, the near-field source terms are amplified in the case when the airfoil is placed at high angles of attack through induced flow separation. The separated shear layer can induce Kelvin-Helmholtz like roller structures, %whose 
the imprint of which are registered by the surface pressure probes. This results in an increased noise content, at a frequency that is associated with the wavelength of these roller structures. Having characterized the near-field source terms and its correlation with far-field acoustics, the diffracted acoustic pressure field around the airfoil %will be 
is then quantified. Finally attempts %will be 
are made to identify the equivalent source images responsible for the separation noise. 

%In the sections that follow will quantify the diffractred acoustic pressure field around the airfoil for an airfoil close to stall.

\section{Far-field acoustic pressure analysis}

The far-field acoustic pressure %were 
has been measured around the airfoil mid-chord to compare the influence of angles of attack on the acoustic directivity patterns. This %was 
has been done at several frequencies, and hence at several Helmholtz numbers $kc$, where $k$ %and $c$ are 
is the acoustic wavenumber. %and airfoil chord length respectively. 
The results are shown in figure \ref{fig:directivity}. While there %seems to be 
is an overall increase in absolute levels of the measured sound pressure levels, the overall sound directivity pattern is similar between the $8^{\circ}$ and $15^{\circ}$ angles of attack cases, where the former is known to emit noise through an equivalent dipole at the trailing edge \citep{wu2020noise}. As such, classical dipole noise at the airfoil trailing-edge seems to be the driver of separation noise. In contrast, at higher Mach numbers ($M_\infty=0.3-0.4$) than the ones reported in present study, \cite{turner2022quadrupole} had reported a significant contribution from the quadrupole noise sources.

\begin{figure*}[ht!]
  \centering
  \begin{tabular}{@{}p{0.5\linewidth}@{\quad}p{0.5\linewidth}@{}}
    \subfigimg[width=80 mm,pos=ul,vsep=21pt,hsep=38pt]{(a)}{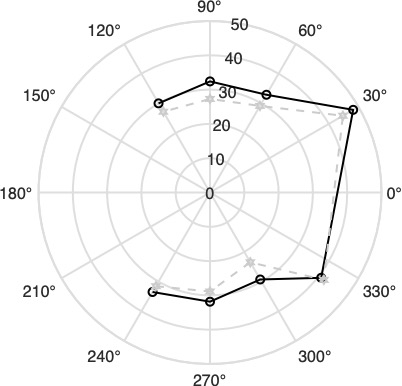} &
    \subfigimg[width=80 mm,pos=ul,vsep=21pt,hsep=38pt]{(b)}{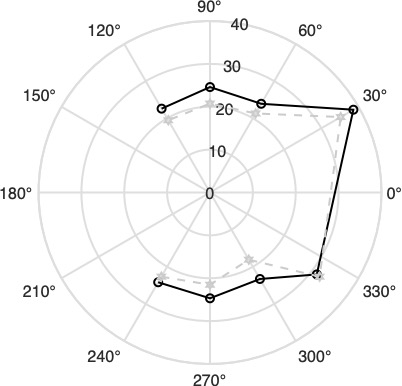} \\
     \subfigimg[width=80 mm,pos=ul,vsep=21pt,hsep=38pt]{(c)}{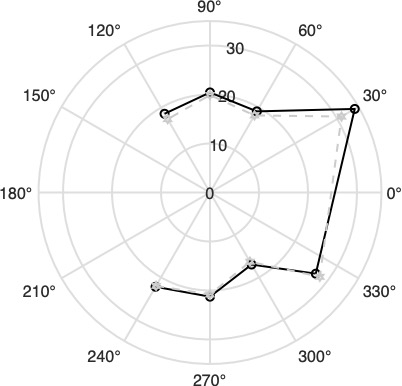} &
    \subfigimg[width=80 mm,pos=ul,vsep=21pt,hsep=38pt]{(d)}{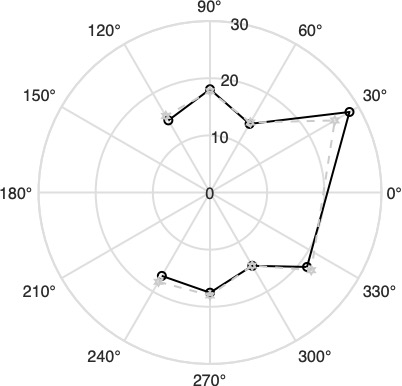} \\
  \end{tabular}
\caption{Sound Pressure Level directivity measured at 1.21~m from the trailing-edge. Microphone locations are shown with respect to the wind tunnel axis. %Legends: 
Solid black lines for airfoil at 15$^\circ$ Grey broken lines for airfoil at 8$^\circ$. (a) 100~Hz ($kC=0.24$) (b) 300~Hz ($kC=0.74$) (c) 500~Hz ($kC=1.23$) (d) 1000~Hz ($kC=2.46$).   }
\label{fig:directivity}
\end{figure*}

\begin{figure*}[ht!]
  \centering
  \begin{tabular}{@{}p{0.45\linewidth}@{\quad}p{0.45\linewidth}@{}}
    \subfigimg[width=75 mm,pos=ul,vsep=21pt,hsep=42pt]{(a)}{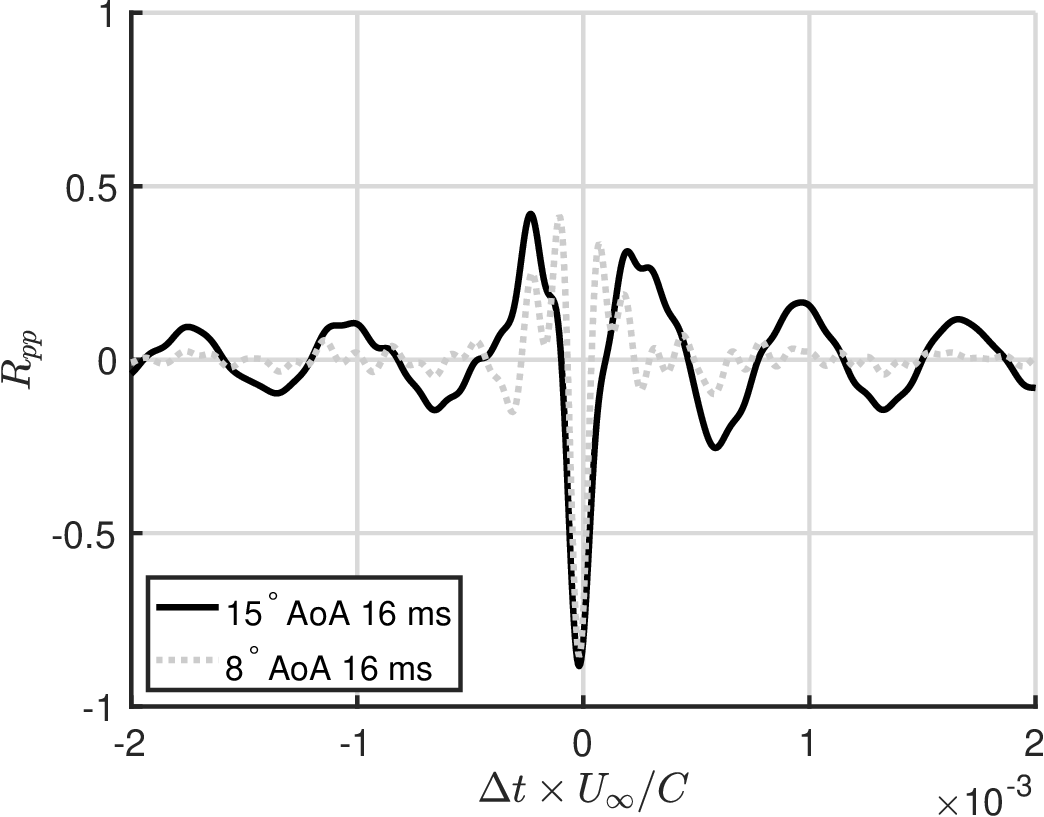} &
    \subfigimg[width=75 mm,pos=ul,vsep=21pt,hsep=42pt]{(b)}{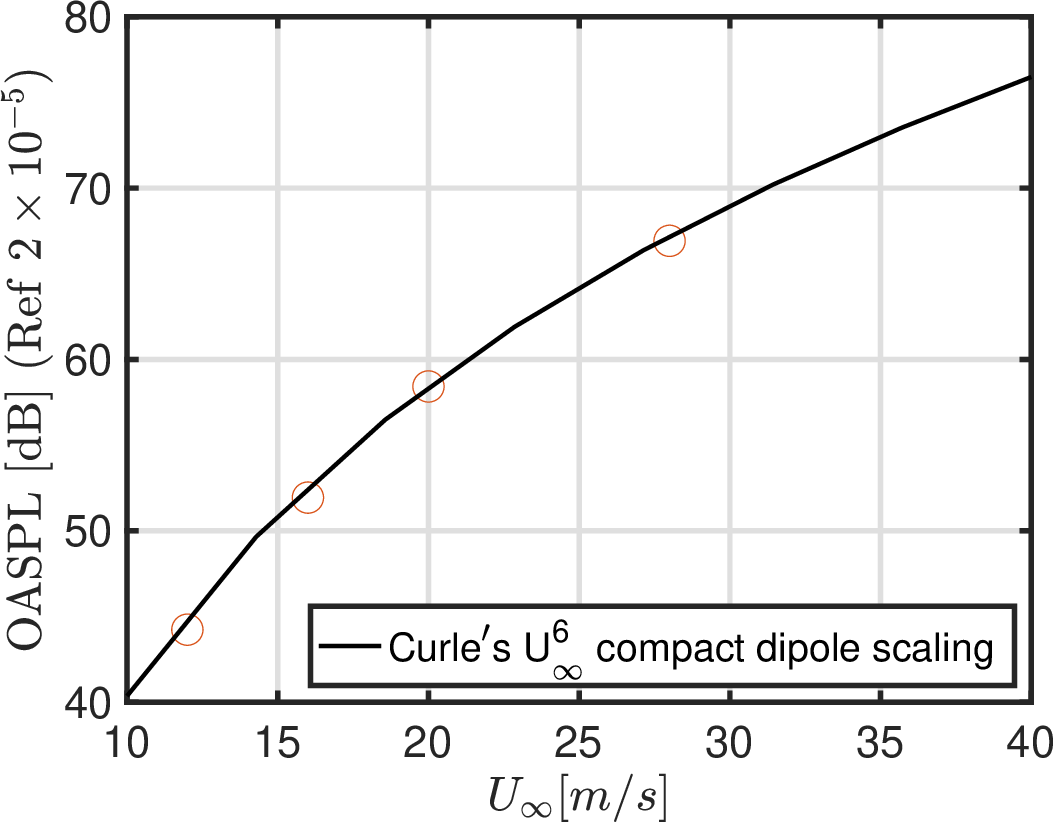} \\
  \end{tabular}
\caption{Far-field sound pressure correlation and scaling %for CD airfoil 
at $15^{\circ}$. %angle of attack. 
The sound pressure signals are filtered between 80 and 1000~Hz to single out flow separation contributions. %Legends: 
(a) Cross-correlation between far-field microphones located perpendicular to airfoil chord. (b) OverAll Sound Pressure Level (OASPL)~%dB 
as a function of $U_{\infty}$. }%for CD airfoil at $15^{\circ}$ angle of attack. }%Solid black line represents OASPL as function $U^6_{\infty}$, e.g. compact dipole scaling \citep[see][for instance]{curle1955influence}.}
\label{Xcorr_farfield}       % Give a unique label
\end{figure*}

To further investigate the overall contribution of quadrupole noise generated, due to separated shear layers, the cross-correlation between two far-field microphones located on either side of the airfoil mid-chord was performed. To isolate the influence of separation noise, the far-field noise signals were band-passed filtered between 80 and 1000~Hz. The %results are compared 
comparison at 16~m/s between the two angles of attacks, $8^{\circ}$ and $15^{\circ}$ are shown in figure \ref{Xcorr_farfield}~(a). %, for fixed inlet velocity of 16~m/s. 
The clear phase opposition decisively demonstrates that the dominant noise source is dipolar in nature. To further reinforce these findings, the OverAll Sound Pressure Level (OASPL) as a function of free-stream velocity $U_{\infty}$ is shown in figure \ref{Xcorr_farfield}~(b). Once again, to isolate overall influence of separation noise, the sound pressure levels have been integrated between $80-1000$~Hz, where the separation noise dominates. The results clearly show that the OASPL due to noise separation follows the classical compact dipole scaling $U^6_{\infty}$, which was first proposed by \cite{curle1955influence}.

\begin{figure*}[ht!]
\centering
  \begin{tabular}{@{}p{0.45\linewidth}@{\quad}p{0.45\linewidth}@{}}
    \subfigimg[width=85 mm,pos=ul,vsep=21pt,hsep=42pt]{}{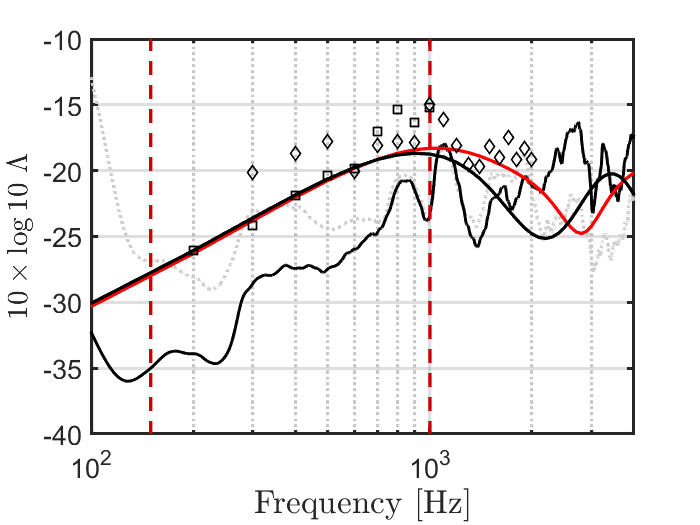} \\
  \end{tabular}
\caption{%Legend: %Non-dimensionalised 
Dimensionless radiation ratio %for measurements performed 
at 16~m/s. Legend: Dotted and solid black line for $15^{\circ}$; %angle of attack case %while the 
dotted gray line for $8^{\circ}$. % angle of attack case. 
Red dashed lines correspond the range where the experimental quantification of spanwise correlation was possible. Solid black and red lines are the theoretical predictions for an observer at $258^{\circ}$ and $265^{\circ}$ with respect to airfoil chord at a distance of 1.21 m. Hollow diamond and square correspond to measurements performed by \cite{moreau2005effect}}\label{Radiation_ratio}       % Give a unique label
\end{figure*}

%\begin{figure*}[h!]
%\centering
 % \begin{tabular}{@{}p{0.45\linewidth}@{\quad}p{0.45\linewidth}@{}}
  %  \subfigimg[width=75 mm,pos=ul,vsep=21pt,hsep=42pt]{}{spl_scaling_prms.eps} &
   % \subfigimg[width=75 mm,pos=ul,vsep=21pt,hsep=42pt]{}{spl_scaling.eps}
 %   \\
 % \end{tabular}
%\caption{Legend: Normalised power spectral density of far-field acoustic pressure measured at a distance of ${\sim 1.21}$~m by a microphone perpendicular to the airfoil chord. Legend: Solid black and red lines correspond to airfoil placed at $15^{\circ}$ angle-of-attack and at an inlet velocity of 16 and 28 [m/s] respectively. Dotted grey line corresponds to airfoil placed at $15^{\circ}$ angle-of-attack and at an inlet velocity of 16 [m/s]. }
%\label{Radiation_ratio}       % Give a unique label
%\end{figure*}

Having shown that the separation noise can be represented by an equivalent compact  dipole source, %the manuscript will 
we now attempt to %show as to weather 
check whether it can be quantified using the diffraction theory outlined above in equation~(\ref{amiet_final})~\citep{amiet1976noise}. The success of %\cite{amiet1976noise} 
Amiet's model and its extension relies on the fact that the response of the airfoil to an incident gust can be predicted using the linearized thin-airfoil theory. Following \cite{roger2004broadband,moreau2009back}, %non-dimensionalised
the dimensionless radiation ratio $\Lambda$ is plotted in figure \ref{Radiation_ratio}. %In the figure \ref{Radiation_ratio}, r
Red dashed lines correspond the frequency range for which the estimation of the spanwise correlation length was performed. The high frequency region is limited to 1~kHz because the spanwise coherence for the $15^{\circ}$ angle-of-attack case drops drastically below the measurement uncertainty beyond this frequency range. The radiation ratio quantifies the diffraction efficacy for an airfoil trailing-edge that is subjected to an unsteady pressure gust. To recall, the radiation ratio is defined as the ratio of far-field and near-field spectra normalised by spanwise length scales, far-field observer distance and the airfoil span length, in the following manner:
\begin{equation} \label{radiation}
\Lambda = \frac{S_{pp}}{\Phi_{pp} \times l_z} \times \frac{\textbf{x}^2}{L}
\end{equation}
As argued by \cite{roger2004broadband,moreau2009back}, a good collapse between various cases should be expected for the same airfoil. 
Figure~\ref{Radiation_ratio} compares the present measurements with the theoretical prediction with Amiet's model (solid lines) and the previous measurements at ECL (diamond and square symbols) as reported in~\cite{moreau2009back}. The two theoretical curves stress the effect of directivity in $\Lambda$ and the actual microphone position in the present experiment, $258^{\circ}$ (on the airfoil pressure side to provide laser access on the suction side), reproduces better the experimental trend compared to the ECL measurements at about $270^{\circ}$ (or equivalently $90^{\circ}$ on the airfoil suction side). The 5~dB spread in the experimental data is consistent with the data of figure 16~(a) in~\cite{moreau2009back}, and can be attributed to both the saturation in the Electret microphones at low frequencies shown in figure~\ref{Wall_pressure_spectra} and a less accurate calibration methods in 2005. Yet, in both data sets,
%While 
while a good collapse between the two angles of attack $15^{\circ}$ and $8^{\circ}$ %degrees angle of attack, 
is %archived 
achieved between 500 and 2000~Hz, the collapse is relatively poor at lower frequencies ($80-500$~Hz range), for the high incidence where the separation noise dominates. On the other hand, the newer $8^{\circ}$ case shows a good match with Amiet's prediction.

\section{Discussion}

% Opening statement:
The low-frequency content of the airfoil self noise increases as the angle of attack is increased from $8^{\circ}$ to $15^{\circ}$. This increase noise is also accompanied by an increase in the amplitude of wall-pressure and velocity disturbances.

% OBSERVATION ON PRESSURE
The wall-pressure field shows an increased amplitude, spanwise extent and the velocity at which pressure gusts convect past the trailing-edge at frequencies where the separation noise is dominant. % OBSERVATIONS ON VELOCITY 
The genesis of the increased disturbances can be linked to the late transition of boundary layer. In particular, as the angle of attack is increased, it was found that the LSB covers at least $40\%$ of the airfoil chord consistently with previous LES results~\citep{christophe2009trailing,Christophe2011}, which leads to a delayed flow transition and re-attachment, somewhere between 40 and 65\% of the chord. As such, the magnitude of flow disturbances represented by the dimensionless Reynolds stress components $\overline{u_1 u_1}/U^2_\infty$, $\overline{u_2 u_2}/U^2_\infty$, and $-\overline{u_1 u_2}/U^2_\infty$, increases substantially compared to the airfoil at $8^{\circ}$ attack, especially close to the airfoil trailing edge.%Adverse pressure gradient due to higher incidence leads to higher values of shape factor close to the trailing edge. As the inlet velocity is increased, the shape factor decreases slightly but remains elevated compared to when airfoil is placed at a lower incidence. The higher incidence also leads to a five-fold increase in the boundary layer thickness.  

% General observations on flow struc
While the time-averaged flow is found to be attached, large-scale flow distortions in form of rollers, that are reminiscent of KH-type instability, are present. These roller structures are similar to the ones that were previously reported on the CD airfoil numerically by \cite{christophe2008} at the same incidence and experimentally by~\cite{jaiswal2022experimental} at a lower angle-of-attack. As the wall-normal spatial extent of these structures can be substantially larger than the mean boundary layer thickness, they have access to higher momentum flow. This explains why an increase in low-frequency convection velocity was observed despite a strong adverse pressure gradient %at 
in the trailing-edge region. 

% Increase in disturbances roll up the shear layer

The velocity %disturbances
fluctuations, $-\overline{u_1 u_2}$, increases steadily before eventually getting saturated close to the trailing-edge region of the airfoil. The peak values of $-\overline{u_1 u_2}$ are shown to scale the wall-pressure spectra for two angles of attack.  As such, it clearly demonstrates that the increase in the magnitude in wall-pressure statistics can be linked to an increase in $-\overline{u_1 u_2}$. The
amplification of flow disturbance, such as ${\overline{-u_1 u_2}}$, is known to %roll up 
yield KH-type instability and vortex pairing in the shear layer, producing %these 
the observed roller structures \citep{yarusevych2006coherent,yarusevych2009vortex,watmuff1999evolution,huang1990small}. In summary, these rollers are present due to the late amplification ($x>0.4C$) of the LSB instability, and its subsequent roll-up, which ensures %rollers 
that large eddies reach the trailing edge of the airfoil.

% Rollers increase the spanwise content
\cite{jaiswal2022experimental} showed that these rollers have large coherence in the spanwise direction. Furthermore, the mode associated with roller structures correlates with the wall-pressure fluctuations at frequencies that correspond to the maximum levels of spanwise coherence. In addition, the spanwise coherence of wall-pressure and HWA spectrogram peak at the same frequency. In the absence of any alternative frequency-centred activities in the flow, it may be concluded that these rollers are responsible for an increase in the spanwise correlation length.

Finally, \cite{lacagnina2019mechanisms} had shown that flapping of the shear yields an increase in low-frequency noise. While the flapping of LSB may result in flapping of shear-layer, no evidence for its contribution to separation noise is found in the absence of rollers. This is because no modal structures associated exclusively with shear-layer flapping were identified. The noise mechanism due to shear-layer flapping is thus not universally present for an airfoil at near stall conditions. In the present case, the increase in flow disturbances, and the associated rolling-up of the shear layer are the only dominant flow mechanisms that contribute to separation noise.

% Furthermore, previous experiments have shown that these coherent structures have high spanwise coherence \citep{christophe2008,jaiswal2022experimental}. As such they increase both the magnitude and spanwise extent of the wall-pressure field, which ultimately leads to an increase in noise.

%separation noise can be linked directly to the increase in Reynolds shear stress close to the trailing edge. 

Therefore, the question naturally arises: is the increase in the magnitude of $-\overline{u_1 u_2}$ sufficient to nullify %the \cite{amiet1976noise} 
Amiet's diffraction theory that depends on the thin-airfoil linearized theory? To answer this question, the radiation ratio is calculated for cases with variable flow incidence at the same Reynolds number based on chord. The results confirm that the diffraction efficacy of an airfoil subjected to higher angles of attack is substantially attenuated at frequencies associated with separation noise. This is because the overall increase in sound pressure level is comparatively small compared to the rise in spanwise correlation $l_z$. In particular, the energy conversion from near-field pressure to far-field pressure should be more effective as $l_z$ increases; however, this is not achieved. Furthermore, the roller structures imply that the unsteady Kutta condition may not be valid, as its validity hinges on the flow leaving the airfoil trailing edge smoothly. As such, the diffraction efficacy for an airfoil trailing-edge that is subjected to an unsteady pressure gust due to flow separation is substantially attenuated. Nevertheless, the separation noise can be fully quantified using a compact dipole. The far-field microphones located on the either side of the airfoil confirm this dipolar behaviour along with $U_\infty^{6}$ scaling of the OASPL measured in the far-field. % location. 
Furthermore, the far-field acoustic pressure field directivity pattern is similar for both $8^{\circ}$ and $15^{\circ}$ angles of attack, which reinforces the dipolar directivity pattern. These observations partly explain why \cite{christophe2008} obtained a more favourable estimate of acoustic noise while using the \cite{williams1970aerodynamic} analogy compared to diffraction theory \citep{amiet1976noise} at frequencies where the separation noise dominates.

\section{Conclusions}
% opening 
The present paper is a detailed aeroacoustic investigation of a CD airfoil at near stall condition. This is achieved by placing the CD airfoil at high angles of attack in an open jet anechoic wind tunnel. Two sets of experiments are performed at $Re_c \simeq 140,000$ and $Re_c \simeq 245,000$ based on airfoil chord for an airfoil placed at $15^{\circ}$ angle of attack. For the airfoil at $Re_c \simeq 140,000$, synchronized PIV, RMP and far-field microphone measurements were performed. 

The present study is driven by two fundamental research questions. 

1) What is the mechanism that is responsible for separation noise for an airfoil near stall conditions ? If so, is it universal ?

2) Is the noise due to flow separation generated by a dipole for airfoil close to stall ? If so, can it be quantified by \cites{amiet1976noise} diffraction theory ?

The present study shows that when the CD airfoil is placed at a higher angle of attack compared to $8^{\circ}$, such as $15^{\circ}$ in the present study, strong amplification of flow disturbance, up to an order of magnitude higher is seen in the trailing-edge region. In fact, noise due to flow separation can be linked to increase in flow disturbances, like $\overline{-u_1u_2}$, which scale up the wall-pressure fluctuations. %, a
This increased Reynolds stress triggers the roll up 
of the separated shear layer. %giving rise to rollers. 
These rollers are %possibly 
linked to the flow transition triggered by the Kelvin-Helmoltz %type flow 
instability. %These rollers are l
They are also linked to an increase in spanwise coherence of the wall-pressure fluctuations, as they convect past the trailing-edge. The modal decomposition obtained by POD shows that the modes associated with these roller structures correlate with near and far-field pressure. This correlation is observed only at frequencies where the separation noise dominates, i.e. frequency at which the $\overline{E_{11}}$ peaks. As such, rollers and associated Kelvin-Helmholtz type flow instability play a central role in the increase in noise due to flow separation.~Lastly, in the present study, no contributions coming exclusively from the flapping of the shear layer were observed.

The present study conclusively shows that separation noise is dipolar in nature, therefore, quadrupole contribution for low-speed airfoils at near-stall conditions can be neglected, at least for flows up to a Mach number of about 0.1. %$\sim 0.1$ . 
Yet the increase in flow disturbances measured close to the trailing-edge of the airfoil implies that the assumption of small amplitude disturbance are no longer valid, which is the central premise of the thin-airfoil~linearized theory used to estimate the response of the airfoil to an incoming pressure gust. Furthermore, passage of large roller structures past the trailing edge may invalidate the unsteady Kutta condition. Yet outside the frequency range at which flow separation operates, \cites{amiet1976noise} theory should be able to predict the far-field noise even at high angles of attack as previously shown by~\cite{christophe2009trailing}.

 \section*{Acknowledgments}
{The authors would like to acknowledge the help of Sidharth Krishnan Kalyani and Yann Pasco during the PIV measurements. Authors are thankful for computational time in supercomputer Graham, managed by the Digital research alliance of Canada.}

\section*{Funding}{This work was supported by the Canadian NSERC Discovery grant (no.RGPIN-2014-04111).}

\section*{Declaration of interests}{The authors report no conflict of interest.}
 
\section*{Data availability statement}{Raw data of PIV were processed %at the 
on the Digital research alliance of Canada's HPC center. Derived data supporting the findings of this study are available from the first author upon reasonable request.}

\section*{Appendix}
\label{sec:appendix}

In this appendix, the influence of the total length of RMP signals on wall-pressure statistics is studied, particularly at low frequencies where this parameter is known to define the lowest achievable frequency in power spectral densities.

Figure \ref{convection_velocity_length} first shows that this total length %of RMP signals 
is an important metric when it is used to estimate the convection velocity. This in part explains as to why previous studies \citep{kalyani2022flow} have reported slightly different values of $U_c$ and $l_z$. 
%While the plot \ref{convection_velocity_length} confirms that 
However, the uncertainty in the estimation of $U_c/U_{\infty}$ is less than $10\%$, which yields a marginal uncertainty in 
the radiation ratio, $\Lambda$, when plotted on a logarithmic scale. In turn, this has no significant impact on the efficacy of diffraction theory for separation noise.
%one of the main objectives of the paper is to verify efficacy of diffraction theory for separation noise. Therefore, when the radiation ratio, $\Lambda$, is plotted on a logarithmic scale, the uncertainty in $\Lambda$ due to the uncertainty in convection velocity should be marginal.
\begin{figure*}[ht!]
\centering
  \begin{tabular}{@{}p{0.45\linewidth}@{\quad}p{0.45\linewidth}@{}}
    \subfigimg[width=85 mm,pos=ul,vsep=21pt,hsep=42pt]{}{
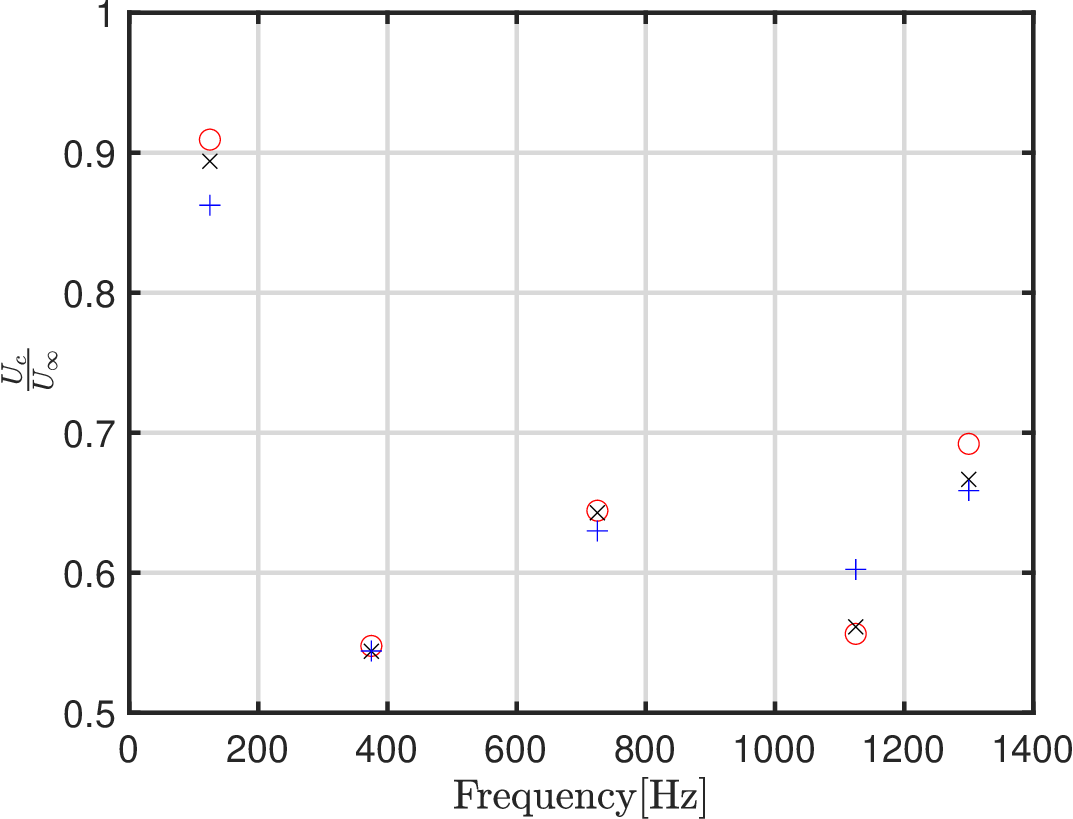} \\
  \end{tabular}
\caption{Convection velocity for $15^{\circ}$ angle of attack and $U_{\infty}=16$~m/s case. Red circles corresponds to full signal length of 1 minute while blue and black crosses denote signal lengths of 20 and 30 seconds respectively.}
\label{convection_velocity_length}       % Give a unique label
\end{figure*}

\begin{figure*}[ht!]
\centering
  \begin{tabular}{@{}p{0.45\linewidth}@{\quad}p{0.45\linewidth}@{}}
    \subfigimg[width=85 mm,pos=ul,vsep=21pt,hsep=42pt]{}{
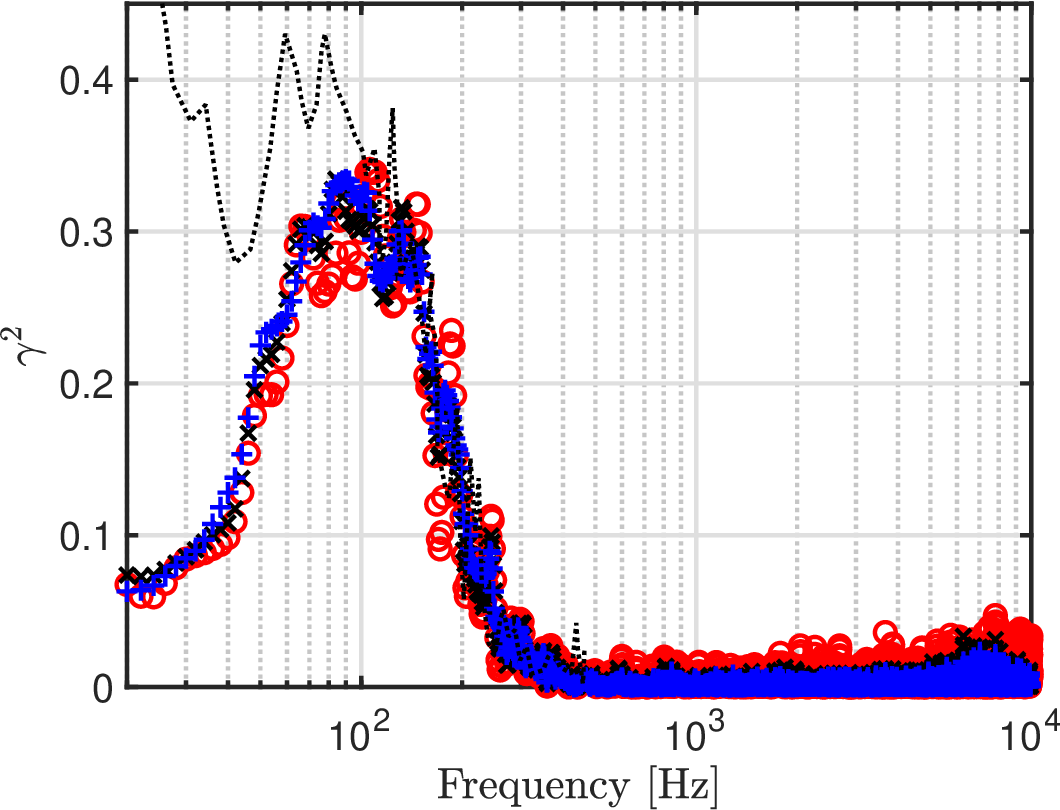} \\
  \end{tabular}
\caption{Legend: $\gamma^2$ between RMP $25$ and RMP $27$ for $15^{\circ}$ angle of attack and U$_{\infty}=16$~m/s case. Legend for signal length same as in figure \ref{convection_velocity_length}. Black dotted line correspond to \cites{kalyani2022flow} data.}
\label{spanwise_sid}       % Give a unique label
\end{figure*}

In order to understand the impact of the signal length on the spanwise correlation length ($l_z$), the spanwise coherence ($\gamma^2$) is plotted in figure \ref{spanwise_sid} between two spanwise probes 25 and 27. The results are also compared with the data reported by \citet{kalyani2022flow}. %in the same figure. 
Figure \ref{spanwise_sid} shows that at low-frequency oscillations, which represent uncertainty in the estimate of $\gamma^2$, are higher for cases where the signal length is truncated below 30 seconds. Consequently, \citet{kalyani2022flow}, who estimated the spanwise coherence ($\gamma^2$) with a signal length of 15 seconds, have a higher uncertainty in the estimate of $\gamma^2$. Furthermore, \citet{kalyani2022flow} took one-tenth of the number of points to estimate the PSD compared to the present case. As such, the low-frequency part of $\gamma^2$ shows an erroneous double peak in their results (see Figure 4(a) of \citet{kalyani2022flow}), which is absent in the present case as well as in one reported earlier by \citet{moreau2005effect}. %These differences explain why the estimate of $l_z$ in the present paper is different from those reported earlier by \citet{kalyani2022flow}.

\FloatBarrier

% tikz legend 

%%====================================================================

% legends for the pictures
\scalebox{0}{%
\begin{tikzpicture}
    \begin{axis}[%
        hide axis
        ]
        %% blue triangle
        \addplot [
            draw={rgb,255:red,55;green,126;blue,184},
            mark=triangle*,
            mark options={fill={rgb,255:red,55;green,126;blue,184}, draw=black, scale=2, line width=0.0pt, solid},
            mark size=1.5pt,
            line width=1pt,
            ]{0}; \label{leg:triblue}
        %% blue triangle
        \addplot [
            draw={rgb,255:red,55;green,126;blue,184},
            mark=square*,
            mark options={fill={rgb,255:red,55;green,126;blue,184}, draw={rgb,255:red,55;green,126;blue,184}, scale=2, line width=0.0pt, solid},
            mark size=1.5pt,
            line width=1pt,
            solid,
            ]{0}; \label{leg:squareblue}
        %% blue triangle
        \addplot [
            draw={rgb,255:red,55;green,126;blue,184},
            dashed,
            mark=square*,
            mark options={fill={rgb,255:red,55;green,126;blue,184}, draw={rgb,255:red,55;green,126;blue,184}, scale=2, line width=0.0pt, solid},
            mark size=1.5pt,
            line width=1pt,
            ]{0}; \label{leg:squarebluedashed}       
%                  
               
        %% blue triangle
        \addplot [
            draw=red,
            mark=*,
            mark options={fill=red, draw=red, scale=2, line width=0.0pt, solid},
            mark size=1.5pt,
            line width=1pt,
            solid,
            ]{0}; \label{leg:circlered}
        %% blue triangle
        \addplot [
            draw=red,
            mark=*,
            mark options={fill=red, draw=red, scale=2, line width=0.0pt, solid},
            mark size=1.5pt,
            line width=1pt,
            dashed,
            ]{0}; \label{leg:circlereddashed}                        
        %% red triangle down
        \addplot [
            draw=red,
            mark=triangle*,
            mark options={fill=red, draw=black, scale=2, line width=0.0pt, solid},
            every mark/.append style={rotate=180},
            mark size=1.5pt,
            line width=1pt,
            ]{0}; \label{leg:trired}

        %% green diamond
        \addplot [
            draw=ForestGreen,
            mark=diamond*,
            mark options={fill=ForestGreen, draw=black, scale=2, line width=0.0pt, solid},
            mark size=1.5pt,
            line width=1pt,
            ]{0}; \label{leg:diagreen}

        % %% black dash
        % \addplot [
        %     color=black,
        %     solid,
        %     line width=1.pt,
        %     dashed,
        %     ]{0}; \label{leg:blackdash}
        %% grey dash
        \addplot [
            color=lightgray,
            solid,
            line width=1.pt,
            dashed,
            ]{0}; \label{leg:greydashed}
		
			 \addplot [
            color=gray,
            mark options={solid, scale=1.3},
            line width=1.pt,
            solid,
            ]{0}; \label{leg:greysolid}		
            
			 \addplot [
            color=lightgray,
            mark=triangle,
            mark options={fill=lightgray, draw=lightgray, scale=1.5, line width=0.5pt, solid, rotate=0},
            line width=1.pt,
            solid,
            ]{0}; \label{leg:greyuptrianglesolid}            
            
             \addplot [
            color=lightgray,
            mark options={solid, scale=1.3},
            line width=1.5pt,
            solid,
            ]{0}; \label{leg:greysolidthick}

        \addplot [
            color=lightgray,
            dotted,
            line width=1.pt,
            ]{0}; \label{leg:greydotted}
		
        %% red
        \addplot [
            color=red,
            solid,
            line width=1.pt,
            ]{0}; \label{leg:red}
        %% red
        \addplot [
            color=black,
            solid,
            line width=1.pt,
            ]{0}; \label{leg:black}
        \addplot [
            color=black,
            dotted,
            line width=1.pt,
            ]{0}; \label{leg:blackdotted}

        %% red
        \addplot [
            color=red,
            dashdotted,
            line width=1.pt,
            ]{0}; \label{leg:reddashdotted}            

        %% red
        \addplot [
            color={rgb,255:red,55;green,126;blue,184},
            dashdotted,
            line width=1.pt,
            ]{0}; \label{leg:bluedashdotted}            

        %% blue
        \addplot [
            color={rgb,255:red,55;green,126;blue,184},
            solid,
            line width=1.pt,
            ]{0}; \label{leg:blue}

            % orange line 
            \addplot [
            color={rgb,1:red,1;green,0.49;blue,0},
            solid,
            line width=1.pt,
            ]{0}; \label{leg:orange}
            
             % black dashed line 
            \addplot [
            color={rgb,1:red,0;green,0.0;blue,0},
            dashed,
            line width=1.pt,
            ]{0}; \label{leg:blackdashline}

            \addplot [
            color={rgb,1:red,0;green,0.0;blue,0},
            dashdotted,
            line width=1.pt,
            ]{0}; \label{leg:black_dashdotted}

             % kind of blue dashed line 
            \addplot [
            color={rgb,1:red,0;green,0.2;blue,1},
            dashed,
            line width=1.pt,
            ]{0}; \label{leg:bluedashline}
            
             % purple dashed line 
            \addplot [
            color={rgb,1:red,1;green,0.1;blue,1},
            dashed,
            line width=1.pt,
            ]{0}; \label{leg:purpledashedline}
            
             % Green dashed line 
            \addplot [
            color={rgb,1:red,0.1;green,0.5;blue,0},
            dashed,
            line width=1.pt,
            ]{0}; \label{leg:Greendashedline}
            
              % red  dashed line
            \addplot [
            color={rgb,1:red,0.8;green,0.0;blue,0.0},
            dashed,
            line width=1.pt,
            ]{0}; \label{leg:reddashedline}

             % 50 shades of grey dashed line 
            \addplot [
            color={rgb,1:red,0.8;green,0.8;blue,0.8},
            dashed,
            line width=1.pt,
            ]{0}; \label{leg:greydashedline}
            
              \addplot [
            color={rgb,1:red,0.8;green,0.8;blue,0.8},
            dashed,
            line width=2.pt,
            ]{0}; \label{leg:greydashedline-thick}

        %% black dash
        \addplot [
            color=black,
            mark options={solid, scale=1.3},
            line width=0.5pt,
            dashed,
            ]{0}; \label{leg:blackdashed}
        \addplot [
            color=red,
            mark options={solid, scale=1.3},
            line width=1.pt,
            dashed,
            ]{0}; \label{leg:reddashed}

        %% black dash + square
        \addplot [
            color=black,
            mark=square,
            mark options={solid, scale=1.2},
            line width=1.pt,
            dashed,
            ]{0}; \label{leg:blacksquaredash}
        %% black square
        \addplot [
            color=black,
            mark=square*,
            mark options={fill=black, draw=black, scale=1.3, line width=0.pt, solid},
            line width=0.pt,
            only marks,
            ]{0}; \label{leg:blacksquare}

			    \addplot [
            color=black,
            mark=square*,
            mark options={fill=black, draw=black, scale=1.3, line width=0.pt, solid},
            line width=1.pt,
            solid,
            ]{0}; \label{leg:blacksquareline}
    		
			\addplot [
            color=black,
            mark=square*,
            mark options={fill=white, draw=black, scale=1.3, line width=0.pt, solid},
            line width=1.pt,
            solid,
            ]{0}; \label{leg:blacksquarelineempty}	
		
        %% grey dash + square
        \addplot [
            color=lightgray,
            mark=o,
            mark options={solid, scale=1.2},
            line width=1.pt,
            dashed,
            ]{0}; \label{leg:greycircledash}
            
             \addplot [
            color=lightgray,
            mark=o,
            mark options={solid, scale=1.2},
            line width=1.pt,
            solid,
            ]{0}; \label{leg:greycirclesolid}

        %% black dash + circle
        \addplot [
            color=black,
            mark=o,
            mark options={solid, scale=1.3},
            line width=1.pt,
            dashed,
            ]{0}; \label{leg:blackcircledash}

        %% red + dashdotted + square
        \addplot [
            color=red,
            dashdotted,
            mark=square*,
            mark options={fill=red, draw=red, scale=1.3, line width=0.pt, solid},
            line width=1.pt,
            ]{0}; \label{leg:reddashdottedsquare}
        %% red + dashdotted + square
        \addplot [
            color={rgb,255:red,55;green,126;blue,184},
            dashdotted,
            mark=square*,
            mark options={fill={rgb,255:red,55;green,126;blue,184}, draw={rgb,255:red,55;green,126;blue,184}, scale=1.3, line width=0.pt, solid},
            line width=1.pt,
            ]{0}; \label{leg:bluedashdottedsquare}

        %% blue + square
        \addplot [
            color={rgb,255:red,55;green,126;blue,184},
            solid,
            mark=square*,
            mark options={fill={rgb,255:red,55;green,126;blue,184}, draw={rgb,255:red,55;green,126;blue,184}, scale=1.3, line width=0.pt, solid},
            line width=1.pt,
            ]{0}; \label{leg:bluesquare}
            
            \addplot [
            color={rgb,255:red,0;green,0;blue,255},
            solid,
            mark=square*,
            mark options={fill=white, draw={rgb,255:red,0;green,0;blue,255}, scale=1.3, line width=0.pt, solid},
            line width=1.pt,
            ]{0}; \label{leg:bluesquare_empty}

        %% blue + square
        \addplot [
            color=red,
            solid,
            mark=square*,
            mark options={fill=red, draw=red, scale=1.3, line width=0.0pt, solid},
            line width=1.pt,
            ]{0}; \label{leg:redsquare}

        %% blue + circle
        \addplot [
            color={rgb,255:red,55;green,126;blue,184},
            mark=*,
            mark options={fill={rgb,255:red,55;green,126;blue,184}, draw={rgb,255:red,55;green,126;blue,184}, scale=1.3, line width=0.0pt, solid},
            line width=1.pt,
            ]{0}; \label{leg:bluecircle}
        %% blue + circle
        \addplot [
            color=red,
            mark=*,
            mark options={fill=red, draw=red, scale=1.3, line width=0.0pt, solid},
            line width=1.pt,
            ]{0}; \label{leg:redcircle}

        %% red + dashdotted + circle
        \addplot [
            color=red,
            dashdotted,
            mark=*,
            mark options={fill=red, draw=red, scale=1.3, line width=0.0pt, solid},
            line width=1.pt,
            ]{0}; \label{leg:reddashdottedcircle}

            \addplot [
            color=red,
            solid,
            mark=*,
            mark options={fill=white, draw=red, scale=1.3, line width=0.5pt, solid},
            line width=1.pt,
            ]{0}; \label{leg:redcontdottedcircle}
            
            \addplot [
            color=red,
            only marks, % this makes the line disappear
            mark=*,
            mark options={fill=red, draw=red, scale=1.3, line width=0.0pt, solid},
            line width=1.pt,
            ]{0}; \label{leg:redcirclenoline}
        %% red + dashdotted + circle
        \addplot [
            color={rgb,255:red,55;green,126;blue,184},
            dashdotted,
            mark=*,
            mark options={fill={rgb,255:red,55;green,126;blue,184}, draw={rgb,255:red,55;green,126;blue,184}, scale=1.3, line width=0.0pt, solid},
            line width=1.pt,
            ]{0}; \label{leg:bluedashdottedcircle}

        \addplot [
            color=black,
            mark options={solid, scale=1.3},
            line width=1.pt,
            solid,
            ]{0}; \label{leg:blacksolid}

        \addplot [
            color=red,
            mark=square,
            mark options={fill=red, draw=red, scale=1.0, line width=0.5pt, solid, rotate=45},
            line width=1.pt,
            solid,
            ]{0}; \label{leg:reddiamondsolid}

        \addplot [
            color=red,
            mark=square,
            mark options={fill=red, draw=red, scale=1.0, line width=0.5pt, solid, rotate=45},
            line width=1.pt,
            dashed,
            ]{0}; \label{leg:reddiamonddashed}
            
            \addplot [
            color=red,
            mark=*,
            mark options={fill=white, draw=red, scale=1.0, line width=0.5pt, solid, rotate=0},
            line width=1.pt,
            dashed,
            ]{0}; \label{leg:redcircledashedline}

        \addplot [
            color={rgb,255:red,77;green,175;blue,74},
            mark=triangle,
            mark options={fill={rgb,255:red,77;green,175;blue,74}, draw={rgb,255:red,77;green,175;blue,74}, scale=1.5, line width=0.5pt, solid, rotate=0},
            line width=1.pt,
            solid,
            ]{0}; \label{leg:greenuptrianglesolid}

        \addplot [
            color={rgb,255:red,77;green,175;blue,74},
            mark=triangle,
            mark options={fill={rgb,255:red,77;green,175;blue,74}, draw={rgb,255:red,77;green,175;blue,74}, scale=1.5, line width=0.5pt, solid, rotate=0},
            line width=1.pt,
            dashed,
            ]{0}; \label{leg:greenuptriangledashed}

        \addplot [
            color={rgb,255:red,55;green,126;blue,184},
            mark=triangle,
            mark options={fill={rgb,255:red,55;green,126;blue,184}, draw={rgb,255:red,55;green,126;blue,184}, scale=1.5, line width=0.5pt, solid, rotate=180},
            line width=1.pt,
            solid,
            ]{0}; \label{leg:bluedowntrianglesolid}
			
			 \addplot [
            color={rgb,255:red,0;green,0;blue,255},
            mark=diamond*,
            mark options={fill=white, draw={rgb,255:red,0;green,0;blue,255}, scale=1.5, line width=0.5pt, solid, rotate=0},
            line width=1.pt,
            solid,
            ]{0}; \label{leg:bluedowndiamondsolid}

        \addplot [
            color={rgb,255:red,55;green,126;blue,184},
            mark=triangle,
            mark options={fill={rgb,255:red,55;green,126;blue,184}, draw={rgb,255:red,55;green,126;blue,184}, scale=1.5, line width=0.5pt, solid, rotate=180},
            line width=1.pt,
            dashed,
            ]{0}; \label{leg:bluedowntriangledashed}  
            
            \addplot [
            color={rgb,255:red,255;green,127;blue,0},
            mark=triangle,
            mark options={fill={rgb,255:red,255;green,127;blue,0}, draw={rgb,255:red,255;green,127;blue,0}, scale=1.3, line width=0.5pt, solid, rotate=0},
            line width=1.pt,
            only marks,
            ]{0}; \label{leg:orangetriangle} 
            
             \addplot [
            color={rgb,255:red,255;green,127;blue,0},
            mark=x,
            mark options={fill={rgb,255:red,255;green,127;blue,0}, draw={rgb,255:red,255;green,127;blue,0}, scale=1.3, line width=0.5pt, solid, rotate=0},
            line width=1.pt,
            only marks,
            ]{0}; \label{leg:orange_cross}

			 \addplot [
            color={rgb,255:red,0;green,0;blue,255},
            mark=x,
            mark options={fill={rgb,255:red,0;green,0;blue,255}, draw={rgb,255:red,0;green,0;blue,255}, scale=1.3, line width=0.5pt, solid, rotate=0},
            line width=1.pt,
            only marks,
            ]{0}; \label{leg:blue_cross}

             \addplot [
            color={rgb,255:red,0;green,0;blue,255},
            mark=x,
            mark options={fill={rgb,255:red,0;green,0;blue,255}, draw={rgb,255:red,0;green,0;blue,255}, scale=1.3, line width=0.5pt, solid, rotate=0},
            line width=1.pt,
            solid,
            ]{0}; \label{leg:bluecrossline}

            \addplot [
            color={rgb,255:red,0;green,0;blue,255},
            mark=*,
            mark options={fill=white, draw={rgb,255:red,0;green,0;blue,255}, scale=1.7, line width=0.5pt, solid, rotate=0},
            line width=1.pt,
            only marks,
            ]{0}; \label{leg:blue-circle}

 \addplot [
            color={rgb,255:red,255;green,0;blue,0},
            mark=square,
            mark options={fill={rgb,255:red,255;green,0;blue,0}, draw={rgb,255:red,255;green,0;blue,0}, scale=1.7, line width=0.5pt, solid, rotate=0},
            line width=1.pt,
            only marks,
            ]{0}; \label{leg:orange-square}

        \addplot [
            color={rgb,255:red,255;green,127;blue,0},
            mark=square,
            mark options={fill={rgb,255:red,255;green,127;blue,0}, draw={rgb,255:red,255;green,127;blue,0}, scale=1.3, line width=0.5pt, solid, rotate=0},
            line width=1.pt,
            solid,
            ]{0}; \label{leg:orangesquaresolid}
            \addplot [
            color={rgb,255:red,255;green,127;blue,0},
            mark=diamond*,
            mark options={fill={rgb,255:red,255;green,127;blue,0}, draw={rgb,255:red,255;green,127;blue,0}, scale=1.3, line width=0.5pt, solid, rotate=0},
            line width=1.pt,
            solid,
            ]{0}; %\label{leg:orangediamond}
	  
        \addplot [
            color={rgb,255:red,255;green,127;blue,0},
            mark=square,
            mark options={fill={rgb,255:red,255;green,127;blue,0}, draw={rgb,255:red,255;green,127;blue,0}, scale=1.3, line width=0.5pt, solid, rotate=0},
            line width=1.pt,
            dashed,
            ]{0}; \label{leg:orangesquaredashed}           
            %% orange square
        \addplot [
            color={rgb,255:red,255;green,127;blue,0},
            mark=diamond*,
            mark options={fill=white, draw={rgb,255:red,255;green,127;blue,0}, scale=1.3, line width=0.pt, solid},
            line width=0.pt,
            only marks,
            ]{0}; \label{leg:orangediamond}  
            \addplot [
            color=blue,
            mark=triangle*,
            mark options={fill=white, draw=blue, scale=1.3, line width=0.pt, solid},
            line width=0.pt,
            only marks,
            ]{0}; \label{leg:triangle_blue}  
            
			\addplot [
            color=lightgray,
            mark=diamond*,
            mark options={fill=white, draw=lightgray, scale=1.7, line width=0.pt, solid},
            line width=1.pt,
            only marks,
            ]{0}; \label{leg:greydiamond}

             \addplot [
            color=red,
            mark=square*,
            mark options={fill=white, draw=red, scale=1.3, line width=0.pt, solid},
            line width=0.pt,
            only marks,
            ]{0}; \label{leg:square_red}  
            \addplot [
            color=black,
            mark=*,
            mark options={fill=white, draw=black, scale=1.3, line width=0.pt, solid},
            line width=0.pt,
            only marks,
            ]{0}; \label{leg:black_circle}  
            
              \addplot [
            color=blue,
            mark=square,
            mark options={fill=white, draw=blue, scale=1.7, line width=0.1pt, solid},
            line width=0.pt,
            only marks,
            ]{0}; \label{leg:blue_square_marks}  
                 
    \end{axis}
    
\end{tikzpicture}
}

\bibliography{Bibliography.bib}

%merlin.mbs aipauth4-1.bst 2010-07-25 4.21a (PWD, AO, DPC) hacked
%Control: author (9) reversed initials
\begin{thebibliography}{53}%
\makeatletter
\providecommand \@ifxundefined [1]{%
 \@ifx{#1\undefined}
}%
\providecommand \@ifnum [1]{%
 \ifnum #1\expandafter \@firstoftwo
 \else \expandafter \@secondoftwo
 \fi
}%
\providecommand \@ifx [1]{%
 \ifx #1\expandafter \@firstoftwo
 \else \expandafter \@secondoftwo
 \fi
}%
\providecommand \natexlab [1]{#1}%
\providecommand \enquote  [1]{``#1''}%
\providecommand \bibnamefont  [1]{#1}%
\providecommand \bibfnamefont [1]{#1}%
\providecommand \citenamefont [1]{#1}%
\providecommand \href@noop [0]{\@secondoftwo}%
\providecommand \href [0]{\begingroup \@sanitize@url \@href}%
\providecommand \@href[1]{\@@startlink{#1}\@@href}%
\providecommand \@@href[1]{\endgroup#1\@@endlink}%
\providecommand \@sanitize@url [0]{\catcode `\\12\catcode `\$12\catcode
  `\&12\catcode `\#12\catcode `\^12\catcode `\_12\catcode `\%12\relax}%
\providecommand \@@startlink[1]{}%
\providecommand \@@endlink[0]{}%
\providecommand \url  [0]{\begingroup\@sanitize@url \@url }%
\providecommand \@url [1]{\endgroup\@href {#1}{\urlprefix }}%
\providecommand \urlprefix  [0]{URL }%
\providecommand \Eprint [0]{\href }%
\providecommand \doibase [0]{http://dx.doi.org/}%
\providecommand \selectlanguage [0]{\@gobble}%
\providecommand \bibinfo  [0]{\@secondoftwo}%
\providecommand \bibfield  [0]{\@secondoftwo}%
\providecommand \translation [1]{[#1]}%
\providecommand \BibitemOpen [0]{}%
\providecommand \bibitemStop [0]{}%
\providecommand \bibitemNoStop [0]{.\EOS\space}%
\providecommand \EOS [0]{\spacefactor3000\relax}%
\providecommand \BibitemShut  [1]{\csname bibitem#1\endcsname}%
\let\auto@bib@innerbib\@empty
%</preamble>
\bibitem [{\citenamefont {Abe}(2017)}]{abe2017reynolds}%
  \BibitemOpen
  \bibfield  {author} {\bibinfo {author} {\bibnamefont {Abe}, \bibfnamefont
  {H.}},\ }\bibfield  {title} {\enquote {\bibinfo {title} {Reynolds-number
  dependence of wall-pressure fluctuations in a pressure-induced turbulent
  separation bubble},}\ }\href@noop {} {\bibfield  {journal} {\bibinfo
  {journal} {Journal of Fluid Mechanics}\ }\textbf {\bibinfo {volume} {833}},\
  \bibinfo {pages} {563--598} (\bibinfo {year} {2017})}\BibitemShut {NoStop}%
\bibitem [{\citenamefont {Amiet}(1976)}]{amiet1976noise}%
  \BibitemOpen
  \bibfield  {author} {\bibinfo {author} {\bibnamefont {Amiet}, \bibfnamefont
  {R.~K.}},\ }\bibfield  {title} {\enquote {\bibinfo {title} {Noise due to
  turbulent flow past a trailing edge},}\ }\href@noop {} {\bibfield  {journal}
  {\bibinfo  {journal} {Journal of sound and vibration}\ }\textbf {\bibinfo
  {volume} {47}},\ \bibinfo {pages} {387--393} (\bibinfo {year}
  {1976})}\BibitemShut {NoStop}%
\bibitem [{\citenamefont {Bertagnolio}\ \emph {et~al.}(2017)\citenamefont
  {Bertagnolio}, \citenamefont {Madsen}, \citenamefont {Fischer},\ and\
  \citenamefont {Bak}}]{bertagnolio2017semi}%
  \BibitemOpen
  \bibfield  {author} {\bibinfo {author} {\bibnamefont {Bertagnolio},
  \bibfnamefont {F.}}, \bibinfo {author} {\bibnamefont {Madsen}, \bibfnamefont
  {H.~A.}}, \bibinfo {author} {\bibnamefont {Fischer}, \bibfnamefont {A.}}, \
  and\ \bibinfo {author} {\bibnamefont {Bak}, \bibfnamefont {C.}},\ }\bibfield
  {title} {\enquote {\bibinfo {title} {A semi-empirical airfoil stall noise
  model based on surface pressure measurements},}\ }\href@noop {} {\bibfield
  {journal} {\bibinfo  {journal} {Journal of Sound and Vibration}\ }\textbf
  {\bibinfo {volume} {387}},\ \bibinfo {pages} {127--162} (\bibinfo {year}
  {2017})}\BibitemShut {NoStop}%
\bibitem [{\citenamefont {Bossomaier}\ \emph {et~al.}(2016)\citenamefont
  {Bossomaier}, \citenamefont {Barnett}, \citenamefont {Harr{\'e}},\ and\
  \citenamefont {Lizier}}]{bossomaierintroduction}%
  \BibitemOpen
  \bibfield  {author} {\bibinfo {author} {\bibnamefont {Bossomaier},
  \bibfnamefont {T.}}, \bibinfo {author} {\bibnamefont {Barnett}, \bibfnamefont
  {L.}}, \bibinfo {author} {\bibnamefont {Harr{\'e}}, \bibfnamefont {M.}}, \
  and\ \bibinfo {author} {\bibnamefont {Lizier}, \bibfnamefont {J.~T.}},\
  }\href@noop {} {\emph {\bibinfo {title} {An Introduction to Transfer Entropy:
  Information Flow in Complex Systems}}}\ (\bibinfo  {publisher} {Springer
  International Publishing},\ \bibinfo {year} {2016})\BibitemShut {NoStop}%
\bibitem [{\citenamefont {Brooks}, \citenamefont {Pope},\ and\ \citenamefont
  {Marcolini}(1989)}]{brooks1989airfoil}%
  \BibitemOpen
  \bibfield  {author} {\bibinfo {author} {\bibnamefont {Brooks}, \bibfnamefont
  {T.~F.}}, \bibinfo {author} {\bibnamefont {Pope}, \bibfnamefont {D.~S.}}, \
  and\ \bibinfo {author} {\bibnamefont {Marcolini}, \bibfnamefont {M.~A.}},\
  }\href@noop {} {\enquote {\bibinfo {title} {Airfoil self-noise and
  prediction},}\ }\bibinfo {type} {Tech. Rep.}\ (\bibinfo  {institution}
  {"NASA"},\ \bibinfo {year} {1989})\BibitemShut {NoStop}%
\bibitem [{\citenamefont {Caiazzo}\ \emph {et~al.}(2023)\citenamefont
  {Caiazzo}, \citenamefont {Pargal}, \citenamefont {Wu}, \citenamefont
  {Sanjos{\'e}}, \citenamefont {Yuan},\ and\ \citenamefont {Moreau}}]{Caiazzo}%
  \BibitemOpen
  \bibfield  {author} {\bibinfo {author} {\bibnamefont {Caiazzo}, \bibfnamefont
  {A.}}, \bibinfo {author} {\bibnamefont {Pargal}, \bibfnamefont {S.}},
  \bibinfo {author} {\bibnamefont {Wu}, \bibfnamefont {H.}}, \bibinfo {author}
  {\bibnamefont {Sanjos{\'e}}, \bibfnamefont {M.}}, \bibinfo {author}
  {\bibnamefont {Yuan}, \bibfnamefont {J.}}, \ and\ \bibinfo {author}
  {\bibnamefont {Moreau}, \bibfnamefont {S.}},\ }\bibfield  {title} {\enquote
  {\bibinfo {title} {On the effect of adverse pressure gradients on
  wall-pressure statistics in a controlled-diffusion aerofoil turbulent
  boundary layer},}\ }\href@noop {} {\bibfield  {journal} {\bibinfo  {journal}
  {Journal of Fluid Mechanics}\ }\textbf {\bibinfo {volume} {960}},\ \bibinfo
  {pages} {A17} (\bibinfo {year} {2023})}\BibitemShut {NoStop}%
\bibitem [{\citenamefont {Christophe}(2011)}]{Christophe2011}%
  \BibitemOpen
  \bibfield  {author} {\bibinfo {author} {\bibnamefont {Christophe},
  \bibfnamefont {J.}},\ }\emph {\bibinfo {title} {{Application of Hybrid
  Methods to High Frequency Aeroacoustics}}},\ \href@noop {} {Ph.D. thesis},\
  \bibinfo  {school} {von Karman Institute for Fluid Dynamics / Universite
  Libre de Bruxelles} (\bibinfo {year} {2011})\BibitemShut {NoStop}%
\bibitem [{\citenamefont {Christophe}, \citenamefont {Anthoine},\ and\
  \citenamefont {Moreau}(2008)}]{christophe2008trailing}%
  \BibitemOpen
  \bibfield  {author} {\bibinfo {author} {\bibnamefont {Christophe},
  \bibfnamefont {J.}}, \bibinfo {author} {\bibnamefont {Anthoine},
  \bibfnamefont {J.}}, \ and\ \bibinfo {author} {\bibnamefont {Moreau},
  \bibfnamefont {S.}},\ }\bibfield  {title} {\enquote {\bibinfo {title}
  {Trailing edge noise computation of a fan blade profile},}\ }\href@noop {}
  {\bibfield  {journal} {\bibinfo  {journal} {The Journal of the Acoustical
  Society of America}\ }\textbf {\bibinfo {volume} {123}},\ \bibinfo {pages}
  {3539--3539} (\bibinfo {year} {2008})}\BibitemShut {NoStop}%
\bibitem [{\citenamefont {Christophe}, \citenamefont {Anthoine},\ and\
  \citenamefont {Moreau}(2009)}]{christophe2009trailing}%
  \BibitemOpen
  \bibfield  {author} {\bibinfo {author} {\bibnamefont {Christophe},
  \bibfnamefont {J.}}, \bibinfo {author} {\bibnamefont {Anthoine},
  \bibfnamefont {J.}}, \ and\ \bibinfo {author} {\bibnamefont {Moreau},
  \bibfnamefont {S.}},\ }\bibfield  {title} {\enquote {\bibinfo {title}
  {{Trailing} edge noise of a controlled-diffusion airfoil at moderate and high
  angle of attack},}\ }in\ \href@noop {} {\emph {\bibinfo {booktitle} {15th
  AIAA/CEAS Aeroacoustics Conference (30th AIAA Aeroacoustics Conference)}}}\
  (\bibinfo {address} {Miami, Florida},\ \bibinfo {year} {2009})\ p.\ \bibinfo
  {pages} {3196}\BibitemShut {NoStop}%
\bibitem [{\citenamefont {Christophe}\ and\ \citenamefont
  {Moreau}(2008)}]{christophe2008}%
  \BibitemOpen
  \bibfield  {author} {\bibinfo {author} {\bibnamefont {Christophe},
  \bibfnamefont {J.}}\ and\ \bibinfo {author} {\bibnamefont {Moreau},
  \bibfnamefont {S.}},\ }\bibfield  {title} {\enquote {\bibinfo {title} {{{LES}
  of the trailing-edge flow and noise of a controlled-diffusion airfoil at high
  angle of attack}},}\ }in\ \href@noop {} {\emph {\bibinfo {booktitle}
  {Proceedings of the Summer Program}}}\ (\bibinfo {address} {Stanford,
  California, U.S.A},\ \bibinfo {year} {2008})\ p.\ \bibinfo {pages}
  {305}\BibitemShut {NoStop}%
\bibitem [{\citenamefont {Corcos}(1964)}]{corcos_JFM1964}%
  \BibitemOpen
  \bibfield  {author} {\bibinfo {author} {\bibnamefont {Corcos}, \bibfnamefont
  {G.}},\ }\bibfield  {title} {\enquote {\bibinfo {title} {The structure of the
  turbulent pressure field in boundary-layer flows},}\ }\href@noop {}
  {\bibfield  {journal} {\bibinfo  {journal} {Journal of Fluid Mechanics}\
  }\textbf {\bibinfo {volume} {18}},\ \bibinfo {pages} {353--378} (\bibinfo
  {year} {1964})}\BibitemShut {NoStop}%
\bibitem [{\citenamefont {Curle}(1955)}]{curle1955influence}%
  \BibitemOpen
  \bibfield  {author} {\bibinfo {author} {\bibnamefont {Curle}, \bibfnamefont
  {N.}},\ }\bibfield  {title} {\enquote {\bibinfo {title} {The influence of
  solid boundaries upon aerodynamic sound},}\ }\href@noop {} {\bibfield
  {journal} {\bibinfo  {journal} {Proceedings of the Royal Society of London.
  Series A. Mathematical and Physical Sciences}\ }\textbf {\bibinfo {volume}
  {231}},\ \bibinfo {pages} {505--514} (\bibinfo {year} {1955})}\BibitemShut
  {NoStop}%
\bibitem [{\citenamefont {Efimtsov}(1982)}]{efimtsov1982characteristics}%
  \BibitemOpen
  \bibfield  {author} {\bibinfo {author} {\bibnamefont {Efimtsov},
  \bibfnamefont {B.}},\ }\bibfield  {title} {\enquote {\bibinfo {title}
  {Characteristics of the field of turbulent wall pressure-fluctuations at
  large {R}eynolds-numbers},}\ }\href@noop {} {\bibfield  {journal} {\bibinfo
  {journal} {Soviet Physics Acoustics-USSR}\ }\textbf {\bibinfo {volume}
  {28}},\ \bibinfo {pages} {289--292} (\bibinfo {year} {1982})}\BibitemShut
  {NoStop}%
\bibitem [{\citenamefont {Ffowcs~Williams}\ and\ \citenamefont
  {Hall}(1970)}]{williams1970aerodynamic}%
  \BibitemOpen
  \bibfield  {author} {\bibinfo {author} {\bibnamefont {Ffowcs~Williams},
  \bibfnamefont {J.~E.}}\ and\ \bibinfo {author} {\bibnamefont {Hall},
  \bibfnamefont {L.~H.}},\ }\bibfield  {title} {\enquote {\bibinfo {title}
  {Aerodynamic sound generation by turbulent flow in the vicinity of a
  scattering half plane},}\ }\href@noop {} {\bibfield  {journal} {\bibinfo
  {journal} {Journal of fluid mechanics}\ }\textbf {\bibinfo {volume} {40}},\
  \bibinfo {pages} {657--670} (\bibinfo {year} {1970})}\BibitemShut {NoStop}%
\bibitem [{\citenamefont {Ffowcs~Williams}\ and\ \citenamefont
  {Hawkings}(1969)}]{williams1969sound}%
  \BibitemOpen
  \bibfield  {author} {\bibinfo {author} {\bibnamefont {Ffowcs~Williams},
  \bibfnamefont {J.~E.}}\ and\ \bibinfo {author} {\bibnamefont {Hawkings},
  \bibfnamefont {D.~L.}},\ }\bibfield  {title} {\enquote {\bibinfo {title}
  {Sound generation by turbulence and surfaces in arbitrary motion},}\
  }\href@noop {} {\bibfield  {journal} {\bibinfo  {journal} {Philosophical
  Transactions for the Royal Society of London. Series A, Mathematical and
  Physical Sciences}\ ,\ \bibinfo {pages} {321--342}} (\bibinfo {year}
  {1969})}\BibitemShut {NoStop}%
\bibitem [{\citenamefont {Goldstein}(1976)}]{goldstein1976aeroacoustics}%
  \BibitemOpen
  \bibfield  {author} {\bibinfo {author} {\bibnamefont {Goldstein},
  \bibfnamefont {M.~E.}},\ }\href@noop {} {\emph {\bibinfo {title}
  {Aeroacoustics}}}\ (\bibinfo  {publisher} {McGraw-Hill Inc.,US},\ \bibinfo
  {year} {1976})\BibitemShut {NoStop}%
\bibitem [{\citenamefont {Goody}(2004)}]{goody2004empirical}%
  \BibitemOpen
  \bibfield  {author} {\bibinfo {author} {\bibnamefont {Goody}, \bibfnamefont
  {M.}},\ }\bibfield  {title} {\enquote {\bibinfo {title} {Empirical spectral
  model of surface pressure fluctuations},}\ }\href@noop {} {\bibfield
  {journal} {\bibinfo  {journal} {AIAA journal}\ }\textbf {\bibinfo {volume}
  {42}},\ \bibinfo {pages} {1788--1794} (\bibinfo {year} {2004})}\BibitemShut
  {NoStop}%
\bibitem [{\citenamefont {Grasso}\ \emph {et~al.}(2019)\citenamefont {Grasso},
  \citenamefont {Jaiswal}, \citenamefont {Wu}, \citenamefont {Moreau},\ and\
  \citenamefont {Roger}}]{grasso2019analytical}%
  \BibitemOpen
  \bibfield  {author} {\bibinfo {author} {\bibnamefont {Grasso}, \bibfnamefont
  {G.}}, \bibinfo {author} {\bibnamefont {Jaiswal}, \bibfnamefont {P.}},
  \bibinfo {author} {\bibnamefont {Wu}, \bibfnamefont {H.}}, \bibinfo {author}
  {\bibnamefont {Moreau}, \bibfnamefont {S.}}, \ and\ \bibinfo {author}
  {\bibnamefont {Roger}, \bibfnamefont {M.}},\ }\bibfield  {title} {\enquote
  {\bibinfo {title} {Analytical models of the wall-pressure spectrum under a
  turbulent boundary layer with adverse pressure gradient},}\ }\href@noop {}
  {\bibfield  {journal} {\bibinfo  {journal} {Journal of Fluid Mechanics}\
  }\textbf {\bibinfo {volume} {877}},\ \bibinfo {pages} {1007--1062} (\bibinfo
  {year} {2019})}\BibitemShut {NoStop}%
\bibitem [{\citenamefont {Henning}\ \emph {et~al.}(2008)\citenamefont
  {Henning}, \citenamefont {Kaepernick}, \citenamefont {Ehrenfried},
  \citenamefont {Koop},\ and\ \citenamefont
  {Dillmann}}]{henning2008investigation}%
  \BibitemOpen
  \bibfield  {author} {\bibinfo {author} {\bibnamefont {Henning}, \bibfnamefont
  {A.}}, \bibinfo {author} {\bibnamefont {Kaepernick}, \bibfnamefont {K.}},
  \bibinfo {author} {\bibnamefont {Ehrenfried}, \bibfnamefont {K.}}, \bibinfo
  {author} {\bibnamefont {Koop}, \bibfnamefont {L.}}, \ and\ \bibinfo {author}
  {\bibnamefont {Dillmann}, \bibfnamefont {A.}},\ }\bibfield  {title} {\enquote
  {\bibinfo {title} {Investigation of aeroacoustic noise generation by
  simultaneous particle image velocimetry and microphone measurements},}\
  }\href@noop {} {\bibfield  {journal} {\bibinfo  {journal} {Experiments in
  fluids}\ }\textbf {\bibinfo {volume} {45}},\ \bibinfo {pages} {1073--1085}
  (\bibinfo {year} {2008})}\BibitemShut {NoStop}%
\bibitem [{\citenamefont {Holmes}\ \emph {et~al.}(2012)\citenamefont {Holmes},
  \citenamefont {Lumley}, \citenamefont {Berkooz},\ and\ \citenamefont
  {Rowley}}]{holmes2012turbulence}%
  \BibitemOpen
  \bibfield  {author} {\bibinfo {author} {\bibnamefont {Holmes}, \bibfnamefont
  {P.}}, \bibinfo {author} {\bibnamefont {Lumley}, \bibfnamefont {J.~L.}},
  \bibinfo {author} {\bibnamefont {Berkooz}, \bibfnamefont {G.}}, \ and\
  \bibinfo {author} {\bibnamefont {Rowley}, \bibfnamefont {C.~W.}},\
  }\href@noop {} {\emph {\bibinfo {title} {Turbulence, coherent structures,
  dynamical systems and symmetry}}}\ (\bibinfo  {publisher} {Cambridge
  university press},\ \bibinfo {year} {2012})\BibitemShut {NoStop}%
\bibitem [{\citenamefont {Huang}\ and\ \citenamefont
  {Ho}(1990)}]{huang1990small}%
  \BibitemOpen
  \bibfield  {author} {\bibinfo {author} {\bibnamefont {Huang}, \bibfnamefont
  {L.-S.}}\ and\ \bibinfo {author} {\bibnamefont {Ho}, \bibfnamefont {C.-M.}},\
  }\bibfield  {title} {\enquote {\bibinfo {title} {Small-scale transition in a
  plane mixing layer},}\ }\href@noop {} {\bibfield  {journal} {\bibinfo
  {journal} {Journal of Fluid Mechanics}\ }\textbf {\bibinfo {volume} {210}},\
  \bibinfo {pages} {475--500} (\bibinfo {year} {1990})}\BibitemShut {NoStop}%
\bibitem [{\citenamefont {Jaiswal}(2020)}]{Prateek:diss}%
  \BibitemOpen
  \bibfield  {author} {\bibinfo {author} {\bibnamefont {Jaiswal}, \bibfnamefont
  {P.}},\ }\emph {\bibinfo {title} {"Etude exp{\'e}rimentale du bruit propre de
  profil aérodynamique"}},\ \href@noop {} {Ph.D. thesis},\ \bibinfo  {school}
  {University of Sherbrooke}, \bibinfo {address}
  {"http://hdl.handle.net/11143/17867"} (\bibinfo {year} {2020})\BibitemShut
  {NoStop}%
\bibitem [{\citenamefont {Jaiswal}\ \emph {et~al.}(2020)\citenamefont
  {Jaiswal}, \citenamefont {Moreau}, \citenamefont {Avallone}, \citenamefont
  {Ragni},\ and\ \citenamefont {Pr{\"o}bsting}}]{jaiswal2020use}%
  \BibitemOpen
  \bibfield  {author} {\bibinfo {author} {\bibnamefont {Jaiswal}, \bibfnamefont
  {P.}}, \bibinfo {author} {\bibnamefont {Moreau}, \bibfnamefont {S.}},
  \bibinfo {author} {\bibnamefont {Avallone}, \bibfnamefont {F.}}, \bibinfo
  {author} {\bibnamefont {Ragni}, \bibfnamefont {D.}}, \ and\ \bibinfo {author}
  {\bibnamefont {Pr{\"o}bsting}, \bibfnamefont {S.}},\ }\bibfield  {title}
  {\enquote {\bibinfo {title} {On the use of two-point velocity correlation in
  wall-pressure models for turbulent flow past a trailing edge under adverse
  pressure gradient},}\ }\href@noop {} {\bibfield  {journal} {\bibinfo
  {journal} {Physics of Fluids}\ }\textbf {\bibinfo {volume} {32}},\ \bibinfo
  {pages} {105105} (\bibinfo {year} {2020})}\BibitemShut {NoStop}%
\bibitem [{\citenamefont {Jaiswal}\ \emph {et~al.}(2022)\citenamefont
  {Jaiswal}, \citenamefont {Pasco}, \citenamefont {Yakhina},\ and\
  \citenamefont {Moreau}}]{jaiswal2022experimental}%
  \BibitemOpen
  \bibfield  {author} {\bibinfo {author} {\bibnamefont {Jaiswal}, \bibfnamefont
  {P.}}, \bibinfo {author} {\bibnamefont {Pasco}, \bibfnamefont {Y.}}, \bibinfo
  {author} {\bibnamefont {Yakhina}, \bibfnamefont {G.}}, \ and\ \bibinfo
  {author} {\bibnamefont {Moreau}, \bibfnamefont {S.}},\ }\bibfield  {title}
  {\enquote {\bibinfo {title} {Experimental investigation of aerofoil tonal
  noise at low mach number},}\ }\href@noop {} {\bibfield  {journal} {\bibinfo
  {journal} {Journal of Fluid Mechanics}\ }\textbf {\bibinfo {volume} {932}},\
  \bibinfo {pages} {A37} (\bibinfo {year} {2022})}\BibitemShut {NoStop}%
\bibitem [{\citenamefont {Kalyani}, \citenamefont {Moreau},\ and\ \citenamefont
  {Ragni}(2022)}]{kalyani2022flow}%
  \BibitemOpen
  \bibfield  {author} {\bibinfo {author} {\bibnamefont {Kalyani}, \bibfnamefont
  {S.~K.}}, \bibinfo {author} {\bibnamefont {Moreau}, \bibfnamefont {S.}}, \
  and\ \bibinfo {author} {\bibnamefont {Ragni}, \bibfnamefont {D.}},\
  }\bibfield  {title} {\enquote {\bibinfo {title} {Flow-field and noise
  characterization of a controlled-diffusion airfoil},}\ }\href@noop {}
  {\bibfield  {journal} {\bibinfo  {journal} {28th AIAA/CEAS Aeroacoustics 2022
  Conference. Paper number 2894.}\ } (\bibinfo {year} {2022})}\BibitemShut
  {NoStop}%
\bibitem [{\citenamefont {Lacagnina}\ \emph {et~al.}(2019)\citenamefont
  {Lacagnina}, \citenamefont {Chaitanya}, \citenamefont {Berk}, \citenamefont
  {Kim}, \citenamefont {Joseph}, \citenamefont {Ganapathisubramani},
  \citenamefont {Hasheminejad}, \citenamefont {Chong}, \citenamefont {Stalnov},
  \citenamefont {Choi} \emph {et~al.}}]{lacagnina2019mechanisms}%
  \BibitemOpen
  \bibfield  {author} {\bibinfo {author} {\bibnamefont {Lacagnina},
  \bibfnamefont {G.}}, \bibinfo {author} {\bibnamefont {Chaitanya},
  \bibfnamefont {P.}}, \bibinfo {author} {\bibnamefont {Berk}, \bibfnamefont
  {T.}}, \bibinfo {author} {\bibnamefont {Kim}, \bibfnamefont {J.-H.}},
  \bibinfo {author} {\bibnamefont {Joseph}, \bibfnamefont {P.}}, \bibinfo
  {author} {\bibnamefont {Ganapathisubramani}, \bibfnamefont {B.}}, \bibinfo
  {author} {\bibnamefont {Hasheminejad}, \bibfnamefont {S.~M.}}, \bibinfo
  {author} {\bibnamefont {Chong}, \bibfnamefont {T.~P.}}, \bibinfo {author}
  {\bibnamefont {Stalnov}, \bibfnamefont {O.}}, \bibinfo {author} {\bibnamefont
  {Choi}, \bibfnamefont {K.-S.}},  \emph {et~al.},\ }\bibfield  {title}
  {\enquote {\bibinfo {title} {Mechanisms of airfoil noise near stall
  conditions},}\ }\href@noop {} {\bibfield  {journal} {\bibinfo  {journal}
  {Physical Review Fluids}\ }\textbf {\bibinfo {volume} {4}},\ \bibinfo {pages}
  {123902} (\bibinfo {year} {2019})}\BibitemShut {NoStop}%
\bibitem [{\citenamefont {Le~Floc'h}\ \emph {et~al.}(2020)\citenamefont
  {Le~Floc'h}, \citenamefont {Weiss}, \citenamefont {Mohammed-Taifour},\ and\
  \citenamefont {Dufresne}}]{le2020measurements}%
  \BibitemOpen
  \bibfield  {author} {\bibinfo {author} {\bibnamefont {Le~Floc'h},
  \bibfnamefont {A.}}, \bibinfo {author} {\bibnamefont {Weiss}, \bibfnamefont
  {J.}}, \bibinfo {author} {\bibnamefont {Mohammed-Taifour}, \bibfnamefont
  {A.}}, \ and\ \bibinfo {author} {\bibnamefont {Dufresne}, \bibfnamefont
  {L.}},\ }\bibfield  {title} {\enquote {\bibinfo {title} {Measurements of
  pressure and velocity fluctuations in a family of turbulent separation
  bubbles},}\ }\href@noop {} {\bibfield  {journal} {\bibinfo  {journal}
  {Journal of Fluid Mechanics}\ }\textbf {\bibinfo {volume} {902}},\ \bibinfo
  {pages} {A13} (\bibinfo {year} {2020})}\BibitemShut {NoStop}%
\bibitem [{\citenamefont {Ligrani}\ and\ \citenamefont
  {Bradshaw}(1987)}]{ligrani1987}%
  \BibitemOpen
  \bibfield  {author} {\bibinfo {author} {\bibnamefont {Ligrani}, \bibfnamefont
  {P.~M.}}\ and\ \bibinfo {author} {\bibnamefont {Bradshaw}, \bibfnamefont
  {P.}},\ }\bibfield  {title} {\enquote {\bibinfo {title} {{S}ubminiature
  hot-wire sensors: development and use},}\ }\href@noop {} {\bibfield
  {journal} {\bibinfo  {journal} {J. Phys. E: Sci. Instrum.}\ }\textbf
  {\bibinfo {volume} {20}},\ \bibinfo {pages} {323--332} (\bibinfo {year}
  {1987})}\BibitemShut {NoStop}%
\bibitem [{\citenamefont {Moreau}, \citenamefont {Christophe},\ and\
  \citenamefont {Roger}(2008)}]{moreau2008trailing}%
  \BibitemOpen
  \bibfield  {author} {\bibinfo {author} {\bibnamefont {Moreau}, \bibfnamefont
  {S.}}, \bibinfo {author} {\bibnamefont {Christophe}, \bibfnamefont {J.}}, \
  and\ \bibinfo {author} {\bibnamefont {Roger}, \bibfnamefont {M.}},\
  }\bibfield  {title} {\enquote {\bibinfo {title} {{LES} of the trailing-edge
  flow and noise of a {NACA}~0012 airfoil near stall},}\ }in\ \href@noop {}
  {\emph {\bibinfo {booktitle} {Proceedings of the Summer Program}}}\ (\bibinfo
  {organization} {Stanford University, Center for Turbulence Research Stanford,
  CA, USA},\ \bibinfo {year} {2008})\ pp.\ \bibinfo {pages}
  {317--329}\BibitemShut {NoStop}%
\bibitem [{\citenamefont {Moreau}\ and\ \citenamefont
  {Roger}(2005)}]{moreau2005effect}%
  \BibitemOpen
  \bibfield  {author} {\bibinfo {author} {\bibnamefont {Moreau}, \bibfnamefont
  {S.}}\ and\ \bibinfo {author} {\bibnamefont {Roger}, \bibfnamefont {M.}},\
  }\bibfield  {title} {\enquote {\bibinfo {title} {Effect of airfoil
  aerodynamic loading on trailing edge noise sources},}\ }\href@noop {}
  {\bibfield  {journal} {\bibinfo  {journal} {AIAA journal}\ }\textbf {\bibinfo
  {volume} {43}},\ \bibinfo {pages} {41--52} (\bibinfo {year}
  {2005})}\BibitemShut {NoStop}%
\bibitem [{\citenamefont {Moreau}\ and\ \citenamefont
  {Roger}(2009)}]{moreau2009back}%
  \BibitemOpen
  \bibfield  {author} {\bibinfo {author} {\bibnamefont {Moreau}, \bibfnamefont
  {S.}}\ and\ \bibinfo {author} {\bibnamefont {Roger}, \bibfnamefont {M.}},\
  }\bibfield  {title} {\enquote {\bibinfo {title} {Back-scattering correction
  and further extensions of amiet's trailing-edge noise model. part ii:
  Application},}\ }\href@noop {} {\bibfield  {journal} {\bibinfo  {journal}
  {Journal of Sound and vibration}\ }\textbf {\bibinfo {volume} {323}},\
  \bibinfo {pages} {397--425} (\bibinfo {year} {2009})}\BibitemShut {NoStop}%
\bibitem [{\citenamefont {Moreau}, \citenamefont {Roger},\ and\ \citenamefont
  {Christophe}(2009)}]{stall}%
  \BibitemOpen
  \bibfield  {author} {\bibinfo {author} {\bibnamefont {Moreau}, \bibfnamefont
  {S.}}, \bibinfo {author} {\bibnamefont {Roger}, \bibfnamefont {M.}}, \ and\
  \bibinfo {author} {\bibnamefont {Christophe}, \bibfnamefont {J.}},\
  }\bibfield  {title} {\enquote {\bibinfo {title} {Flow features and self-noise
  of airfoils near stall or in stall},}\ }\href@noop {} {\bibfield  {journal}
  {\bibinfo  {journal} {15th AIAA/CEAS Aeroacoustics Conference (30th AIAA
  Aeroacoustics Conference), paper number 2009-3198}\ } (\bibinfo {year}
  {2009})}\BibitemShut {NoStop}%
\bibitem [{\citenamefont {Na}\ and\ \citenamefont
  {Moin}(1998)}]{na1998structure}%
  \BibitemOpen
  \bibfield  {author} {\bibinfo {author} {\bibnamefont {Na}, \bibfnamefont
  {Y.}}\ and\ \bibinfo {author} {\bibnamefont {Moin}, \bibfnamefont {P.}},\
  }\bibfield  {title} {\enquote {\bibinfo {title} {The structure of
  wall-pressure fluctuations in turbulent boundary layers with adverse pressure
  gradient and separation},}\ }\href@noop {} {\bibfield  {journal} {\bibinfo
  {journal} {Journal of Fluid Mechanics}\ }\textbf {\bibinfo {volume} {377}},\
  \bibinfo {pages} {347--373} (\bibinfo {year} {1998})}\BibitemShut {NoStop}%
\bibitem [{\citenamefont {Neal}(2010)}]{Neal:diss}%
  \BibitemOpen
  \bibfield  {author} {\bibinfo {author} {\bibnamefont {Neal}, \bibfnamefont
  {D.}},\ }\emph {\bibinfo {title} {{Effects of rotation on the flow field over
  a Controlled-Diffusion airfoil}}},\ \href@noop {} {Ph.D. thesis},\ \bibinfo
  {school} {Michigan State University}, \bibinfo {address} {East Lansing, MI}
  (\bibinfo {year} {2010})\BibitemShut {NoStop}%
\bibitem [{\citenamefont {Pargal}(2023)}]{pargal2023non}%
  \BibitemOpen
  \bibfield  {author} {\bibinfo {author} {\bibnamefont {Pargal}, \bibfnamefont
  {S.}},\ }\emph {\bibinfo {title} {Non-equilibrium wall-bounded turbulence and
  associated noise generation}},\ \href@noop {} {Ph.D. thesis},\ \bibinfo
  {school} {University of Sherbrooke Quebec, Canada} (\bibinfo {year}
  {2023})\BibitemShut {NoStop}%
\bibitem [{\citenamefont {Perennes}\ and\ \citenamefont
  {Roger}(1998)}]{perennes_rmp}%
  \BibitemOpen
  \bibfield  {author} {\bibinfo {author} {\bibnamefont {Perennes},
  \bibfnamefont {S.}}\ and\ \bibinfo {author} {\bibnamefont {Roger},
  \bibfnamefont {M.}},\ }\bibfield  {title} {\enquote {\bibinfo {title}
  {Aerodynamic noise of a two-dimensional wing with high-lift devices},}\ }in\
  \href@noop {} {\emph {\bibinfo {booktitle} {4th AIAA/CEAS aeroacoustics
  conference}}}\ (\bibinfo {address} {Toulouse,France},\ \bibinfo {year}
  {1998})\ p.\ \bibinfo {pages} {2338}\BibitemShut {NoStop}%
\bibitem [{\citenamefont {Raus}\ \emph {et~al.}(2022)\citenamefont {Raus},
  \citenamefont {Cott{\'e}}, \citenamefont {Monchaux}, \citenamefont {Jondeau},
  \citenamefont {Souchotte},\ and\ \citenamefont {Roger}}]{Raus2022}%
  \BibitemOpen
  \bibfield  {author} {\bibinfo {author} {\bibnamefont {Raus}, \bibfnamefont
  {D.}}, \bibinfo {author} {\bibnamefont {Cott{\'e}}, \bibfnamefont {B.}},
  \bibinfo {author} {\bibnamefont {Monchaux}, \bibfnamefont {R.}}, \bibinfo
  {author} {\bibnamefont {Jondeau}, \bibfnamefont {E.}}, \bibinfo {author}
  {\bibnamefont {Souchotte}, \bibfnamefont {S.}}, \ and\ \bibinfo {author}
  {\bibnamefont {Roger}, \bibfnamefont {M.}},\ }\bibfield  {title} {\enquote
  {\bibinfo {title} {Experimental study of the dynamic stall noise on an
  oscillating airfoil},}\ }\href@noop {} {\bibfield  {journal} {\bibinfo
  {journal} {Journal of Sound and Vibration}\ }\textbf {\bibinfo {volume}
  {537}},\ \bibinfo {pages} {117144} (\bibinfo {year} {2022})}\BibitemShut
  {NoStop}%
\bibitem [{\citenamefont {Roger}\ and\ \citenamefont
  {Moreau}(2004)}]{roger2004broadband}%
  \BibitemOpen
  \bibfield  {author} {\bibinfo {author} {\bibnamefont {Roger}, \bibfnamefont
  {M.}}\ and\ \bibinfo {author} {\bibnamefont {Moreau}, \bibfnamefont {S.}},\
  }\bibfield  {title} {\enquote {\bibinfo {title} {Broadband self noise from
  loaded fan blades},}\ }\href@noop {} {\bibfield  {journal} {\bibinfo
  {journal} {AIAA journal}\ }\textbf {\bibinfo {volume} {42}},\ \bibinfo
  {pages} {536--544} (\bibinfo {year} {2004})}\BibitemShut {NoStop}%
\bibitem [{\citenamefont {Roger}\ and\ \citenamefont
  {Moreau}(2005)}]{roger_moreau}%
  \BibitemOpen
  \bibfield  {author} {\bibinfo {author} {\bibnamefont {Roger}, \bibfnamefont
  {M.}}\ and\ \bibinfo {author} {\bibnamefont {Moreau}, \bibfnamefont {S.}},\
  }\bibfield  {title} {\enquote {\bibinfo {title} {{Back-scattering correction
  and further extensions of Amiet's trailing-edge noise model. Part 1:
  theory}},}\ }\href@noop {} {\bibfield  {journal} {\bibinfo  {journal}
  {Journal of Sound and vibration}\ }\textbf {\bibinfo {volume} {286}},\
  \bibinfo {pages} {477--506} (\bibinfo {year} {2005})}\BibitemShut {NoStop}%
\bibitem [{\citenamefont {Roger}\ and\ \citenamefont
  {Moreau}(2012)}]{Roger2012}%
  \BibitemOpen
  \bibfield  {author} {\bibinfo {author} {\bibnamefont {Roger}, \bibfnamefont
  {M.}}\ and\ \bibinfo {author} {\bibnamefont {Moreau}, \bibfnamefont {S.}},\
  }\bibfield  {title} {\enquote {\bibinfo {title} {{Addendum to the
  back-scattering correction of Amiet's trailing-edge noise model}},}\ }\href
  {\doibase 10.1016/j.jsv.2012.06.019} {\bibfield  {journal} {\bibinfo
  {journal} {Journal of Sound and vibration}\ }\textbf {\bibinfo {volume}
  {331}},\ \bibinfo {pages} {5383--5385} (\bibinfo {year} {2012})}\BibitemShut
  {NoStop}%
\bibitem [{\citenamefont {Sanjose}\ \emph {et~al.}(2019)\citenamefont
  {Sanjose}, \citenamefont {Towne}, \citenamefont {Jaiswal}, \citenamefont
  {Moreau}, \citenamefont {Lele},\ and\ \citenamefont
  {Mann}}]{sanjose2019modal}%
  \BibitemOpen
  \bibfield  {author} {\bibinfo {author} {\bibnamefont {Sanjose}, \bibfnamefont
  {M.}}, \bibinfo {author} {\bibnamefont {Towne}, \bibfnamefont {A.}}, \bibinfo
  {author} {\bibnamefont {Jaiswal}, \bibfnamefont {P.}}, \bibinfo {author}
  {\bibnamefont {Moreau}, \bibfnamefont {S.}}, \bibinfo {author} {\bibnamefont
  {Lele}, \bibfnamefont {S.}}, \ and\ \bibinfo {author} {\bibnamefont {Mann},
  \bibfnamefont {A.}},\ }\bibfield  {title} {\enquote {\bibinfo {title} {Modal
  analysis of the laminar boundary layer instability and tonal noise of an
  airfoil at {R}eynolds number 150,000},}\ }\href@noop {} {\bibfield  {journal}
  {\bibinfo  {journal} {International Journal of Aeroacoustics}\ }\textbf
  {\bibinfo {volume} {18}},\ \bibinfo {pages} {317--350} (\bibinfo {year}
  {2019})}\BibitemShut {NoStop}%
\bibitem [{\citenamefont {Schloemer}(1967)}]{schloemer1967effects}%
  \BibitemOpen
  \bibfield  {author} {\bibinfo {author} {\bibnamefont {Schloemer},
  \bibfnamefont {H.~H.}},\ }\bibfield  {title} {\enquote {\bibinfo {title}
  {Effects of pressure gradients on turbulent-boundary-layer wall-pressure
  fluctuations},}\ }\href@noop {} {\bibfield  {journal} {\bibinfo  {journal}
  {The journal of the acoustical society of America}\ }\textbf {\bibinfo
  {volume} {42}},\ \bibinfo {pages} {93--113} (\bibinfo {year}
  {1967})}\BibitemShut {NoStop}%
\bibitem [{\citenamefont {Sirovich}(1987)}]{sirovich1987turbulence}%
  \BibitemOpen
  \bibfield  {author} {\bibinfo {author} {\bibnamefont {Sirovich},
  \bibfnamefont {L.}},\ }\bibfield  {title} {\enquote {\bibinfo {title}
  {Turbulence and the dynamics of coherent structures. i. coherent
  structures},}\ }\href@noop {} {\bibfield  {journal} {\bibinfo  {journal}
  {Quarterly of applied mathematics}\ }\textbf {\bibinfo {volume} {45}},\
  \bibinfo {pages} {561--571} (\bibinfo {year} {1987})}\BibitemShut {NoStop}%
\bibitem [{\citenamefont {Spalart}\ \emph {et~al.}(2019)\citenamefont
  {Spalart}, \citenamefont {Belyaev}, \citenamefont {Shur}, \citenamefont
  {Kh~Strelets},\ and\ \citenamefont {Travin}}]{spalart2019differences}%
  \BibitemOpen
  \bibfield  {author} {\bibinfo {author} {\bibnamefont {Spalart}, \bibfnamefont
  {P.~R.}}, \bibinfo {author} {\bibnamefont {Belyaev}, \bibfnamefont {K.~V.}},
  \bibinfo {author} {\bibnamefont {Shur}, \bibfnamefont {M.~L.}}, \bibinfo
  {author} {\bibnamefont {Kh~Strelets}, \bibfnamefont {M.}}, \ and\ \bibinfo
  {author} {\bibnamefont {Travin}, \bibfnamefont {A.~K.}},\ }\bibfield  {title}
  {\enquote {\bibinfo {title} {On the differences in noise predictions based on
  solid and permeable surface ffowcs williams--hawkings integral solutions},}\
  }\href@noop {} {\bibfield  {journal} {\bibinfo  {journal} {International
  Journal of Aeroacoustics}\ }\textbf {\bibinfo {volume} {18}},\ \bibinfo
  {pages} {621--646} (\bibinfo {year} {2019})}\BibitemShut {NoStop}%
\bibitem [{\citenamefont {Turner}\ and\ \citenamefont
  {Kim}(2022)}]{turner2022quadrupole}%
  \BibitemOpen
  \bibfield  {author} {\bibinfo {author} {\bibnamefont {Turner}, \bibfnamefont
  {J.~M.}}\ and\ \bibinfo {author} {\bibnamefont {Kim}, \bibfnamefont
  {J.~W.}},\ }\bibfield  {title} {\enquote {\bibinfo {title} {Quadrupole noise
  generated from a low-speed aerofoil in near-and full-stall conditions},}\
  }\href@noop {} {\bibfield  {journal} {\bibinfo  {journal} {Journal of Fluid
  Mechanics}\ }\textbf {\bibinfo {volume} {936}},\ \bibinfo {pages} {A34}
  (\bibinfo {year} {2022})}\BibitemShut {NoStop}%
\bibitem [{\citenamefont {Venkatraman}\ \emph {et~al.}(2023)\citenamefont
  {Venkatraman}, \citenamefont {Moreau}, \citenamefont {Christophe},\ and\
  \citenamefont {Schram}}]{venkatraman2023numerical}%
  \BibitemOpen
  \bibfield  {author} {\bibinfo {author} {\bibnamefont {Venkatraman},
  \bibfnamefont {K.}}, \bibinfo {author} {\bibnamefont {Moreau}, \bibfnamefont
  {S.}}, \bibinfo {author} {\bibnamefont {Christophe}, \bibfnamefont {J.}}, \
  and\ \bibinfo {author} {\bibnamefont {Schram}, \bibfnamefont {C.}},\
  }\bibfield  {title} {\enquote {\bibinfo {title} {Numerical investigation of
  h-darrieus wind turbine aerodynamics at different tip speed ratios},}\
  }\href@noop {} {\bibfield  {journal} {\bibinfo  {journal} {International
  Journal of Numerical Methods for Heat \& Fluid Flow}\ } (\bibinfo {year}
  {2023})}\BibitemShut {NoStop}%
\bibitem [{\citenamefont {Watmuff}(1999)}]{watmuff1999evolution}%
  \BibitemOpen
  \bibfield  {author} {\bibinfo {author} {\bibnamefont {Watmuff}, \bibfnamefont
  {J.~H.}},\ }\bibfield  {title} {\enquote {\bibinfo {title} {Evolution of a
  wave packet into vortex loops in a laminar separation bubble},}\ }\href@noop
  {} {\bibfield  {journal} {\bibinfo  {journal} {Journal of Fluid Mechanics}\
  }\textbf {\bibinfo {volume} {397}},\ \bibinfo {pages} {119--169} (\bibinfo
  {year} {1999})}\BibitemShut {NoStop}%
\bibitem [{\citenamefont {Wu}\ \emph {et~al.}(2016)\citenamefont {Wu},
  \citenamefont {Laffay}, \citenamefont {Idier}, \citenamefont {Jaiswal},
  \citenamefont {Sanjos{\'e}},\ and\ \citenamefont {Moreau}}]{wu2016numerical}%
  \BibitemOpen
  \bibfield  {author} {\bibinfo {author} {\bibnamefont {Wu}, \bibfnamefont
  {H.}}, \bibinfo {author} {\bibnamefont {Laffay}, \bibfnamefont {P.}},
  \bibinfo {author} {\bibnamefont {Idier}, \bibfnamefont {A.}}, \bibinfo
  {author} {\bibnamefont {Jaiswal}, \bibfnamefont {P.}}, \bibinfo {author}
  {\bibnamefont {Sanjos{\'e}}, \bibfnamefont {M.}}, \ and\ \bibinfo {author}
  {\bibnamefont {Moreau}, \bibfnamefont {S.}},\ }\bibfield  {title} {\enquote
  {\bibinfo {title} {Numerical study of the installed controlled diffusion
  airfoil at transitional {R}eynolds number},}\ }in\ \href@noop {} {\emph
  {\bibinfo {booktitle} {Mathematical and Computational Approaches in Advancing
  Modern Science and Engineering}}}\ (\bibinfo {organization} {Springer},\
  \bibinfo {year} {2016})\ pp.\ \bibinfo {pages} {505--515}\BibitemShut
  {NoStop}%
\bibitem [{\citenamefont {Wu}, \citenamefont {Moreau},\ and\ \citenamefont
  {Sandberg}(2019)}]{Wu2019}%
  \BibitemOpen
  \bibfield  {author} {\bibinfo {author} {\bibnamefont {Wu}, \bibfnamefont
  {H.}}, \bibinfo {author} {\bibnamefont {Moreau}, \bibfnamefont {S.}}, \ and\
  \bibinfo {author} {\bibnamefont {Sandberg}, \bibfnamefont {R.}},\ }\bibfield
  {title} {\enquote {\bibinfo {title} {{Effects of pressure gradient on the
  evolution of velocity-gradient tensor invariant dynamics on a
  controlled-diffusion aerofoil at $Re_c=150000$}},}\ }\href@noop {} {\bibfield
   {journal} {\bibinfo  {journal} {Journal of Fluid Mechanics}\ }\textbf
  {\bibinfo {volume} {868}},\ \bibinfo {pages} {584--610} (\bibinfo {year}
  {2019})}\BibitemShut {NoStop}%
\bibitem [{\citenamefont {Wu}, \citenamefont {Moreau},\ and\ \citenamefont
  {Sandberg}(2020)}]{wu2020noise}%
  \BibitemOpen
  \bibfield  {author} {\bibinfo {author} {\bibnamefont {Wu}, \bibfnamefont
  {H.}}, \bibinfo {author} {\bibnamefont {Moreau}, \bibfnamefont {S.}}, \ and\
  \bibinfo {author} {\bibnamefont {Sandberg}, \bibfnamefont {R.~D.}},\
  }\bibfield  {title} {\enquote {\bibinfo {title} {On the noise generated by a
  controlled-diffusion aerofoil at {Re}$_c~= 1.5\times 10^5$},}\ }\href@noop {}
  {\bibfield  {journal} {\bibinfo  {journal} {Journal of Sound and Vibration}\
  }\textbf {\bibinfo {volume} {487}},\ \bibinfo {pages} {115620} (\bibinfo
  {year} {2020})}\BibitemShut {NoStop}%
\bibitem [{\citenamefont {Yarusevych}, \citenamefont {Sullivan},\ and\
  \citenamefont {Kawall}(2006)}]{yarusevych2006coherent}%
  \BibitemOpen
  \bibfield  {author} {\bibinfo {author} {\bibnamefont {Yarusevych},
  \bibfnamefont {S.}}, \bibinfo {author} {\bibnamefont {Sullivan},
  \bibfnamefont {P.~E.}}, \ and\ \bibinfo {author} {\bibnamefont {Kawall},
  \bibfnamefont {J.~G.}},\ }\bibfield  {title} {\enquote {\bibinfo {title}
  {Coherent structures in an airfoil boundary layer and wake at low {R}eynolds
  numbers},}\ }\href@noop {} {\bibfield  {journal} {\bibinfo  {journal}
  {Physics of Fluids}\ }\textbf {\bibinfo {volume} {18}},\ \bibinfo {pages}
  {044101} (\bibinfo {year} {2006})}\BibitemShut {NoStop}%
\bibitem [{\citenamefont {Yarusevych}, \citenamefont {Sullivan},\ and\
  \citenamefont {Kawall}(2009)}]{yarusevych2009vortex}%
  \BibitemOpen
  \bibfield  {author} {\bibinfo {author} {\bibnamefont {Yarusevych},
  \bibfnamefont {S.}}, \bibinfo {author} {\bibnamefont {Sullivan},
  \bibfnamefont {P.~E.}}, \ and\ \bibinfo {author} {\bibnamefont {Kawall},
  \bibfnamefont {J.~G.}},\ }\bibfield  {title} {\enquote {\bibinfo {title} {On
  vortex shedding from an airfoil in low-{R}eynolds-number flows},}\
  }\href@noop {} {\bibfield  {journal} {\bibinfo  {journal} {Journal of Fluid
  Mechanics}\ }\textbf {\bibinfo {volume} {632}},\ \bibinfo {pages} {245--271}
  (\bibinfo {year} {2009})}\BibitemShut {NoStop}%
\bibitem [{\citenamefont {Zang}, \citenamefont {Mayer},\ and\ \citenamefont
  {Azarpeyvand}(2021)}]{zang2021experimental}%
  \BibitemOpen
  \bibfield  {author} {\bibinfo {author} {\bibnamefont {Zang}, \bibfnamefont
  {B.}}, \bibinfo {author} {\bibnamefont {Mayer}, \bibfnamefont {Y.~D.}}, \
  and\ \bibinfo {author} {\bibnamefont {Azarpeyvand}, \bibfnamefont {M.}},\
  }\bibfield  {title} {\enquote {\bibinfo {title} {Experimental investigation
  of near-field aeroacoustic characteristics of a pre-and post-stall naca
  65-410 airfoil},}\ }\href@noop {} {\bibfield  {journal} {\bibinfo  {journal}
  {Journal of Aerospace Engineering}\ }\textbf {\bibinfo {volume} {34}},\
  \bibinfo {pages} {04021080} (\bibinfo {year} {2021})}\BibitemShut {NoStop}%
\end{thebibliography}%

\end{document}